\documentclass[a4paper,10pt,reqno]{amsart}
\newtheorem{theorem}{\bf Theorem}[section]

\usepackage{xcolor}
\usepackage[colorlinks=true,linkcolor=blue,citecolor=orange]{hyperref}
\usepackage[left=1.4in, right=1.4in, top=1.4in, bottom=1.4in]{geometry}
\newtheorem{remark}[theorem]{Remark}
\usepackage{amsmath}
\usepackage{amsfonts}
\usepackage{calrsfs}
\usepackage{mathrsfs}
\usepackage{upgreek}
\usepackage{cancel}
\usepackage{chngcntr}
\numberwithin{equation}{section}
\usepackage{mathtools}
\usepackage{amssymb}
\usepackage{amsfonts}

\DeclareMathAlphabet{\mathpzc}{OT1}{pzc}{m}{it}
\renewcommand{\Re}[1]{\textrm{Re~}#1}
\renewcommand{\Im}[1]{\textrm{Im~}#1}

\begin{document}
\title{Scattering of surface waves by inhomogeneities in crystalline structures}
\author{Basant Lal Sharma}
\address{Department of Mechanical Engineering, Indian Institute of Technology Kanpur, Kanpur, 208016 UP, India}
\email[Corresponding author]{bls@iitk.ac.in}

\begin{abstract}
In current scientific and technological scenario, studies of transmittance of surface waves across structured interfaces have gained some wind
amidst applications to metasurfaces, electronic edge-waves, crystal grain boundaries, etc.
The results presented in the present article shed a light on the influence of material inhomogeneities on propagation of surface waves. 
Within the framework of classical mechanics, an analog of Gurtin-Murdoch model is employed where elastic properties on surface are assumed to be distinct from bulk.
Restricting to scalar waves on prototype square lattice half-plane, particles on considered structured surface have piecewise-constant mass and surface force-constants across an interfacial point.
Particles in bulk lattice interact with nearest neighbours in a way 
that involves unequal force-constants parallel to surface versus normal to it.
A surface wave band exists for such lattice structure wherein the waveform decays exponentially inside half-plane.
A formula for surface wave transmittance is given based on
an exact solution on half-plane,
and, thus, previous work of {\em Sharma \& Eremeyev 2019 Int. J. Eng. Sci. 143, 33--38} is extended.
An explicit expression for fraction of energy influx leaked via bulk waves is a highlight. 
Included are graphical results for several illustrative values of 
surface structure
parameters.

{\it subject: scattering theory, coatings, surface physics, phononics}\\
{\it keywords: lattice dynamics, surface interfaces, transmittance, Wiener-Hopf}

\end{abstract}
\maketitle

\section*{Introduction}
\label{secInt}
Researches in mechanics and physics at nano-scale are becoming increasingly relevant, and so is the importance of surfaces and interfaces, in the functional nanomaterials and miniaturized structures
\cite{Lieber,Schaefer,Duan,javili2013thermomechanics}. 
As a consequence of the developed understanding of these effects, a deluge of novel applications have come forth, such as 
nanowires, metasurfaces, nanoscale 
semiconductors, etc, which are 
confined in certain dimensions leading to an expanding and rich set of theoretical problems \cite{Chen,Holloway,Zheludev,Kadic,Besenbacher,Torun,Cahill}.

The investigations of surface effects in dynamics and equilibrium
go back to the times of Laplace and Gibbs, and still constitute the basis of multi-scale physics and many diverse fields in modern times. 
Surface waves, 
named after Rayleigh, Lamb, Love, and, in general, addressed as localized waves, continue to play a significant role across a wide spectrum of scales ranging from seismic to nano scales 
\cite{Tiersten,Freundsurfacewave,
Nieves20,Carta} even though formulated in linear elastodynamics. 
Besides the canonical scattering 
by an edge in continuous media \cite{Sommerfeld1}, and equivalent problems involving elastic, acoustic, and electromagnetic waves, the scattering associated with discontinuities or steps/serrations on crystal surfaces also gives rise to related challenges \cite{LuLagally,MartinJ,Zijlstra,Schiller}
under different types of environmental conditions.
In general, the identification and nature of surface wave modes is in itself an important question
and there is a long history of different approaches. 

In continuum models of surface elasticity, an independent set of constitutive relations on the surface are needed in addition to ones defined in the bulk such as 
the one proposed by \cite{GurtinMurdoch1975a} via an elastic membrane glued on its surface (also known as the Gurtin-Murdoch model). 
In case of linear waves, as known several decades before, within the special kinematics of antiplane shear motion, the presence of a single layer on half-space can allow a dispersive surface wave (Love wave) with velocity smaller than shear wave velocity of the bulk.
In the absence of any such layer, more structure near the free surface \cite{Balogun,Vardoulakis1997,Rosi} ensures the existence of such wave.
On the other hand, it is quite common to find surface wave bands in lattice models \cite{Brillouin,Maradudinbook} which incorporate more interactions between particles than just nearest neighbours and/or additional structure \cite{Wallis,Blstgsurf}.
Recently a comparison of the surface waves between the continuous framework of Gurtin-Murdoch model and the discrete framework of lattice dynamics was provided in \cite{Victor_Bls_surf1}. 
Aside these theoretical aspects, during manufacturing of actual physical surfaces at nano-scale for diverse applications \cite{Merlitz,Variola} it is likely that the surfaces and coatings created are not perfect and interfaces as well as surface steps are formed. On other hand, surfaces with such substructures
may be intentionally crafted to transport a signal in a certain way.
In the presence of surface wave excitation along the free surface the waves penetrate the half-plane through the interfaces and inhomogeneities.
In such realistic cases the problem of scattering of surface waves due to surface inhomogeneities becomes relevant. \cite{Victor_Bls_surf2} described one such situation involving two types of structures separated by an atomistically thin interface 
for the geometry of lattice strips.
In general, there are many physical applications of discrete models of wave propagation, such as pure crystals dynamics (simple, complex, etc.) in harmonic approximation \cite{Maradudinbook} and also in the presence of impurities \cite{Lifshitz1}, mechanics of dynamic fracture \cite{Slepyan1981b,Slepyanbook,Marder,Mishuris2}, vibrations of macro-molecules appearing in biological applications \cite{YuLeitner}, chemical applications \cite{Sandor}, as well as transport in phononic \cite{Deymier}, photonic \cite{Rechtsman}, electronic \cite{Schrieffer}, magnetic-spin \cite{Ashokan} systems.
Many times these apparently diverse situations involve governing equations which have mathematical structure similar to the problem tackled in the present article; for example, the discrete Schr\"{o}dinger equation appears from basic viewpoint in physics \cite{Anderson} and mathematics \cite{Eskina1,Shaban}.

A discrete model of scalar surface wave propagation in an elastic half-space, representing surface of a crystalline structure, is analyzed in this article. 
The investigations of \cite{Victor_Bls_surf2} 
are continued and
certain extensions of the analysis provided. 
The problem of scattering of 
surface wave due to 
interface, associated with piece-wise constant surface properties,
is solved in the backdrop of 
discrete scattering problems
involving atomically sharp crack tips and rigid constraints \cite{sK, sC, BlsMaurya3}.
The exact solution of lattice half-plane problem was deferred in \cite{Victor_Bls_surf2} and instead it was provided for a lattice strip (see the paragraph preceding Section 3 in \cite{Victor_Bls_surf2});
naturally, the same can be easily obtained as a special case of the solution presented.
The specific model adopted is a generalization of that in \cite{Victor_Bls_surf2} and incorporates an anisotropy due to unequal force-constants in horizontal (parallel to surface) versus vertical (normal to surface) directions; this is found to affect the nature of surface wave transmittance.
The anisotropy can arise due to geometrical aspects of the constructed lattice or alteration in its material response due to a tuned pre-stretch prior to wave propagation. 
Aside from the construction of exact solution,
the reflection and transmission coefficient for surface wave propagation, from one side of interface to another,
is obtained in a closed form. 
These expressions assist in providing a formula for, the physically important, fraction ${\mathscr{D}}_{{surf}}$ of incident energy flux that is `leaked' at the interface in the form of bulk waves. 
In the present article, the question about the dependence of ${\mathscr{D}}_{{surf}}$ on the surface material parameters is answered by finding a closed form expression of ${\mathscr{D}}_{{surf}}$. This is carried out after solving 
for the scattering solution 
using the limiting absorption principle.
The numerical calculations suggest that the dependence of transmittance, and the leaked energy flux, on the incident wave frequency is non-monotone in surface wave band.
It is found in the article that this dependence is highly susceptible to the choice of values of physical parameters associated with surface structure. 
The assumed physical parameters also allow several limiting cases of model some of which are discussed towards the end of article.
From the viewpoint of multiple scattering, due to a large set of such interfaces of varied sizes on the half space,
an alternative treatment is required \cite{martinbook}.
At the moment several such extensions are underway including an application of stochastic methods \cite{Maurel,Abrahams,Garnierbook,JGBls}.

The present article is organized as follows. 
After notational and mathematical preliminaries, the lattice model with surface structure is formulated in \S\ref{sec2d}, where the physical structure is described, the dimensionless equations of lattice dynamics are presented along with description of surface waves. 
The scattering problem formulation appears
in \S\ref{sec2dsca}.
An 
analysis of geometric part of the scattered field is included in \S\ref{sec2dgeo}, before diving into the exact mathematical solution.
\S\ref{sec2dWH} is at the heart of the present article where, after the Fourier transform is applied
on the equation of motion, the scattering problem is reduced to a Wiener-Hopf equation and its exact solution is found. 
The crucial result on reflection and transmission coefficients in \S\ref{sec2leak} yields the fraction of incident energy flux 
leaked via bulk lattice waves.
Several graphical illustrations and numerical results are presented in \S\ref{secNum}.
\S\ref{secDisc} incorporates the relationship between the Wiener-Hopf kernel and the Green’s function on half-plane, the scattering in one dimensional lattice model with an interface, a brief discussion on continuum limit of the Wiener-Hopf equation,
and connections with lattice strip problem.
After some comments on the obtained results and concluding remarks,
a list of references and four short appendices end the article.

{\bf Mathematical preliminaries:}
Let ${\mathbb{Z}}$ denote the set of integers, ${{{\mathbb{Z}}^2}}$ denote ${\mathbb{Z}}\times{\mathbb{Z}},$ ${{\mathbb{Z}}^+}$ (resp. ${{\mathbb{Z}}^-}$) denote the set of non-negative (resp. negative) integers.
Let ${\mathbb{R}}$ denote the set of real numbers and ${\mathbb{C}}$ denote the set of complex numbers. 
Typical ${{{z}}}\in\mathbb{C}$ is written as 
${z}={{{z}}}_1+i{{{z}}}_2=|{z}|\exp({i\arg {{{z}}}}),$
${{{z}}}_1=\Re {{{z}}}\in{\mathbb{R}}$ denotes real part,
${{{z}}}_2=\Im {{{z}}}\in{\mathbb{R}}$ denotes the imaginary part, $|{{{z}}}|$ denotes the modulus and $\arg {{{z}}}$ denotes the argument.
The square root function
has the usual branch cut in the complex plane running from $-\infty$ to $0$. 
If $f$ is a differentiable, real or complex valued, function then $f'$ denotes the derivative of $f$.
The symbol ${\mathbb{T}}$ denotes the unit circle (as a counterclockwise contour) in complex plane.
Let ${\mathbf{i}}, {\mathbf{j}}$ form the standard basis, i.e. $(1, 0), (0,1)$, of ${\mathbb{R}}^2=\mathbb{R}\times\mathbb{R}$ and `$\cdot$' denote the standard Euclidean dot product in two dimensions.

\section{Square lattice half-plane with structured boundary}
\label{sec2d}

Consider the definitions
\begin{equation}
\mathbb{H}:=\{(\mathtt{x}, \mathtt{y})\in{{{\mathbb{Z}}^2}}: \mathtt{y}\le0\},\quad
\mathring{\mathbb{H}}:=\{(\mathtt{x}, \mathtt{y})\in{{{\mathbb{Z}}^2}}: \mathtt{y}<0\},\quad
\partial{\mathbb{H}}:=\{(\mathtt{x}, \mathtt{y})\in{{{\mathbb{Z}}^2}}: \mathtt{y}=0\}
\label{hlfplane}\tag{H}
\end{equation}
to represent the lattice half-plane with boundary, without boundary and the boundary itself, respectively. 
The key assumptions
on the physical structure of the lattice 
appear in the following. 

\begin{figure}[!ht]
\center
{\includegraphics[width=.7\textwidth]{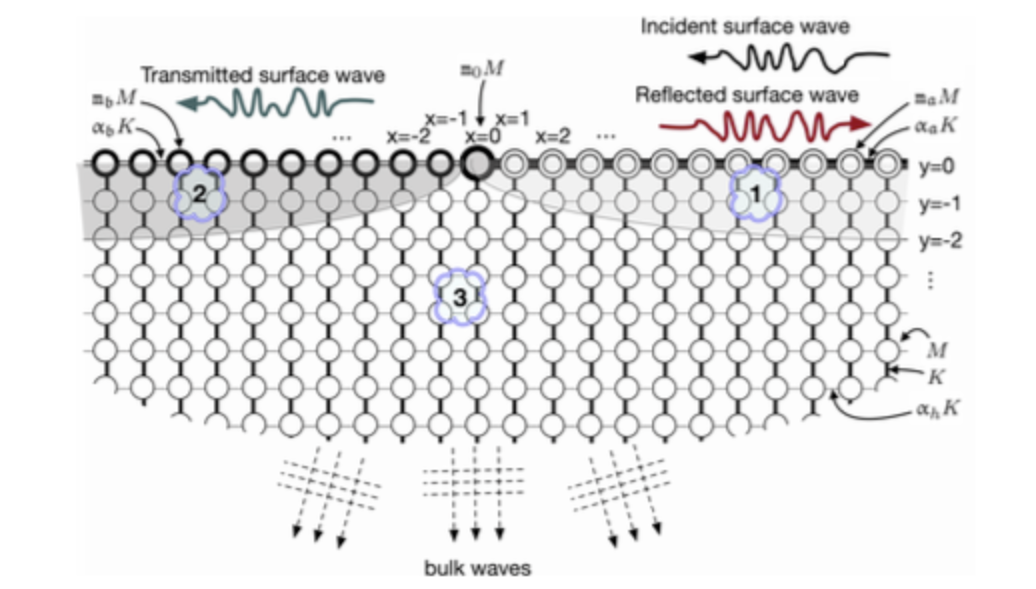}}
\caption[ ]{
Geometry of 
physical structure associated with 
$\mathbb{H}$ ($={\mathring{\mathbb{H}}}\cup{\partial{\mathbb{H}}}$, recall definition \eqref{hlfplane}, with surface interface on 
${\partial{\mathbb{H}}}$).
Pattern of schematic bonds and mass points 
indicates 
dimensionless 
parameters
${\mathit{m}}_{{\mathit A}}, {\mathit{m}}_{{\mathit B}}, {\mathit{m}}_{0}, {\alpha}_{{\mathit A}}, {\alpha}_{{\mathit B}}$, and ${\alpha}_{{}}$.}
\label{SchematicFig2D}
\end{figure}

\subsection{Physical structure of lattice half-plane}
\label{sec2dphy}

Consider a two-dimensional square lattice half-plane $\mathbb{H}={\mathring{\mathbb{H}}}\cup{\partial{\mathbb{H}}}$, with boundary ${\partial{\mathbb{H}}}$, associated with the following planar region in physical space (recall \eqref{hlfplane} for definition of $\mathbb{H}, {\mathring{\mathbb{H}}}, {\partial{\mathbb{H}}}$):
\begin{equation}\begin{split}
\{\tilde{\mathbf{p}}: \tilde{\mathbf{p}}=\mathtt{x}{a}{\mathbf{i}}+\mathtt{y}{a}{\mathbf{j}}, (\mathtt{x}, \mathtt{y})\in\mathbb{H}\}\subset {\mathbb{R}}^2,
\text{ where }a>0\text{ is the lattice parameter}.
\label{physicalcoord1}
\end{split}\end{equation}
The lattice half-plane away from its boundary ${\mathring{\mathbb{H}}}$ 
(see \eqref{hlfplane}${}_2$)
consists of identical particles, with {\em lattice coordinates} $({\mathtt{x}}, {\mathtt{y}})\in {\mathbb{Z}}^2$, of mass $M$ connected to their nearest neighbours by linearly elastic, massless bonds. 
It is assumed that the displacement field of this structure, as part of three dimensional physical space, is along the out-of-plane direction and is thus represented as a scalar $\tilde{u}(\tilde{\mathbf{p}}, \tilde{t})
$ at location $\tilde{\mathbf{p}}$ where $\tilde{t}$ is physical time.
It is assumed that particle at location $\tilde{\mathbf{p}}$, equivalently at $({\mathtt{x}}, {\mathtt{y}})\in {\mathbb{Z}}^2$, bonds with particles at $({\mathtt{x}}, {\mathtt{y}}\pm1)\in {\mathbb{Z}}^2$ using a force-constant $K$ 
and bonds with particles at $({\mathtt{x}}\pm1, {\mathtt{y}})\in {\mathbb{Z}}^2$
using ${{\alpha}_{{}}} K$;
see schematic illustration in Fig. \ref{SchematicFig2D} corresponding to the lattice rows at $\mathtt{y}\in{{\mathbb{Z}}^-}$. 
Let the dimensionless displacement ${\mathtt{u}}^{\textrm{to}}$
and time $t$ be defined by 
\begin{equation}\begin{split}
{\mathtt{u}}^{\textrm{to}}: \mathbb{H}\times\mathbb{R}\to\mathbb{R},\quad
\tilde{u}(\tilde{\mathbf{p}}, \tilde{t})=a~{\mathtt{u}}^{\textrm{to}}_{\mathtt{x}, \mathtt{y}}(t), \quad
\tilde{t}=a~t/c, \quad
\text{ where }
c=a\sqrt{K/M},
\label{physicalcoord2}
\end{split}\end{equation}
along with the relations $\mathtt{x}=\tilde{\mathbf{p}}\cdot{\mathbf{i}}/a, \mathtt{y}=\tilde{\mathbf{p}}\cdot{\mathbf{j}}/a$ from \eqref{physicalcoord1}.
Naturally, as part of the usual conditions in framework of Newtonian mechanics, it is assumed that $M, K, {{\alpha}_{{}}}\ge0.$

Particles at sites in $\partial{\mathbb{H}}$ 
(see \eqref{hlfplane}${}_3$)
belong to half-plane boundary which is assumed to be `traction-free' surface of the physical structure. Each particle in $\partial{\mathbb{H}}$, with coordinates $({\mathtt{x}}, {0})\in {\mathbb{Z}}^2$, has mass ${\mathit{m}}_{{\mathit A}} M$ and bonds with its horizontal nearest neighbours through force-constant ${\alpha}_{{\mathit A}} K$ when $\mathtt{x}\in{\mathbb{Z}}^-$, whereas it has mass ${\mathit{m}}_{{\mathit B}} M$ and horizontal bonds with force-constant ${\alpha}_{{\mathit B}} K$ when $\mathtt{x}\in{\mathbb{Z}}^+\setminus\{0\}$.
The particle at $(0, 0)$ 
has mass ${\mathit{m}}_{0} M$ while its horizontal bond has force-constant ${\alpha}_{{\mathit A}}K$ due to interaction with the site $(1, 0)$ ahead of the interface and ${\alpha}_{{\mathit B}}K$ due to interaction with site $(-1, 0)$ behind.
It is assumed that all particles in $\partial{\mathbb{H}}$ interact with the penultimate row 
via force-constant $K$.
It is also a physical assumption that ${\mathit{m}}_{{\mathit A}}, {\mathit{m}}_{{\mathit B}}, {\mathit{m}}_{0}, {\alpha}_{{\mathit A}}, {\alpha}_{{\mathit B}}\ge0.$

\subsection{Lattice dynamics in dimensionless form}
\label{sec2ddim}

Suppose that $\ddot{\mathtt{u}}^{\textrm{to}}_{{\mathtt{x}},{\mathtt{y}}}$ denotes the second order derivative of ${\mathtt{u}}^{\textrm{to}}_{{\mathtt{x}},{\mathtt{y}}}$ with respect to time $t$.
Suppressing the explicit presence of time $t$ for brevity, the equation of motion for a particle on ${\mathring{\mathbb{H}}}$
is
\begin{subequations}
\begin{equation}\begin{split}\label{eqmotion}
\ddot{\mathtt{u}}^{\textrm{to}}_{{\mathtt{x}},{\mathtt{y}}}={{\alpha}_{{}}}({\mathtt{u}}^{\textrm{to}}_{{\mathtt{x}}+1,{\mathtt{y}}}+{\mathtt{u}}^{\textrm{to}}_{{\mathtt{x}}-1,{\mathtt{y}}})+{\mathtt{u}}^{\textrm{to}}_{{\mathtt{x}},{\mathtt{y}}+1}+{\mathtt{u}}^{\textrm{to}}_{{\mathtt{x}},{\mathtt{y}}-1}-2(1+{{\alpha}_{{}}}){\mathtt{u}}^{\textrm{to}}_{{\mathtt{x}},{\mathtt{y}}},
\end{split}\end{equation}
\begin{align}
\label{eqmotionBCa}
\text{and on $\partial{\mathbb{H}}$,
for $\mathtt{x}\in{\mathbb{Z}}^+\setminus\{0\}$:}\quad {\mathit{m}}_{{\mathit A}} \ddot{\mathtt{u}}^{\textrm{to}}_{{\mathtt{x}},{\mathtt{y}}}={\alpha}_{{\mathit A}} ({\mathtt{u}}^{\textrm{to}}_{{\mathtt{x}}+1,{\mathtt{y}}}+{\mathtt{u}}^{\textrm{to}}_{{\mathtt{x}}-1,{\mathtt{y}}}-2{\mathtt{u}}^{\textrm{to}}_{{\mathtt{x}},{\mathtt{y}}})+{\mathtt{u}}^{\textrm{to}}_{{\mathtt{x}},{\mathtt{y}}-1}-{\mathtt{u}}^{\textrm{to}}_{{\mathtt{x}},{\mathtt{y}}},\\
\text{for $\mathtt{x}\in{\mathbb{Z}}^-$:}\quad
\label{eqmotionBCb}
{\mathit{m}}_{{\mathit B}} \ddot{\mathtt{u}}^{\textrm{to}}_{{\mathtt{x}},{\mathtt{y}}}={\alpha}_{{\mathit B}} ({\mathtt{u}}^{\textrm{to}}_{{\mathtt{x}}+1,{\mathtt{y}}}+{\mathtt{u}}^{\textrm{to}}_{{\mathtt{x}}-1,{\mathtt{y}}}-2{\mathtt{u}}^{\textrm{to}}_{{\mathtt{x}},{\mathtt{y}}})+{\mathtt{u}}^{\textrm{to}}_{{\mathtt{x}},{\mathtt{y}}-1}-{\mathtt{u}}^{\textrm{to}}_{{\mathtt{x}},{\mathtt{y}}},\\
\text{for ${\mathtt{x}}=0$:}\quad
\label{eqmotionBC0}
{\mathit{m}}_{0} \ddot{\mathtt{u}}^{\textrm{to}}_{{\mathtt{x}},{\mathtt{y}}}= {\alpha}_{{\mathit A}} ({\mathtt{u}}^{\textrm{to}}_{{\mathtt{x}}+1,{\mathtt{y}}}-{\mathtt{u}}^{\textrm{to}}_{{\mathtt{x}},{\mathtt{y}}})+{\alpha}_{{\mathit B}} ({\mathtt{u}}^{\textrm{to}}_{{\mathtt{x}}-1,{\mathtt{y}}}-{\mathtt{u}}^{\textrm{to}}_{{\mathtt{x}},{\mathtt{y}}})+{\mathtt{u}}^{\textrm{to}}_{{\mathtt{x}},{\mathtt{y}}-1}-{\mathtt{u}}^{\textrm{to}}_{{\mathtt{x}},{\mathtt{y}}}.
\end{align}
\label{eqmotionandBCs}
\end{subequations}
In the rest of the article, 
the subscript `${\mathit A}$' refers to entities defined on the portion of surface ahead of the interface while `${\mathit B}$' refers to the entities behind the interface.

\subsection{Surface waves on $\mathbb{H}$}
\label{sec2dsurf}

In line with the 
recent analysis of surface waves in certain simple lattice models \cite{Victor_Bls_surf1,Victor_Bls_surf2}, consider the structure {\em without} an interface, i.e. \eqref{eqmotionandBCs} with 
\begin{equation}\begin{split}
{\mathit{m}}_{{\mathit A}}={\mathit{m}}_{{\mathit B}}={\mathit{m}}_{0}={\mathit{m}}_{{sf}}\text{ and }{\alpha}_{{\mathit A}}={\alpha}_{{\mathit B}}={\alpha}_{{sf}},
\label{sec2dsurfcase}
\end{split}\end{equation}
where ${\mathit{m}}_{{sf}}, {\alpha}_{{sf}}>0$ are specified.
The lattice model in this case can be addressed as a discrete analog of Gurtin-Murdoch model \cite{GurtinMurdoch1975a};
a comparison of the surface waves 
in such discrete vs continuum model
was provided in \cite{Victor_Bls_surf1} for the choice ${{\alpha}_{{}}}=1$.
A typical time harmonic lattice wave in the bulk of the physical structure can be expressed 
as $\exp(i\tilde{\mathbf{k}}\cdot\tilde{\mathbf{p}}-i\tilde{\omega}\tilde{t}),$ where $\tilde{\mathbf{k}}:=(\tilde{k}_x, \tilde{k}_y)$ is the wave vector, while a surface wave on the assumed half plane structure has the form $\exp(i\tilde{{\upxi}}{\mathbf{i}}\cdot\tilde{\mathbf{p}}+\tilde{\eta}{\mathbf{j}}\cdot\tilde{\mathbf{p}}-i\tilde{\omega}\tilde{t})$. 
Suppose that the 
parameters 
\eqref{sec2dsurfcase} and ${{\alpha}_{{}}}$
are such that a surface wave, denoted by ${\mathtt{u}}^{{{sf}}}$, is admissible in certain band of frequencies.
For time harmonic wave motion in the physical structure, in line with \eqref{physicalcoord1}, \eqref{physicalcoord2}${}_2$, it is convenient to work with the dimensionless entities
\begin{equation}\begin{split}\label{deffreq}
{\upomega}=\tilde{\omega}{a}/{c}, \quad
{\upxi}=a~\tilde{{\upxi}},\quad
\eta=a~\tilde{\eta}.
\end{split}\end{equation}
It is easy to see that after the substitution of
\begin{equation}\begin{split}
{\mathtt{u}}^{\textrm{to}}_{{\mathtt{x}},{\mathtt{y}}}(t)={\mathtt{u}}^{{{sf}}}_{{\mathtt{x}},{\mathtt{y}}}\exp(-i\upomega t),\quad
\text{where }
{\mathtt{u}}^{{{sf}}}_{{\mathtt{x}},{\mathtt{y}}}:=\exp({ i{{\upxi}} {\mathtt{x}}})\exp({{\eta} {\mathtt{y}}}),
\quad (\mathtt{x},\mathtt{y})\in\mathbb{H},
\label{uincSwave}
\end{split}\end{equation}
\eqref{eqmotionandBCs} is satisfied,
provided that
\begin{equation}\begin{split}
{\upomega}\equiv{\upomega}({{\upxi}})\text{ and }{\eta}\equiv{\eta}({{\upxi}})
\label{funcomega}
\end{split}\end{equation}
are implicitly defined by the two coupled equations, as a result of \eqref{eqmotionandBCs},
\begin{align}\label{funcomegaeq}
{\upomega}^2&=4{{\alpha}_{{}}}\sin^2({{{\upxi}}}/{2})+2-2\cosh{\eta},\quad
{\mathit{m}}_{{sf}}{\upomega}^2=4{\alpha}_{{sf}} \sin^2({{{\upxi}}}/{2})+1-\exp(-{\eta}).
\end{align}
Thus, ${\mathtt{u}}^{{{sf}}}$ in \eqref{uincSwave}${}_2$ represents a {\em surface wave mode} of dimensionless wavenumber ${{\upxi}}\in(-\pi,\pi)\setminus\{0\}$ that
decays exponentially into 
$\mathbb{H}$ with an attenuation coefficient 
${{\eta}({{\upxi}})}$. 
A natural requirement of a propagating surface wave on $\mathbb{H}$ is that ${{\eta}({{\upxi}})}>0$ so it is localized at $\partial{\mathbb{H}}$.
As a result of this condition in \eqref{funcomegaeq}, it can be found that the surface wave exists if
${\alpha}_{{sf}}<{\mathit{m}}_{{sf}}\min\{{\alpha}_{{}}, 1\}$.
Note that $\upomega({\upxi})\in(0, \upomega_{\max})$, ${{\upxi}}\in(-\pi,\pi)\setminus\{0\}$, where $\upomega_{\max}$ depends on ${\mathit{m}}_{{sf}}, {\alpha}_{{sf}}$.
A 
description of the surface waves in the special case when ${{\alpha}_{{}}}=1$ can be found in \cite{Victor_Bls_surf1}.
Using \eqref{funcomega} and \eqref{funcomegaeq}, by definition \cite{Brillouin}, 
\begin{align}
{{\mathit v}}_{{sf}}({{\upxi}}):=
\frac{d}{d{\upxi}}{\upomega}({\upxi})
=\frac{\sin{\upxi}}{\upomega({{\upxi}})}\frac{{{\alpha}_{{}}}+({\alpha}_{{sf}}-{{\alpha}_{{}}})(1-\exp(-2{{{\eta}({{\upxi}})}}))}{1+({\mathit{m}}-1)(1-\exp(-2{{{\eta}({{\upxi}})}}))}, \quad{\upxi}\in(-\pi, \pi)\setminus\{0\},
\label{surfgpvel}
\end{align}
is the group velocity of \eqref{uincSwave}.
Hence, \eqref{uincSwave} represents a surface wave which carries energy flux towards $\mathtt{x}\to+\infty$ (resp. $-\infty$) when ${{\upxi}}\in(0,\pi)$ (resp. $-{\upxi}\in(0, \pi)$) as ${{\mathit v}}_{{sf}}({{\upxi}})>0$ (resp. ${{\mathit v}}_{{sf}}({{\upxi}})<0$);
see Appendix \ref{appenfixsurfwaveflux} for more details.
Later in the article, a phrase is often used that \eqref{uincSwave} represents wave travelling towards $\mathtt{x}\to+\infty$ (resp. $\mathtt{x}\to-\infty$) when ${\upxi}\in(0, \pi)$ (resp. $-{\upxi}\in(0, \pi)$).

\begin{remark}
When ${{\alpha}_{{}}}=0, {\alpha}_{{sf}}\ne0$, half-plane corresponds to one dimensional `strings' attached to each site in the lattice row at $\mathtt{y}=0.$ 
However this limit does not admit propagating waves except in case of certain choice of material parameters when the frequency belongs to the stop band of the `cross' chains.
When ${{\alpha}_{{}}}\to\infty, {\alpha}_{{sf}}\ne0$, the lattice model corresponds to a one dimensional lattice row at $\mathtt{y}=0$ attached to a rigid foundation.
When ${{\alpha}_{{}}}\ne0,$ ${\alpha}_{{sf}}=0$ but ${\mathit{m}}_{{sf}}>0$, then the lattice model corresponds to two dimensional lattice half-plane with (inertial oscillator) mass attached to each site in $\partial{\mathbb{H}}$.
Given any ${{\alpha}_{{}}}\ge0$, when ${\alpha}_{{\mathit A}}=\mathpzc{O}(1/K)\ne0, {\alpha}_{{\mathit B}}=\mathpzc{O}(1/K)\ne0$ and $K\to\infty$, the limiting lattice model corresponds to a purely one dimensional lattice row at $\mathtt{y}=0$.
\label{remMiM}
\end{remark}

\section{Problem formulation of surface wave scattering}
\label{sec2dsca}

Reverting back to the case when ${\mathit{m}}_{{\mathit A}}, {\mathit{m}}_{{\mathit B}},{\mathit{m}}_{0}, {\alpha}_{{\mathit A}}, {\alpha}_{{\mathit B}}, {\alpha}_{{}}$ are arbitrary (i.e. discarding the assumption \eqref{sec2dsurfcase} of a uniform surface),
consider an incident wave ${\mathtt{u}}^{{\mathit A}_{\textrm{in}}}$ to be a surface wave admissible in the region ahead of interface.
In other words, ${\upomega}$ is assumed to satisfy \eqref{funcomegaeq} for a choice of surface parameters corresponding to the side ahead of interface, that is with subscript ${\mathit A}$, so that ${\upomega}$ belongs to the surface wave band of the side ahead.
This is the case discussed above in the context of \eqref{uincSwave} (with `${sf}$' replaced by ${\mathit A}_{\textrm{in}}$) and $-\mathtt{x}\in(0, \pi)$.
Such a wave 
is schematically shown as incident wave on half-plane boundary in Fig. \ref{SchematicFig2D}, while traveling from region $\mathtt{x}\to+\infty$ towards $\mathtt{x}\to-\infty$.
As surface material parameters differ on the side behind interface, starting at $(0, 0)$, the incident surface wave necessarily scatters. 
This leads to excitation of not only surface waves, as transmitted and reflected waves, but also bulk lattice waves; these waves are also shown schematically in Fig. \ref{SchematicFig2D} where the excited 
bulk waves are highlighted. 
In view 
of \eqref{uincSwave}, \eqref{surfgpvel}, the above-mentioned incident surface wave for all $t\in\mathbb{R}$ is written as
\begin{equation}\begin{split}
{\mathtt{u}}^{\textrm{in}}_{{\mathtt{x}},{\mathtt{y}}}(t):={\mathtt{u}}^{{\mathit A}_{\textrm{in}}}_{{\mathtt{x}},{\mathtt{y}}}\exp(-i\upomega t), \qquad
\label{uincawave2d}
{\mathtt{u}}^{{\mathit A}_{\textrm{in}}}_{{\mathtt{x}},{\mathtt{y}}}:={{{\mathtt{u}}^{{\mathit A}_{\textrm{in}}}_{0,0}}}\exp({-i{{\upxi}}_{{\mathit A}} {\mathtt{x}}})\exp({{\eta}({\upxi}_{{\mathit A}}) {\mathtt{y}}}), 
\quad (\mathtt{x},\mathtt{y})\in\mathbb{H},
\end{split}\end{equation}
where ${{{\mathtt{u}}^{{\mathit A}_{\textrm{in}}}_{0,0}}}\in\mathbb{C}$ 
and ${\upxi}_{{\mathit A}}\in(0, \pi)$
while $\upomega=\upomega({\upxi}_{{\mathit A}})$ and $\eta({\upxi}_{{\mathit A}})$ 
according to \eqref{funcomega},\eqref{funcomegaeq}.
The expression in \eqref{uincawave2d}${}_2$ can be construed as the incident surface wave mode incoming from $\mathtt{x}\to+\infty$ and travelling towards $\mathtt{x}\to-\infty$.

When the surface material parameters ${\mathit{m}}_{{\mathit A}}, {\mathit{m}}_{{\mathit B}}$, and/or ${\alpha}_{{\mathit A}}, {\alpha}_{{\mathit B}},$ are unequal the incident wave ansatz \eqref{uincawave2d} does not satisfy the equations of motion \eqref{eqmotionandBCs} everywhere on $\mathbb{H}$; in fact \eqref{eqmotionBC0}, \eqref{eqmotionBCb} are not satisfied by \eqref{uincawave2d}.
However, the actual displacement field ${\mathtt{u}}^{\textrm{to}}_{{\mathtt{x}},{\mathtt{y}}}(t)$ \eqref{physicalcoord2}${}_1$ must satisfy \eqref{eqmotionandBCs} on $\mathbb{H}$. Therefore, it is convenient to define the scattered wave field according to 
\begin{equation}\begin{split}
{\mathtt{u}}^{\textrm{sc}}_{{\mathtt{x}},{\mathtt{y}}}(t):={\mathtt{u}}^{\textrm{to}}_{{\mathtt{x}},{\mathtt{y}}}(t)-{\mathtt{u}}^{\textrm{in}}_{{\mathtt{x}},{\mathtt{y}}}(t),
\label{totincsplit}
\end{split}\end{equation}
for all $({\mathtt{x}},{\mathtt{y}})\in\mathbb{H}, t\in\mathbb{R}$, 
using the definition in \eqref{uincawave2d}. 

The definition \eqref{totincsplit} is the standard {additive composition} of ${\mathtt{u}}^{\textrm{to}}$ into the two components, namely incident wave ${\mathtt{u}}^{\textrm{in}}$ and the scattered field ${\mathtt{u}}^{\textrm{sc}}$, i.e. ${\mathtt{u}}^{\textrm{to}}_{{\mathtt{x}},{\mathtt{y}}}(t)={\mathtt{u}}^{\textrm{in}}_{{\mathtt{x}},{\mathtt{y}}}(t)+{\mathtt{u}}^{\textrm{sc}}_{{\mathtt{x}},{\mathtt{y}}}(t).$ 
As the incident wave is known apriori 
as time harmonic and complex valued, naturally ${\mathtt{u}}^{\textrm{to}}$ and ${\mathtt{u}}^{\textrm{sc}}$ are complex valued too. Moreover, upon attaining the {\em steady state regime}, using the principle of limiting absorption,
it is natural to seek the scattered field ${\mathtt{u}}^{\textrm{sc}}$ such that
\begin{equation}\begin{split}
{\mathtt{u}}^{\textrm{sc}}_{{\mathtt{x}},{\mathtt{y}}}(t)={\mathtt{u}}_{{\mathtt{x}},{\mathtt{y}}}\exp(-i\upomega t),
\label{uscadef}
\end{split}\end{equation}
where ${\mathtt{u}}_{{\mathtt{x}},{\mathtt{y}}}$ is {\em time-independent}
for all $({\mathtt{x}},{\mathtt{y}})\in\mathbb{H}$ and all $t\in\mathbb{R}$ corresponding to {steady state regime}.
Thus, the spatial part of scattered field in {steady state regime} on $\mathbb{H}$, is described by
\begin{equation}\begin{split}
{\mathtt{u}}:\mathbb{H}\to\mathbb{C}
\label{uscafun}
\end{split}\end{equation}

The incident surface wave 
\eqref{uincawave2d}${}_1$
satisfies the equation of motion 
at $({\mathtt{x}}, {\mathtt{y}})\in\mathring{\mathbb{H}}$ and at those sites belonging to the portion of $\partial{\mathbb{H}}$ with index ${{\mathit A}}$, at all times.
In the rest of the article, the letter ${{\mathscr{H}}}$ stands for the Heaviside function defined on integers: 
\begin{equation}\begin{split}
{{\mathscr{H}}}({\mathtt{x}}):=0, {\mathtt{x}}\in{\mathbb{Z}}^-\quad\text{ and }\quad {{\mathscr{H}}}({\mathtt{x}}):=1, {\mathtt{x}}\in{\mathbb{Z}}^+.
\label{defH}
\end{split}\end{equation}
As a result of the additive decomposition \eqref{totincsplit}, the scattered field 
\eqref{uscafun} 
satisfies 
\begin{equation}\begin{split}\label{eqmotionsca}
-\upomega^2{\mathtt{u}}_{{\mathtt{x}},{\mathtt{y}}}={{\alpha}_{{}}}({\mathtt{u}}_{{\mathtt{x}}+1,{\mathtt{y}}}+{\mathtt{u}}_{{\mathtt{x}}-1,{\mathtt{y}}})+{\mathtt{u}}_{{\mathtt{x}},{\mathtt{y}}+1}+{\mathtt{u}}_{{\mathtt{x}},{\mathtt{y}}-1}-2(1+{{\alpha}_{{}}}){\mathtt{u}}_{{\mathtt{x}},{\mathtt{y}}},
\text{ for all }({\mathtt{x}}, {\mathtt{y}})\in\mathring{\mathbb{H}},
\end{split}\end{equation}
\begin{subequations}
\begin{align}
\text{and }
&{\alpha}_{{\mathit A}}({\mathtt{u}}_{{\mathtt{x}}+1, {0}}-{\mathtt{u}}_{{\mathtt{x}}, {0}}){{\mathscr{H}}}({\mathtt{x}})
+{\alpha}_{{\mathit A}}({\mathtt{u}}_{{\mathtt{x}}-1, {0}}-{\mathtt{u}}_{{\mathtt{x}}, {0}}){{\mathscr{H}}}({\mathtt{x}}-1)+{\alpha}_{{\mathit B}}({\mathtt{u}}_{{\mathtt{x}}+1, {0}}-{\mathtt{u}}_{{\mathtt{x}}, {0}}){{\mathscr{H}}}(-{\mathtt{x}}-1)\nonumber\\
&+{\alpha}_{{\mathit B}}({\mathtt{u}}_{{\mathtt{x}}-1, {0}}-{\mathtt{u}}_{{\mathtt{x}}, {0}}){{\mathscr{H}}}(-{\mathtt{x}})+{\mathtt{u}}_{{\mathtt{x}}, -1}-{\mathtt{u}}_{{\mathtt{x}}, {0}}+{\mathit{m}}_{{\mathit A}}{\upomega}^2{\mathtt{u}}_{{\mathtt{x}}, {0}}{{\mathscr{H}}}({\mathtt{x}})+{\mathit{m}}_{{\mathit B}}{\upomega}^2{\mathtt{u}}_{{\mathtt{x}}, {0}}{{\mathscr{H}}}(-{\mathtt{x}}-1)\nonumber\\
&+({\mathit{m}}_{0}-{\mathit{m}}_{{\mathit A}})\upomega^2{\mathtt{u}}_{{\mathtt{x}}, {0}}\delta_{{\mathtt{x}}, {0}}
=-f^{{\mathit A}_{\textrm{in}}}_{\mathtt{x}}-({\mathit{m}}_{0}-{\mathit{m}}_{{\mathit A}})\upomega^2{\mathtt{u}}^{{\mathit A}_{\textrm{in}}}_{{\mathtt{x}}, {0}}\delta_{{\mathtt{x}}, {0}}\quad \text{ on }({\mathtt{x}}, {\mathtt{y}})\in\partial{\mathbb{H}},\label{BCeq}
\end{align}
\begin{align}
\text{where }
f^{{\mathit A}_{\textrm{in}}}_{\mathtt{x}}
&=({\alpha}_{{\mathit B}}-{\alpha}_{{\mathit A}})({\mathtt{u}}^{{\mathit A}_{\textrm{in}}}_{{\mathtt{x}}+1, {0}}-{\mathtt{u}}^{{\mathit A}_{\textrm{in}}}_{{\mathtt{x}}, {0}}){{\mathscr{H}}}(-{\mathtt{x}}-1)
+({\alpha}_{{\mathit B}}-{\alpha}_{{\mathit A}})({\mathtt{u}}^{{\mathit A}_{\textrm{in}}}_{{\mathtt{x}}-1, {0}}-{\mathtt{u}}^{{\mathit A}_{\textrm{in}}}_{{\mathtt{x}}, {0}}){{\mathscr{H}}}(-{\mathtt{x}})\nonumber\\
&+({\mathit{m}}_{{\mathit B}}-{\mathit{m}}_{{\mathit A}}){\upomega}^2{\mathtt{u}}^{{\mathit A}_{\textrm{in}}}_{{\mathtt{x}}, {0}}{{\mathscr{H}}}(-{\mathtt{x}}-1).\label{deffinc}
\end{align}
\label{eqmotionBCFT}
\end{subequations}
In above, so called `discrete Helmholtz equation',
the Kronecker delta is defined by
\begin{equation}\begin{split}
\delta_{{\mathtt{x}}, {n}}:=0, \text{ if }\mathtt{x}\ne n\quad\text{ and }\quad \delta_{{\mathtt{x}}, {n}}:=1, \text{ if }\mathtt{x}=n, \quad{\mathtt{x}}, n\in{\mathbb{Z}}.
\label{defdelta}
\end{split}\end{equation}

\begin{remark}
An incident wave traveling towards the interface from the portion behind it, i.e. from $\mathtt{x}\to-\infty$ to $\mathtt{x}\to+\infty$ for all $t\in\mathbb{R}$, is given by
\begin{equation}\begin{split}
{\mathtt{u}}^{{\mathit B}_{\textrm{in}}}_{{\mathtt{x}},{\mathtt{y}}}\exp(-i\upomega t)={{{\mathtt{u}}^{{\mathit B}_{\textrm{in}}}_{0,0}}}\exp({+i{{\upxi}}_{{\mathit B}} {\mathtt{x}}}-i\upomega t)\exp({{\eta}({\upxi}_{{\mathit B}}) {\mathtt{y}}}), 
\quad (\mathtt{x},\mathtt{y})\in\mathbb{H},
\label{uincbwave2d}
\end{split}\end{equation}
where ${{\upxi}}_{{\mathit B}}\in(0, \pi)$ and ${\eta}({\upxi}_{{\mathit B}})$ satisfy \eqref{funcomega},\eqref{funcomegaeq}.
In this case, \eqref{eqmotionBCa}, \eqref{eqmotionBC0} are certainly not satisfied by \eqref{uincbwave2d} when the surface material parameters ${\mathit{m}}_{{\mathit A}}, {\mathit{m}}_{{\mathit B}}$ and/or ${\alpha}_{{\mathit A}}, {\alpha}_{{\mathit B}},$ are unequal.
Regarding the choice of sign in front of the wavenumber in \eqref{uincbwave2d}, recall the statements 
accompanying \eqref{uincSwave}, \eqref{surfgpvel}.
\label{altcase1}
\end{remark}

\section{Scattered surface waves and their energy flux}
\label{sec2dgeo}

In the context of the discussion presented in the paragraph preceding \eqref{uincawave2d}, as well as \eqref{surfgpvel},
and in view of \eqref{uincSwave}, the reflected surface wave, for all $t\in\mathbb{R}$, corresponds to
\begin{equation}\begin{split}
{\mathtt{u}}^{{\mathit A}}_{{\mathtt{x}},{\mathtt{y}}}\exp(-i\upomega t)={{\mathtt{u}}^{{\mathit A}}_{0,0}}\exp({+i{{\upxi}}_{{\mathit A}} {\mathtt{x}}}-i\upomega t)\exp({{\eta}({\upxi}_{{\mathit A}}) {\mathtt{y}}}),
\quad (\mathtt{x},\mathtt{y})\in\mathbb{H},
\label{urefawave}
\end{split}\end{equation}
while the transmitted surface wave is expressed as
\begin{equation}\begin{split}
{\mathtt{u}}^{{\mathit B}}_{{\mathtt{x}},{\mathtt{y}}}\exp(-i\upomega t)={{\mathtt{u}}^{{\mathit B}}_{0,0}}\exp({-i{{\upxi}}_{{\mathit B}} {\mathtt{x}}}-i\upomega t)\exp({{\eta}({\upxi}_{{\mathit B}}) {\mathtt{y}}}), 
\quad (\mathtt{x},\mathtt{y})\in\mathbb{H},
\label{utransbwave}
\end{split}\end{equation}
where
$\upomega=\upomega({\upxi}_{{\mathit A}})=\upomega({\upxi}_{{\mathit B}}),$ $\eta({\upxi}_{{\mathit B}})$, with ${{\upxi}}_{{\mathit B}}\in(0, \pi)$, are
according to \eqref{funcomega},\eqref{funcomegaeq} while
${\upxi}_{{\mathit A}}$, $\eta({\upxi}_{{\mathit A}})$ are the same as in 
\eqref{uincawave2d}.
Due to \eqref{surfgpvel}, indeed,
following the sign convention discussed with
\eqref{uincSwave}, \eqref{urefawave} (resp. \eqref{utransbwave}) represents a surface wave which carries energy flux towards $\mathtt{x}\to+\infty$ (resp. $\mathtt{x}\to-\infty$) since ${{\upxi}}_{{\mathit A}}\in(0,\pi)$ (resp. ${{\upxi}}_{{\mathit B}}\in(0,\pi)$).
Using \eqref{surfgpvel}, the magnitudes of energy fluxes are \cite{Brillouin}
\begin{subequations}
\begin{align}
\mathcal{E}_{{\mathit A}_{\textrm{in}}}&=\frac{1}{2}{\upomega}^2|{{{\mathtt{u}}^{{\mathit A}_{\textrm{in}}}_{0,0}}}|^2{{{\mathit v}}_{{\mathit A}}}({{\upxi}}_{{\mathit A}}),\\
\mathcal{E}_{{\mathit A}}&=\frac{1}{2}{\upomega}^2|{{{\mathtt{u}}^{{\mathit A}_{\textrm{in}}}_{0,0}}}{\mathcal{C}_{{\mathit A}_{\textrm{in}}}^{{\mathit A}}}|^2{{{\mathit v}}_{{\mathit A}}}({{\upxi}}_{{\mathit A}}),\\
\mathcal{E}_{{\mathit B}}&=\frac{1}{2}{\upomega}^2|{{{\mathtt{u}}^{{\mathit A}_{\textrm{in}}}_{0,0}}}\mathcal{C}_{{\mathit A}_{\textrm{in}}}^{{\mathit B}}|^2{{{\mathit v}}_{{\mathit B}}}({{\upxi}}_{{\mathit B}}),
\end{align}
\label{Efluxall}
\end{subequations}
which are carried by the incident wave \eqref{uincawave2d} (towards $\mathtt{x}\to-\infty$), the reflected wave \eqref{urefawave} (towards $\mathtt{x}\to+\infty$), and the transmitted wave \eqref{utransbwave} (towards $\mathtt{x}\to-\infty$), respectively.
In \eqref{Efluxall},
$\mathcal{C}_{{\mathit A}_{\textrm{in}}}^{{\mathit A}}$ is the reflection coefficient while
$\mathcal{C}_{{\mathit A}_{\textrm{in}}}^{{\mathit A}}$ is the transmission coefficient, such that in \eqref{urefawave}, \eqref{utransbwave}, respectively,
\begin{equation}\begin{split}
{{\mathtt{u}}^{{\mathit A}}_{0,0}}:=\mathcal{C}_{{\mathit A}_{\textrm{in}}}^{{\mathit A}}{{{\mathtt{u}}^{{\mathit A}_{\textrm{in}}}_{0,0}}},\quad {{\mathtt{u}}^{{\mathit B}}_{0,0}}:=\mathcal{C}_{{\mathit A}_{\textrm{in}}}^{{\mathit B}}{{{\mathtt{u}}^{{\mathit A}_{\textrm{in}}}_{0,0}}}.
\label{reftranscoefftdef}
\end{split}\end{equation}

\begin{remark}
In the context of Remark \ref{altcase1},
the reflected wave has the form \eqref{utransbwave} while the transmitted wave is \eqref{urefawave}.
Analogous to \eqref{reftranscoefftdef}, 
${{\mathtt{u}}^{{\mathit B}}_{0,0}}:=\mathcal{C}_{{\mathit B}_{\textrm{in}}}^{{\mathit B}}{{{\mathtt{u}}^{{\mathit B}_{\textrm{in}}}_{0,0}}}, 
{{\mathtt{u}}^{{\mathit A}}_{0,0}}:=\mathcal{C}_{{\mathit B}_{\textrm{in}}}^{{\mathit A}}{{{\mathtt{u}}^{{\mathit B}_{\textrm{in}}}_{0,0}}},$
where $\mathcal{C}_{{\mathit B}_{\textrm{in}}}^{{\mathit B}}$ is the reflection coefficient while
$\mathcal{C}_{{\mathit B}_{\textrm{in}}}^{{\mathit A}}$ is the transmission coefficient.
Due to the symmetry in the mathematical formulation of the physical problem, the solution to the scattered wave field, and consequent surface wave amplitudes, can be alternatively obtained by using the symmetry transformation $({\mathit A}, {\mathit B}, \mathtt{x})\mapsto({\mathit B}, {\mathit A}, -\mathtt{x})$ (at fixed ${\mathit{m}}_{0}$, ${{\alpha}_{{}}}$).
Due to the linearity of problem, an arbitrary incident wave can be considered as a superposition of \eqref{uincawave2d} and \eqref{uincbwave2d}, i.e.
${\mathtt{u}}^{\textrm{in}}_{{\mathtt{x}},{\mathtt{y}}}=
{\mathtt{u}}^{{\mathit A}_{\textrm{in}}}_{{\mathtt{x}},{\mathtt{y}}}
+{\mathtt{u}}^{{\mathit B}_{\textrm{in}}}_{{\mathtt{x}},{\mathtt{y}}}.$
\label{altcase2}
\end{remark}

\begin{remark}
In the case of bulk wave incidence, the anstaz for the incident wave corresponds to 
\begin{equation}\begin{split}
{\mathtt{u}}^{\textrm{in}}_{{\mathtt{x}},{\mathtt{y}}}(t)=
\exp({+i{k}_x^{\textrm{in}} {\mathtt{x}}+i{k}_y^{\textrm{in}} {\mathtt{y}}}-i\upomega t)
\text{ with }
{\upomega}^2=4{{\alpha}_{{}}}\sin^2\frac{1}{2}{k}_x^{\textrm{in}}+4\sin^2\frac{1}{2}{k}_y^{\textrm{in}},
\end{split}\end{equation}
for all $(\mathtt{x},\mathtt{y})\in\mathbb{H}.$
The incident lattice wave in this case arrives from $\mathtt{y}\to-\infty$ travels towards $\partial{\mathbb{H}}$ at some excited frequency in the same way as that formulated in \cite{sharma2017scattering}.
Assuming the surface material parameters differ from the bulk lattice, i.e. ${\mathit{m}}_{{\mathit A}}, {\mathit{m}}_{{\mathit B}} \ne 1, {\alpha}_{{\mathit A}}, {\alpha}_{{\mathit B}}\ne {\alpha}_{{}}$, the scattering of incident bulk wave occurs may lead to excitation of not only bulk lattice wave, such as the wave reflected by half-plane boundary, but also surface waves \eqref{urefawave}, \eqref{utransbwave} on those sides of the interface at $(0,0)$ which support surface wave at the excited frequency; the detailed calculations are however omitted in the present article. 
A related research problem of surface wave excitation due to incident bulk wave on the face of sharp crack in triangular lattice has been recently discussed in \cite{Blstgsurf},
while
discrete scattering problems, without surface waves, have been investigated before in the context of edges of line defects on uniform square lattices, see for example \cite{sK,sC}.
\end{remark}

\begin{remark}
A special situation occurs when ${\mathit{m}}_{{\mathit A}}={\mathit{m}}_{{\mathit B}} \ne {\mathit{m}}_{0}, {\alpha}_{{\mathit A}}={\alpha}_{{\mathit B}}$.
In this case there is no surface interface as the surface is uniform 
except at $(0,0)$. The surface wave scattering occurs, for incidence from either side, due to the mass defect at $(0,0)$ on $\partial{\mathbb{H}}$. This is analyzed 
in the later section titled discussion. 
\end{remark}

Suppose ${\mathscr{R}}_{{surf}}$ and ${\mathscr{T}}_{{surf}}$ denote the surface wave reflectance and transmittance \cite{Brillouin} of the surface interface relative to the given incident surface wave. These entities represent the reflected and transmitted energy flux in the surface wave per unit energy flux of the incident surface wave. In terms of mathematical expressions, the surface wave reflectance and transmittance are, respectively, defined by
\begin{align}
{\mathscr{R}}_{{surf}}:=\frac{\mathcal{E}_{{\mathit A}}}{\mathcal{E}_{{\mathit A}_{\textrm{in}}}}=|\mathcal{C}_{{\mathit A}_{\textrm{in}}}^{{\mathit A}}|^2,\quad
{\mathscr{T}}_{{surf}}:=\frac{\mathcal{E}_{{\mathit B}}}{\mathcal{E}_{{\mathit A}_{\textrm{in}}}}=|\mathcal{C}_{{\mathit A}_{\textrm{in}}}^{{\mathit B}}|^2\frac{{{{\mathit v}}_{{\mathit B}}}({{\upxi}}_{{\mathit B}})}{{{{\mathit v}}_{{\mathit A}}}({{\upxi}}_{{\mathit A}})}.
\label{reftransance}
\end{align}

\section{Scattering solutions on $\mathbb{H}$}
\label{sec2dWH}

Assuming the lattice dynamics in its steady state regime, an approach, equivalent to that described thus far, for example, the one that motivated \eqref{uscadef}, involves the time harmonic solutions of 
\eqref{eqmotionandBCs} where the constant frequency is same as that of the incident wave. This has been employed in the study of scattering of bulk waves by a crack tip in mode III and a rigid constraint tip \cite{sK,sC}, as well as by a special kind of surface interface \cite{sharma2017scattering}, amongst several other researches.
With this background, the construction of scattering solution ${\mathtt{u}}$ \eqref{totincsplit}, technically via the principle of limiting absorption,
is made convenient by assuming a vanishingly small damping so that ${\upomega}$ is complex:
\begin{equation}\begin{split}
{\upomega}={\upomega}_1+i{\upomega}_2,\quad 0<{\upomega}_2\ll1;
\label{complexfq}
\end{split}\end{equation}
see the paragraph preceding (1.8) in \cite{sK}, for example.
For a fixed integer $\mathtt{y}\in{\mathbb{Z}}^-\cup\{0\}$, the Fourier transform (along the ${\mathtt{x}}$ axis) \cite{sK,sC} 
${\mathtt{u}}_{\mathtt{y}}^{\rm F}: {\mathbb{C}}\to{\mathbb{C}}$ of \eqref{uscafun}
is defined by
\begin{equation}\begin{split}
{\mathtt{u}}_{\mathtt{y}}^{\rm F}({{{z}}}):={\mathtt{u}}_{\mathtt{y}}^{+}({{{z}}})+{\mathtt{u}}_{\mathtt{y}}^{-}({{{z}}}), \text{ where }
{\mathtt{u}}_{\mathtt{y}}^{+}({{{z}}}):=\sum\nolimits_{{\mathtt{x}}\in\mathbb{Z}^+} {\mathtt{u}}_{{\mathtt{x}}, {\mathtt{y}}}{{{z}}}^{-{\mathtt{x}}}, {\mathtt{u}}_{\mathtt{y}}^{-}({{{z}}}):=\sum\nolimits_{{\mathtt{x}}\in\mathbb{Z}^-} {\mathtt{u}}_{{\mathtt{x}}, {\mathtt{y}}}{{{z}}}^{-{\mathtt{x}}}.\label{unpm}\end{split}\end{equation}
The symbol ${{{z}}}\in\mathbb{C}$ is exclusively used throughout as a complex variable.
The heuristic discussion below concerns the convergence of series stated in \eqref{unpm} and
follows the analysis of \S2.1 in \cite{sK} and \S2 in \cite{sharma2017scattering} (and 
part of chapter II in \cite{Noble}). 

\subsection{Conditions for well-posedness}
\label{wellposed}

The geometric part of ${\mathtt{u}}$, including the reflected and transmitted surface waves in region (1), (2) of Fig. \ref{SchematicFig2D}, is expected to have the form (with ${{\upkappa}}_{2{\mathit A}}, {{\upkappa}}_{2{\mathit B}}>0$ such that ${{\upkappa}}_{2{\mathit A}}, {{\upkappa}}_{2{\mathit B}}\to0+$ as $\upomega_2\to0+$):
\begin{equation}\begin{split}
{\mathtt{u}}_{\mathtt{x}, \mathtt{y}}^{\textrm{rf}\textrm{tr}}&:={{\mathtt{u}}^{{\mathit A}}_{0,0}}\exp({+i{{\upxi}}_{{\mathit A}} {\mathtt{x}}}-{{\upkappa}}_{2{\mathit A}}{\mathtt{x}})\exp({{\eta}({\upxi}_{{\mathit A}}) {\mathtt{y}}})
{\mathscr{H}}(\mathtt{x})\\
&+{{\mathtt{u}}^{{\mathit B}}_{0,0}}\exp({-i{{\upxi}}_{{\mathit B}} {\mathtt{x}}}+{{\upkappa}}_{2{\mathit B}}{\mathtt{x}})\exp({{\eta}({\upxi}_{{\mathit B}}) {\mathtt{y}}})
{\mathscr{H}}(-\mathtt{x}), \quad (\mathtt{x},\mathtt{y})\in\mathbb{H},
\label{uref}
\end{split}\end{equation}
while the incident wave \eqref{uincawave2d} appears as:
\begin{equation}\begin{split}
{\mathtt{u}}_{\mathtt{x}, \mathtt{y}}^{\textrm{in}}&:={{\mathtt{u}}^{{\mathit A}_{\textrm{in}}}_{0,0}}\exp({-i{{\upxi}}_{{\mathit A}} {\mathtt{x}}}+{{\upkappa}}_{2{\mathit A}}{\mathtt{x}})\exp({{\eta}({\upxi}_{{\mathit A}}) {\mathtt{y}}}), \quad (\mathtt{x},\mathtt{y})\in\mathbb{H}.
\end{split}\end{equation}
The diffracted waves, defined by ${\mathtt{u}}^{\textrm{df}}:={\mathtt{u}}-{\mathtt{u}}^{\textrm{rf}\textrm{tr}}$, can be expressed in polar coordinates $({R}, {\theta})$ (specially in region (3) of Fig. \ref{SchematicFig2D}),
for $(\mathtt{x}, \mathtt{y})\in{\mathbb{H}}$,
specified by the relations
$\mathtt{x}={R}\cos{\theta}, \mathtt{y}={R}\sin{\theta}, \text{ with }{R}>0, {\theta}\in(-\pi, 0).$
${\mathbb{H}}$ is divided into three regions (see Fig. \ref{SchematicFig2D}):
 (1) -- consists of a diffracted wave and a reflected surface wave; 
 (2) -- consists of a diffracted wave and a transmitted wave minus an incident wave;
 (3) -- consists of only a diffracted wave.
As ${R}\to\infty$, since the diffracted wave ${\mathtt{u}}^{\textrm{df}}$ is regarded as produced by a point source at the `tip' of interface, it is expected \cite{Shaban,Martin} to behave, for any fixed ${\theta}$ ($-\pi < {\theta} < 0$) as 
${\mathtt{u}}_{\mathtt{x}, \mathtt{y}}^{\textrm{df}}= {\mathit{C}}_{03}~{R}^{-\frac{1}{2}}e^{i{\upkappa}_{2b1}{R}-{\upkappa}_{2b2}{R}}+\mathpzc{o}({R}^{-\frac{1}{2}}),$
where $
 {\upkappa}_{2b1}, {\upkappa}_{2b2}>0$ and ${\mathit{C}}_{03}$ is a positive constant. 
From these statements, it is deduced that, for any fixed $\mathtt{y}\in\mathbb{Z}^-$, (${\eta}_{{\mathit A}{\mathit B}}:=\min\{{\eta}_{{\mathit A}},{\eta}_{{\mathit B}}\}>0, {{\upkappa}}_{2{\mathit A}{\mathit B}}:=\min\{{{\upkappa}}_{2{\mathit A}},{{\upkappa}}_{2{\mathit B}}\}>0$) the scattered field \eqref{uscadef}, \eqref{uscafun} satisfies
\begin{equation}\begin{split}
|{\mathtt{u}}_{\mathtt{x}, \mathtt{y}}| < ({\mathit{C}}_{01}~\exp(-{{\upkappa}}_{2{\mathit A}}\mathtt{x}){\mathscr{H}}(\mathtt{x})+{\mathit{C}}_{02}~\exp({{\upkappa}}_{2{\mathit A}{\mathit B}}\mathtt{x}){\mathscr{H}}(-\mathtt{x}))\exp({\eta}_{{\mathit A}{\mathit B}}\mathtt{y})\\
+ {\mathit{C}}_{03}~\exp (-{\upkappa}_{2b2}(\mathtt{x}^2+\mathtt{y}^2)^{\frac{1}{2}}), \mathtt{x}\in\mathbb{Z}.
\label{condab}
\end{split}\end{equation}
where ${\mathit{C}}_{01}, {\mathit{C}}_{02}$ are positive. 
Above estimates are used to capture behavior of ${\mathtt{u}}_{\mathtt{x}, \mathtt{y}}$ as $\mathtt{x}\to\pm\infty$. 
Let
${\upkappa}_{21}:=\min\{{{\upkappa}}_{2{\mathit A}}, {{\upkappa}}_{2b2}\}>0$,
${\upkappa}_{22}:=\min\{{{\upkappa}}_{2{\mathit A}}, {{\upkappa}}_{2{\mathit B}}, {{\upkappa}}_{2b2}\}>0$.
Using \eqref{condab}, 
$|{\mathtt{u}}_{\mathtt{x}, \mathtt{y}}| < {\mathit{C}}_1 e^{-{\upkappa}_{21}\mathtt{x}}$ as $\mathtt{x}\to+\infty$ and $|{\mathtt{u}}_{\mathtt{x}, \mathtt{y}}| < {\mathit{C}}_2 e^{{\upkappa}_{22}\mathtt{x}}$ as $\mathtt{x}\to-\infty$, where ${\mathit{C}}_1, {\mathit{C}}_2$ are positive constants; thus, 
\begin{equation}\begin{split}\sum\nolimits_{\mathtt{x}\in\mathbb{Z}^+} |{\mathtt{u}}_{\mathtt{x}, \mathtt{y}}{{z}}^{-\mathtt{x}}| < {\mathit{C}}_1 \sum\nolimits_{\mathtt{x}\in\mathbb{Z}^+} e^{-{\upkappa}_{21}\mathtt{x}}|{{z}}|^{-\mathtt{x}}={\mathit{C}}_1 \sum\nolimits_{\mathtt{x}\in\mathbb{Z}^+}|{{z}} e^{{\upkappa}_{21}}|^{-\mathtt{x}}, \label{C1eq}\end{split}\end{equation}
and similar statement for $\sum\nolimits_{\mathtt{x}\in\mathbb{Z}^-} |{\mathtt{u}}_{\mathtt{x}, \mathtt{y}}{{z}}^{-\mathtt{x}}|$, i.e.
\begin{equation}\begin{split}\sum\nolimits_{\mathtt{x}\in\mathbb{Z}^-} |{\mathtt{u}}_{\mathtt{x}, \mathtt{y}}{{z}}^{-\mathtt{x}}| < {\mathit{C}}_1 \sum\nolimits_{\mathtt{x}\in\mathbb{Z}^-} e^{{\upkappa}_{22}\mathtt{x}}|{{z}}|^{-\mathtt{x}}={\mathit{C}}_1 \sum\nolimits_{\mathtt{x}\in\mathbb{Z}^-}|{{z}} e^{-{\upkappa}_{22}}|^{-\mathtt{x}}.
\label{C2eq}\end{split}\end{equation}
It follows that the series defining
${\mathtt{u}}_{\mathtt{y}}^{+}({{{z}}})$ and ${\mathtt{u}}_{\mathtt{y}}^{-}({{{z}}})$, in \eqref{unpm},
are absolutely convergent, provided that 
$|{{z}}|>{\mathit{R}}_+$ and $|{{z}}|<{\mathit{R}}_-$, respectively, where
\begin{equation}\begin{split}
{\mathit{R}}_+:=e^{-{\upkappa}_{21}}, {\mathit{R}}_-:=e^{{\upkappa}_{22}}.
\label{Rpm}
\end{split}\end{equation}
Note that ${\mathit{R}}_+<1<{\mathit{R}}_-$. Therefore, using \eqref{unpm} for any fixed $\mathtt{y}\in\mathbb{Z}^-\cup\{0\}$, ${\mathtt{u}}_{\mathtt{y};+}({{z}})$ and ${\mathtt{u}}_{\mathtt{y};-}({{z}})$ are analytic at ${{z}}\in\mathbb{C}$ such that $|{{z}}|>{\mathit{R}}_+,$ $|{{z}}|<{\mathit{R}}_-,$ respectively.
Putting both facts together it is concluded that ${\mathtt{u}}_{\mathtt{y}}^{\rm F}$ \eqref{unpm} is analytic inside an annulus in $\mathbb{C}$ (containing $\mathbb{T}$), defined by
\begin{equation}\begin{split}
{{\mathscr{A}}}_u:=\{{{z}}\in\mathbb{C}: {\mathit{R}}_+< |{{z}}|< {\mathit{R}}_-\};
\label{annulusAu}
\end{split}\end{equation}
see Appendix A of \cite{sK} for more details on definitions 
related to the transform \eqref{unpm}. 

\subsection{Description of scattered field on $\mathring{\mathbb{H}}$}
\label{sec2dhalf}

The scattering solution ${\mathtt{u}}:\mathbb{H}\to\mathbb{C}$ \eqref{uscafun} satisfies \eqref{eqmotionsca} away from the boundary $\partial{\mathbb{H}}$ (recall definition \eqref{hlfplane}).
Assuming $\upomega_2>0$, following the analysis of \cite{sK} and \cite{sharma2017scattering,Victor_Bls_surf2}, 
it is found that the Fourier transform ${\mathtt{u}}_{\mathtt{y}}^{\rm F}$ of the scattered wavefield ${\mathtt{u}}$ on $\mathring{\mathbb{H}}$
can be expressed in terms of ${\mathtt{u}}_{0}^{\rm F}$ as
\begin{subequations}
\begin{align}
{\mathtt{u}}^{\rm F}_{\mathtt{y}}({{{z}}})&={\mathtt{u}}^{\rm F}_0({{{z}}}){\lambda}^{-{\mathtt{y}}}({{{z}}}), \quad {\mathtt{y}}\in{\mathbb{Z}}^-\cup\{0\},\label{halfspace}\\
\lambda({{z}})&:= \frac{{\mathtt{r}}({{z}})-{\mathtt{h}}({{z}})}{{\mathtt{r}}({{z}})+{\mathtt{h}}({{z}})},\label{defyzy}\\
\text{where }{\mathtt{h}}({{z}})&:=\sqrt{{\mathtt{Q}}({{z}})-2},\quad
{\mathtt{r}}({{z}}):=\sqrt{{\mathtt{Q}}({{z}})+2},\quad
{\mathtt{Q}}({{z}}):=2+{{\alpha}_{{}}}(2-{{z}}-{{z}}^{-1})-{\upomega}^2.
\label{defQ}
\end{align}
\label{ansatzhalfspace}
\end{subequations}
Assuming $\upomega_2>0$ and following \cite{sK}, as briefly discussed in the beginning of this section,
the Fourier transform as well as other functions stated above (according to the summation \eqref{unpm}) are analytic in (non-empty) annulus ${{{\mathscr{A}}}}$ defined by
\begin{equation}\begin{split}
{{\mathscr{A}}}\subset{{\mathscr{A}}}_u\cap{{\mathscr{A}}}_L, ~{{\mathscr{A}}}_L:=\{{{z}}\in\mathbb{C} : {\mathit{R}}_L< |{{z}}|< {\mathit{R}}_L^{-1}\}, {\mathit{R}}_L:=\max\{|{{z}}_{{\mathtt{h}}}|, |{{z}}_{{\mathtt{r}}}|\},
\label{annulusA}
\end{split}\end{equation}
where ${z}_{\mathtt{h}}, {z}_{\mathtt{h}}^{-1}$ (resp. ${z}_{\mathtt{r}}, {z}_{\mathtt{r}}^{-1}$) are zeros of ${\mathtt{h}}$ (resp. ${\mathtt{r}}$) such that $|{z}_{\mathtt{h}}|, |{z}_{\mathtt{r}}|<1$.
Assuming $\upomega_2>0$, it is easily found that zeros of ${\mathtt{h}}$ and ${\mathtt{r}}$ do not belong to the unit circle $\mathbb{T}$.
Note that ${\mathbb{T}}\subset{{{\mathscr{A}}}}$. 

\begin{remark}
The inner and outer radius of ${{{\mathscr{A}}}}$ can be expressed as
${{\mathit{R}_c}}$, ${{\mathit{R}_c}}^{-1}$, respectively, where ${\mathit{R}_c}<1$.
\label{annulus}
\end{remark}

\begin{remark}
Rigorous arguments behind heuristics of \S\ref{wellposed} and definition \eqref{annulusA} can be founded upon the following.
The complex frequency \eqref{complexfq} analog of surface wave mode in \eqref{uincSwave} 
is 
${z}_{{sf}}^{-\mathtt{x}}\lambda^{-\mathtt{y}}{\mathscr{H}}(-\mathtt{y})$ where $|{z}_{{sf}}|<1$ (resp. $|{z}_{{sf}}|>1$) for waveform that decays (outgoing) 
towards $\mathtt{x}\to-\infty$ (resp. 
$\mathtt{x}\to+\infty$) and
\begin{align}\label{funcomegaeqFT}
{\upomega}^2&=-{{\alpha}_{{}}}({z}_{{sf}}+{z}_{{sf}}^{-1}-2)+2-(\lambda_{{sf}}+\lambda^{-1}_{{sf}}),\quad
{\mathit{m}}_{{sf}}{\upomega}^2=-{\alpha}_{{sf}} ({z}_{{sf}}+{z}_{{sf}}^{-1}-2)+1-\lambda_{{sf}},
\end{align}
where $\lambda_{{sf}}:=\lambda({z}_{{sf}}).$
In 
fact, conditions \eqref{funcomegaeqFT} are equivalent to the surface wave dispersion relation \eqref{funcomegaeq}.
The first equation is satisfied by the definition \eqref{defyzy} of $\lambda$ when $\upomega_1$ (with $\upomega_2$ vanishingly small) in \eqref{complexfq} belongs to the 
surface wave band (with material parameters ${\mathit{m}}_{{sf}}, {\alpha}_{{sf}}$) and the pass band of bulk lattice; see also \cite{sK} for more discussion on $\lambda$, modulo accommodating the parameter ${{\alpha}_{{}}}>0$ that calls minor modifications.
The second equation is the condition that describes ${z}_{{sf}}$ for a given $\upomega$ in surface wave band.
\label{annulusSM}
\end{remark}
With above Remark \ref{annulusSM},
assuming $\upomega_2>0$, 
it is convenient to define
\begin{equation}\begin{split}
{\mathit{m}}_{{\mathit A}}{\upomega}^2=-{\alpha}_{{\mathit A}} ({z}_{{\mathit A}}+{z}_{{\mathit A}}^{-1}-2)+1-\lambda({z}_{{\mathit A}}),~
{\mathit{m}}_{{\mathit B}}{\upomega}^2=-{\alpha}_{{\mathit B}} ({z}_{{\mathit B}}+{z}_{{\mathit B}}^{-1}-2)+1-\lambda({z}_{{\mathit B}}),\\
\text{ such that }{{z}}_{{\mathit A}}, {{z}}_{{\mathit B}}\in\mathbb{C} \text{ and }
|{{z}}_{{\mathit A}}|, |{{z}}_{{\mathit B}}|<1.
\label{defzazb}
\end{split}\end{equation}

\begin{remark}
As $\upomega_2\to0+$, the scattering solution ${\mathtt{u}}$ \eqref{uscafun} is recovered that satisfies radiation conditions \cite{Shaban}.
In view of Remark \ref{annulusSM},
moreover, it can be verified that
${z}_{{\mathit A}}\to\exp(i{{\upxi}}_{{\mathit A}}),$ 
$\lambda(\exp(\pm i{{\upxi}}_{{\mathit A}}))\to\exp(-{{\eta}({\upxi}_{{\mathit A}})})$, 
${z}_{{\mathit B}}\to\exp(i{{\upxi}}_{{\mathit B}}),$
$\lambda(\exp(\pm i{{\upxi}}_{{\mathit B}}))\to\exp(-{{\eta}({\upxi}_{{\mathit B}})})$ as $\upomega_2\to0+$;
recall 
\eqref{uincSwave}, \eqref{surfgpvel}.
\label{annulus2}
\end{remark}

Counterparts of \eqref{uincawave2d}${}_2$, \eqref{urefawave}, \eqref{utransbwave}, for the complex frequency \eqref{complexfq}, are
\begin{align}
{\mathtt{u}}^{{\mathit A}_{\textrm{in}}}_{{\mathtt{x}},{\mathtt{y}}}:={{{\mathtt{u}}^{{\mathit A}_{\textrm{in}}}_{0,0}}}{{z}}_{{\mathit A}}^{-{\mathtt{x}}}\lambda({{z}}_{{\mathit A}})^{-{\mathtt{y}}},
{\mathtt{u}}^{{\mathit A}}_{{\mathtt{x}},{\mathtt{y}}}:={{\mathtt{u}}^{{\mathit A}}_{0,0}}{{z}}_{{\mathit A}}^{{\mathtt{x}}}\lambda({{z}}_{{\mathit A}})^{-{\mathtt{y}}},
{\mathtt{u}}^{{\mathit B}}_{{\mathtt{x}},{\mathtt{y}}}:={{\mathtt{u}}^{{\mathit B}}_{0,0}}{{z}}_{{\mathit B}}^{-{\mathtt{x}}}\lambda({{z}}_{{\mathit B}})^{-{\mathtt{y}}},
\quad (\mathtt{x},\mathtt{y})\in\mathbb{H}.
\label{waveallcomp}
\end{align}

To find the unknown complex valued function ${\mathtt{u}}^{\rm F}_0$ in \eqref{halfspace}, 
the equation of motion on $\partial{\mathbb{H}}$ \eqref{eqmotionBCFT} is used.
This is the task achieved by an application of Wiener-Hopf technique \cite{Wiener,Noble}.

\subsection{Scattered field on $\partial{\mathbb{H}}$: {Wiener-Hopf} equation and its exact solution}
\label{sec2dWH2}

For convenience of the reader, in the following, the symbols in \eqref{halfspace}, and \eqref{unpm} for $\mathtt{y}=0$,
\begin{equation}\begin{split}
{\mathtt{u}}^{\rm F}_0,{\mathtt{u}}^+_0,{\mathtt{u}}^-_0\text{ are replaced by }{\mathtt{u}}^{\rm F},{\mathtt{u}}^+,{\mathtt{u}}^-,
\label{conven}
\end{split}\end{equation}
respectively.
The Fourier transform of ${{\mathscr{H}}}(-{\mathtt{x}}-1)$ (resp. ${{\mathscr{H}}}({\mathtt{x}})$), with ${\mathtt{x}}\in\mathbb{Z}$, is denoted by ${\updelta}_{D}^-$ (resp. ${\updelta}_{D}^+$).
Recall \eqref{defH} for the definition of ${{\mathscr{H}}}$.
The Fourier transform of \eqref{BCeq} (collecting the terms accompanying ${\mathtt{u}}^{\pm}$; detailed form provided in Appendix \ref{appA}, for example \eqref{discWHkerpre}) leads to the following {Wiener-Hopf} equation on ${\mathscr{A}}$:
\begin{subequations}
\begin{align}
{\mathfrak{K}}_{{\mathit A}}({z}){\mathtt{u}}^+({z})&+{\mathfrak{K}}_{{\mathit B}}({z}){\mathtt{u}}{}^-({z})=
\mathtt{p}({z}){{{\mathtt{u}}^{{\mathit A}_{\textrm{in}}}_{0,0}}}{\updelta}_{D}^-({{z}}{{z}}_{{\mathit A}})-({\alpha}_{{\mathit A}}-{\alpha}_{{\mathit B}})(1-{{z}}){({\mathtt{u}}_{0, 0}+{\mathtt{u}}^{{\mathit A}_{\textrm{in}}}_{0, 0})}\notag\\
&-({\mathit{m}}_{0}-{\mathit{m}}_{{\mathit A}})\upomega^2{({\mathtt{u}}_{0, 0}+{\mathtt{u}}^{{\mathit A}_{\textrm{in}}}_{0, 0})},
\label{discWH}\\
\text{where }
{\updelta}_{D}^-({{z}}{{z}}_{{\mathit A}})&
=-{{{z}}}/({{{z}}-{{z}}_{{\mathit A}}^{-1}}),\label{uincF}\\
{\mathfrak{K}}_{\mathfrak{s}}&:={\lambda}+\mathtt{F}_{\mathfrak{s}},\quad
\label{discWHker}
\mathtt{F}_{\mathfrak{s}}({{z}}):={\mathit{m}}_{\mathfrak{s}}{\upomega}^2-1+{\alpha}_{\mathfrak{s}}({{z}}+{{z}}^{-1}-2),
\text{ with }{\mathfrak{s}}={{\mathit A}}, {{\mathit B}},\\
\mathtt{p}({z})&:=({\alpha}_{{\mathit A}}-{\alpha}_{{\mathit B}})({{z}}+{{z}}^{-1}-2)+({\mathit{m}}_{{\mathit A}}-{\mathit{m}}_{{\mathit B}}){\upomega}^2,\label{kAkBidentity}
\text{ and }{z}\in{\mathscr{A}}.
\end{align}
\label{discWHalleq}
\end{subequations}
Note that $0<|\lambda|<1$ and $\lambda({z})=\lambda({z}^{-1}), {\mathfrak{K}}_{{\mathit A}}({z})={\mathfrak{K}}_{{\mathit A}}({z}^{-1}), {\mathfrak{K}}_{{\mathit B}}({z})={\mathfrak{K}}_{{\mathit B}}({z}^{-1}), $ in ${{{\mathscr{A}}}}$ defined by \eqref{annulusA}.
Using 
\eqref{discWHker} and \eqref{kAkBidentity}, 
it can be verified that
$\mathtt{p}={\mathfrak{K}}_{{\mathit A}}-{\mathfrak{K}}_{{\mathit B}}.$
It is emphasized that all functions appearing in \eqref{discWH} are analytic in ${\mathscr{A}}$ \eqref{annulusA}. 
Moreover, ${\mathfrak{K}}_{{\mathit A}}$ and ${\mathfrak{K}}_{{\mathit B}}$ do not vanish in ${\mathscr{A}}$.

\begin{remark}
Suppose that ${\mathfrak{K}}_{{\mathit A}}({z}_\ast)=0$ where ${z}_\ast\in\mathbb{T}\subset{\mathscr{A}}$. By \eqref{discWHker}, this implies $\Im(\lambda({z}_\ast)+{\mathit{m}}_{{\mathit A}}\upomega^2)=0$. According to the definitions \eqref{defyzy} and \eqref{defQ}, 
$-\Im(\upomega^2)=\Im{\mathtt{Q}({z}_\ast)}=\Im(\lambda({z}_\ast)+\lambda^{-1}({z}_\ast))$ but $\Im\upomega^2\ne0$ and $\mathtt{Q}({z}_\ast)$ is bounded so that $|\lambda({z}_\ast)|\ne0,|\lambda({z}_\ast)|\ne1$, in fact $0<|\lambda({z}_\ast)|<1$. 
Hence, 
$(1-{\mathit{m}}_{{\mathit A}}(1-\frac{1}{|\lambda({z}_\ast)|^2}))\Im\upomega^2=0$ but ${\mathit{m}}_{{\mathit A}}>0$ 
which implies 
a contradiction.
Hence, ${\mathfrak{K}}_{{\mathit A}}$ does not vanish on $\mathbb{T}$ when $\upomega_2>0$; same proof holds for ${\mathfrak{K}}_{{\mathit B}}$.
By analyticity the statement follows that ${\mathfrak{K}}_{{\mathit A}}$ and ${\mathfrak{K}}_{{\mathit B}}$ do not vanish on certain annulus containing $\mathbb{T}$.
\end{remark}

\begin{remark}
Using \eqref{defzazb} and the definitions \eqref{discWHker},
and taking the context of Remark \ref{annulus2}, 
Remark \ref{altcase1}, and the surface wave dispersion relation \eqref{funcomegaeq}, by inspection, it can be verified that 
${\mathfrak{K}}_{{\mathit A}}({z}_{{\mathit A}}^{\pm1})=0$ and 
${\mathfrak{K}}_{{\mathit B}}({z}_{{\mathit B}}^{\pm1})=0$.
\eqref{defzazb}. In view of the Remark \ref{annulus} and definition \eqref{annulusA}, ${{z}}_{{\mathit A}}^{\pm1}, {{z}}_{{\mathit B}}^{\pm1}$ lie outside the annulus ${\mathscr{A}}$; in fact, when $\upomega_2>0$, annulus ${\mathscr{A}}$ exists such that $|{{z}}_{{\mathit A}}|,|{{z}}_{{\mathit B}}|<{{\mathit{R}_c}}<1$, regarding ${{\mathit{R}_c}}$ 
in Remark \ref{annulus}.
\label{annulus3}
\end{remark}

The {Wiener-Hopf} kernel in \eqref{discWH} can be identified with the function 
\begin{equation}
\mathfrak{L}:={{\mathfrak{K}}_{{{\mathit B}}}}/{{\mathfrak{K}}_{{{\mathit A}}}},
\label{defKerL}
\end{equation}
which is analytic and does not vanish in the annulus ${\mathscr{A}}$ (recall Remark \ref{annulus3} and 
\eqref{annulusA}).
The multiplicative {Wiener-Hopf} factorization of 
$\mathfrak{L}$ exists in the annulus ${\mathscr{A}}$ \cite{Wiener,Noble} such that
\begin{equation}\begin{split}
{\mathfrak{L}}({z})={\mathfrak{L}}_+({z}){\mathfrak{L}}_-({z}), {z}\in{\mathscr{A}},
\quad
{\mathfrak{L}}_\pm({z})=\exp(\pm\frac{1}{2\pi i}\oint_{{\mathbb{T}}}\frac{\log\mathfrak{L}({\alpha})}{{z}-{\alpha}}\,d{\alpha}),{z}\in\mathbb{C},\text{with}~|{z}|\lessgtr {{\mathit{R}_c}}^{\mp1},
\label{Lpmfacs}
\end{split}\end{equation}
where ${{\mathbb{T}}}$ can replaced by any rectifiable, closed, counterclockwise contour that lies in the same annulus ${\mathscr{A}}$ and ${{\mathit{R}_c}}$ is defined in Remark \ref{annulus}. As ${\mathfrak{L}}({z})={\mathfrak{L}}({z}^{-1}), {z}\in{\mathscr{A}}$, it is convenient to choose above factors in \eqref{Lpmfacs}, so that ${\mathfrak{L}}_+({z}^{-1})={\mathfrak{L}}_-({z}), {z}\in{\mathscr{A}}$. Additionally, due to the assumption $\upomega_2>0$, ${\mathfrak{L}}$ is continuous and bounded on ${\mathscr{A}}$, so that
\eqref{Lpmfacs} implies that the following limit exists:
\begin{align}
\lim_{{z}\to0}{\mathfrak{L}}_-({z})=\lim_{{z}\to\infty}{\mathfrak{L}}_+({z})=:\ell_\infty\in\mathbb{C}.
\label{LpLmlim}
\end{align}
\begin{remark}
As a result of the definition \eqref{Lpmfacs} and similar factorization for numerator and denominator in \eqref{defKerL}, that is ${\mathfrak{K}}_{{\mathit A}}={\mathfrak{K}}_{{{\mathit A}}{}+}{\mathfrak{K}}_{{{\mathit A}}{}-}, {\mathfrak{K}}_{{\mathit B}}={\mathfrak{K}}_{{{\mathit B}}{}+}{\mathfrak{K}}_{{{\mathit B}}{}-},$ and
\begin{align}
{\mathfrak{L}}_+={{\mathfrak{K}}_{{{\mathit B}}{}+}}/{{\mathfrak{K}}_{{{\mathit A}}{}+}},\quad {\mathfrak{L}}_-={{\mathfrak{K}}_{{{\mathit B}}{}-}}/{{\mathfrak{K}}_{{{\mathit A}}{}-}}\text{ on }{\mathscr{A}}.
\label{discWHLpLm}
\end{align}
By Remark \ref{annulus3}, using 
definitions \eqref{defzazb}:
${\mathfrak{K}}_{{\mathit A}\pm}({{z}}_{{\mathit A}}^{\pm1})=0$,
${\mathfrak{K}}_{{\mathit B}\pm}({{z}}_{{\mathit B}}^{\pm1})=0$.
Note that ${\mathfrak{L}}_+, {\mathfrak{K}}_{{\mathit A}+}, {\mathfrak{K}}_{{\mathit B}+}$ (resp. ${\mathfrak{L}}_-, {\mathfrak{K}}_{{\mathit A}-}, {\mathfrak{K}}_{{\mathit B}-}$) are analytic,
non-vanishing on
$\{{z}\in\mathbb{C}: |{z}|>{{\mathit{R}_c}}\}$ (resp. $\{{z}\in\mathbb{C}: |{z}|<{{\mathit{R}_c}}^{-1}\}$).
\label{annulus4}
\end{remark}

\begin{equation}\begin{split}
\text{Assuming }
{\alpha}_{{\mathit A}}-{\alpha}_{{\mathit B}}\ne0,
\label{aAaBassumed}
\end{split}\end{equation}
all 
zeros of $\mathtt{p}$ are 
${z}_{\mathtt{p}}^{\pm1}$ with $|{z}_{\mathtt{p}}|<1$ (due to \eqref{complexfq}).
This leads to 
{Wiener-Hopf} factorization
of $\mathtt{p}$:
\begin{equation}\begin{split}
\mathtt{p}=\mathtt{p}_+\mathtt{p}_-, ~\mathtt{p}_+({z}):={\mathtt{C}}_{\mathtt{p}}(1-{z}_{\mathtt{p}}/{z}), \mathtt{p}_-({z}):=\mathtt{p}_+({z}^{-1}), ~{z}\in{\mathscr{A}},\quad
{\mathtt{C}}_{\mathtt{p}}&:=\sqrt{-{z}_{\mathtt{p}}^{-1}({\alpha}_{{\mathit A}}-{\alpha}_{{\mathit B}})}.
\label{defpppm}
\end{split}\end{equation}
In light of above observation and Remark \ref{annulus}, it is convenient to define
\begin{equation}
{{\mathit{R}_c}}:=\max\{{\mathit{R}}_+, {\mathit{R}}_-^{-1}, {\mathit{R}}_L, |{z}_{\mathtt{p}}|\}<1,
\label{defannulus}
\end{equation}
using the definitions of radii defined in \eqref{Rpm} and \eqref{annulusA}.
Due to Remark \ref{annulus4}, upon division of both sides in \eqref{discWH} by ${\mathfrak{K}}_{{{\mathit B}}{}+}{\mathfrak{K}}_{{{\mathit A}}{}-}$, and using \eqref{defpppm}, \eqref{defzazb}, and \eqref{uincF}, 
it becomes
\begin{align}
{\mathfrak{L}}_+^{-1}({z}){\mathtt{u}}^+({z})+{\mathfrak{L}}_-({z}){\mathtt{u}}^-({z})&={\mathfrak{C}}({z}),\quad {z}\in{\mathscr{A}},\label{discWHN2}
\end{align}
\begin{align}
\text{with }
{\mathfrak{C}}({z})&:=({\mathfrak{L}}_-({z})-{\mathfrak{L}}_+^{-1}({z}))\bigg(-{{{\mathtt{u}}^{{\mathit A}_{\textrm{in}}}_{0,0}}}{\updelta}_{D}^-({{z}}{{z}}_{{\mathit A}})+({\alpha}_{{\mathit A}}-{\alpha}_{{\mathit B}})\frac{(1-{{z}})+\mathtt{M}}{\mathtt{p}_+({z})\mathtt{p}_-({z})}{({\mathtt{u}}_{0, 0}+{\mathtt{u}}^{{\mathit A}_{\textrm{in}}}_{0, 0})}\bigg),\label{discWHdefC}\\
\text{ and }\mathtt{M}&:={({\mathit{m}}_{0}-{\mathit{m}}_{{\mathit A}})\upomega^2}/{({\alpha}_{{\mathit A}}-{\alpha}_{{\mathit B}})}.
\label{defArat}
\end{align}
The additive {Wiener-Hopf} factorization ${\mathfrak{C}}={\mathfrak{C}}^++{\mathfrak{C}}^-$ on ${\mathscr{A}}$ holds \cite{sK,Noble} with
\begin{align}
{\mathfrak{C}}^\pm({{z}})
&={{{\mathtt{u}}^{{\mathit A}_{\textrm{in}}}_{0,0}}}\frac{(\mp{\mathfrak{L}}_\pm^{\mp1}({{z}})\pm{\mathfrak{L}}_+^{-1}({{z}}_{{\mathit A}}^{-1}))}{1-{{z}}_{{\mathit A}}^{-1}{{z}}^{-1}}\pm({\alpha}_{{\mathit A}}-{\alpha}_{{\mathit B}}){({\mathtt{u}}_{0, 0}+{\mathtt{u}}^{{\mathit A}_{\textrm{in}}}_{0, 0})}\bigg({\mathfrak{L}}_-({z}_{\mathtt{p}})\frac{(1+\mathtt{M}-{{z}_{\mathtt{p}}})}{{\mathtt{p}_+({z})}\mathtt{p}_-({z}_{\mathtt{p}})}\notag\\&
-{\mathfrak{L}}_\pm^{\mp1}({z})\frac{(1+\mathtt{M}-{z})}{\mathtt{p}_+({z})\mathtt{p}_-({z})}-{\mathfrak{L}}_+^{-1}({z}_{\mathtt{p}}^{-1})\frac{(1+\mathtt{M}-{{z}_{\mathtt{p}}^{-1}})}{\mathtt{p}_+({z}_{\mathtt{p}}^{-1})}(\frac{1}{\mathtt{p}_-(0)}-\frac{1}{\mathtt{p}_-({z})})\bigg),\quad {z}\in{\mathscr{A}}.
\label{discCfacalleq}
\end{align}
Note that ${\mathfrak{L}}_+^{-1}{\mathtt{u}}^+-{\mathfrak{C}}^+$ (resp. $-{\mathfrak{L}}_-{\mathtt{u}}^-+{\mathfrak{C}}^-$) is analytic on the region $\{{z}\in\mathbb{C}: |{z}|>{{\mathit{R}_c}}\}$ (resp. $\{{z}\in\mathbb{C}: |{z}|<{{\mathit{R}_c}}^{-1}\}$) in the complex plane.
Further, 
$\lim_{{z}\to0}{\mathfrak{L}}_-({{z}}){\mathtt{u}}^-({{z}})=0, \lim_{{z}\to0}{\mathfrak{C}}^-({{z}})
=0,$
while
$\lim_{{z}\to\infty}{\mathfrak{L}}_{+}^{-1}({{z}}){\mathtt{u}}^+({{z}})$
and
$\lim_{{z}\to\infty}{\mathfrak{C}}^+({{z}})$
exists.
It is useful to note that $\lim_{{z}\to0}1/\mathtt{p}_+({z})=0$
in \eqref{discCfacalleq}.
As a consequence of the Lioville's theorem
\begin{equation}\begin{split}
{\mathtt{u}}^+({{z}})={\mathfrak{L}}_{+}({{z}}){\mathfrak{C}}^+({{z}}), \quad {\mathtt{u}}^-({{z}})={\mathfrak{L}}_-^{-1}({{z}}){\mathfrak{C}}^-({{z}}),\quad {z}\in{\mathscr{A}}. 
\label{discWHNsolhalf}
\end{split}\end{equation}

\begin{remark}
In violation of \eqref{aAaBassumed}, assuming ${\alpha}_{{\mathit A}}={\alpha}_{{\mathit B}}$ but ${\mathit{m}}_{{\mathit A}}\ne{\mathit{m}}_{{\mathit B}}$, according to 
definition 
\eqref{kAkBidentity},
$\mathtt{p}({z})=({\mathit{m}}_{{\mathit A}}-{\mathit{m}}_{{\mathit B}}){\upomega}^2$, which is a constant (independent of ${z}$),
so that
${\mathfrak{C}}({{z}})=({\mathfrak{L}}_-({z})-{\mathfrak{L}}_+^{-1}({z}))(-{{{\mathtt{u}}^{{\mathit A}_{\textrm{in}}}_{0,0}}}{\updelta}_{D}^-({{z}}{{z}}_{{\mathit A}})+\frac{{\mathit{m}}_{0}-{\mathit{m}}_{{\mathit A}}}{{\mathit{m}}_{{\mathit A}}-{\mathit{m}}_{{\mathit B}}}({\mathtt{u}}_{0, 0}+{\mathtt{u}}^{{\mathit A}_{\textrm{in}}}_{0, 0}))={\mathfrak{C}}^+({{z}})+{\mathfrak{C}}^-({{z}})$ holds with
\begin{equation}\begin{split}
{\mathfrak{C}}^\pm({{z}})
&={{{\mathtt{u}}^{{\mathit A}_{\textrm{in}}}_{0,0}}}\frac{(\mp{\mathfrak{L}}_\pm^{\mp1}({{z}})\pm{\mathfrak{L}}_+^{-1}({{z}}_{{\mathit A}}^{-1}))}{1-{{z}}_{{\mathit A}}^{-1}{{z}}^{-1}}\pm({\mathfrak{L}}_-(0)-{\mathfrak{L}}_\pm^{\mp1}({z}))\frac{{\mathit{m}}_{0}-{\mathit{m}}_{{\mathit A}}}{{\mathit{m}}_{{\mathit A}}-{\mathit{m}}_{{\mathit B}}}({\mathtt{u}}_{0, 0}+{\mathtt{u}}^{{\mathit A}_{\textrm{in}}}_{0, 0}),\quad {z}\in{\mathscr{A}},
\label{discCfacalleqeqaAB}
\end{split}\end{equation}
in place of \eqref{discCfacalleq}.
Assuming ${\alpha}_{{\mathit A}}={\alpha}_{{\mathit B}}$ and ${\mathit{m}}_{0}={\mathit{m}}_{{\mathit A}}$ but ${\mathit{m}}_{{\mathit A}}\ne{\mathit{m}}_{{\mathit B}}$, it is even a simpler situation.
The lattice contains only a semi-infinite mass defect on $\partial{\mathbb{H}}$ and the problem concerns the edge scattering due to the defect tip in contrast to \eqref{discCfacalleqeqaAB} where the effect of ${\mathit{m}}_{0}-{\mathit{m}}_{{\mathit A}}$ appears.
Then 
the term containing
${({\mathtt{u}}_{0, 0}+{\mathtt{u}}^{{\mathit A}_{\textrm{in}}}_{0, 0})}$ disappear in the r.h.s. of \eqref{discWH}.
Thus, ${\mathfrak{C}}={\mathfrak{C}}^++{\mathfrak{C}}^-$ with expressions as special case of the ones stated in \eqref{discCfacalleq}; in particular, only the first term in \eqref{discCfacalleq} appears so that
\begin{align}
{\mathfrak{C}}^\pm({{z}})&={{{\mathtt{u}}^{{\mathit A}_{\textrm{in}}}_{0,0}}}(\mp{\mathfrak{L}}_\pm^{\mp1}({{z}})\pm{\mathfrak{L}}_+^{-1}({{z}}_{{\mathit A}}^{-1}))\frac{{{z}}}{{{z}}-{{z}}_{{\mathit A}}^{-1}}, {z}\in{\mathscr{A}}.
\label{discCpmhalfcase2}
\end{align}
Finally, proceeding as above, it is found that
${\mathtt{u}}^+({{z}})={{{\mathtt{u}}^{{\mathit A}_{\textrm{in}}}_{0,0}}}{\updelta}_{D}^-({{z}}{{z}}_{{\mathit A}})(1-{\mathfrak{L}}_{+}({{z}}){\mathfrak{L}}_+^{-1}({{z}}_{{\mathit A}}^{-1})),$ 
${\mathtt{u}}^-({{z}})=-{{{\mathtt{u}}^{{\mathit A}_{\textrm{in}}}_{0,0}}}{\updelta}_{D}^-({{z}}{{z}}_{{\mathit A}})(1-{\mathfrak{L}}_-^{-1}({{z}}){\mathfrak{L}}_+^{-1}({{z}}_{{\mathit A}}^{-1})), {z}\in{\mathscr{A}}.$
\label{speccase2}
\end{remark}

\subsection{Expression for scattered wave field on $\mathbb{H}$}
\label{sec2dsolsca}

After adding both half-Fourier transforms \eqref{discWHNsolhalf}, i.e. ${\mathtt{u}}^{\rm F}({{z}})={\mathtt{u}}^+({{z}})+{\mathtt{u}}^-({{z}}), {z}\in{\mathscr{A}}$, and simplifying further, it is found that
\begin{equation}\begin{split}
{\mathtt{u}}^{\rm F}({{z}})
=\mathcal{T}_1({{z}}){{\mathtt{u}}^{{\mathit A}_{\textrm{in}}}_{0,0}}+\mathcal{T}_2({{z}}){({\mathtt{u}}_{0, 0}+{\mathtt{u}}^{{\mathit A}_{\textrm{in}}}_{0, 0})},\quad
\mathcal{T}_1({{z}}):=-\frac{1}{{\mathfrak{L}}_+
}\frac{1-{\mathfrak{L}}({{z}})}{{\mathfrak{L}}_-({{z}})}\frac{{{z}}}{{{z}}-{{z}}_{{\mathit A}}^{-1}},\\
\mathcal{T}_2({{z}})
:=\frac{({\alpha}_{{\mathit A}}-{\alpha}_{{\mathit B}})(1-{\mathfrak{L}}({{z}}))}{{\mathfrak{L}}_-({{z}})\mathtt{p}_-({z}_{\mathtt{p}})}\bigg(\frac{-{\mathfrak{L}}_-({z}_{\mathtt{p}})}{\mathtt{p}_+({z})}(1+\mathtt{M}-{{z}_{\mathtt{p}}})+\frac{\frac{1}{\mathtt{p}_-(0)}-\frac{1}{\mathtt{p}_-({z})}}{{\mathfrak{L}}_+({z}_{\mathtt{p}}^{-1})}(1+\mathtt{M}-{{z}_{\mathtt{p}}^{-1}})\bigg),
\label{discWHNsolu}
\end{split}\end{equation}
for all ${z}\in{\mathscr{A}}.$
Using the inversion formula for \eqref{unpm} \cite{sK} and \eqref{conven}, it is found that
\begin{equation}\begin{split}
{\mathtt{u}}_{{\mathtt{x}}, \mathtt{y}}=\frac{1}{2\pi i}\oint\nolimits_{\mathbb{T}}{\mathtt{u}}^{\rm F}({{z}})\lambda^{-\mathtt{y}}({{z}}){{z}}^{{\mathtt{x}}-1}d{{z}}, \quad (\mathtt{x},\mathtt{y})\in\mathbb{H},
\label{invuF}
\end{split}\end{equation}
so that the scattering solution can be determined over ${\mathbb{H}}$
(recall definition \eqref{hlfplane}).

\begin{remark}
With $\mathtt{y}=0$ in \eqref{invuF}, 
${\mathtt{u}}_{0, 0}=\frac{1}{2\pi i}\oint\nolimits_{\mathbb{T}}{\mathtt{u}}^{\rm F}({{z}}){{z}}^{-1}d{{z}},$
which, upon simplifying further,
using \eqref{discWHNsolu}, yields
\begin{equation}\begin{split}
{\mathtt{u}}_{0, 0}
&={\mathtt{u}}_{0, 0}^{{\mathit A}_{\textrm{in}}}{\mathfrak{F}}_0
+({\mathtt{u}}_{0, 0}+{\mathtt{u}}_{0, 0}^{{\mathit A}_{\textrm{in}}}){\mathfrak{F}}_1,~
{\mathfrak{F}}_0
:=\frac{1}{2\pi i}\oint_{\mathbb{T}}\mathcal{T}_1({{z}}){{z}}^{-1}d{{z}},
\quad
{\mathfrak{F}}_1:=\frac{1}{2\pi i}\oint_{\mathbb{T}}\mathcal{T}_2({{z}}){{z}}^{-1}d{{z}}.
\label{expu00}
\end{split}\end{equation}
\eqref{expu00} can be easily solved for ${\mathtt{u}}_{0, 0}$ in terms of certain contour integrals on ${{\mathscr{A}}}$. 
Formally,
\begin{align}
{\mathtt{u}}_{0, 0}=\frac{{\mathfrak{F}}_0
+{\mathfrak{F}}_1}{1-{\mathfrak{F}}_1}{\mathtt{u}}_{0, 0}^{{\mathit A}_{\textrm{in}}}.
\label{u00valintg}
\end{align}
\label{u00exp}
\end{remark}

In contrast to the expression \eqref{u00valintg} obtained in above Remark \ref{u00exp} using the inverse transform \eqref{invuF}, in this problem it is serendipitous that 
a closed form expression exists for the same.
Taking the limit ${z}\to\infty$ in \eqref{discWHNsolhalf}${}_1$, it is directly found that
\begin{equation}\begin{split}
{\mathtt{u}}_{0,0}&=\lim_{{z}\to\infty}{\mathfrak{L}}_{+}({z}){\mathfrak{C}}^+({z}).
\label{altu00}
\end{split}\end{equation}
Using 
\eqref{LpLmlim} in \eqref{discCfacalleq}, it is found that $\lim_{{z}\to\infty}{\mathfrak{C}}^+({z})$ equals
\begin{equation}\begin{split}
\ell_\infty^{-1}\bigg({{{\mathtt{u}}^{{\mathit A}_{\textrm{in}}}_{0,0}}}(-1+\frac{\ell_\infty}{{\mathfrak{L}}_+({{z}}_{{\mathit A}}^{-1})})
+{({\mathtt{u}}_{0, 0}+{\mathtt{u}}^{{\mathit A}_{\textrm{in}}}_{0, 0})}\big(1+\ell_\infty(\frac{(1+\mathtt{M}-{{z}_{\mathtt{p}}})(-{z}_{\mathtt{p}})}{{\mathfrak{L}}_-^{-1}({z}_{\mathtt{p}})(1-{z}_{\mathtt{p}}^2)}
-\frac{(1+\mathtt{M}-{{z}_{\mathtt{p}}^{-1}})(-{z}_{\mathtt{p}})}{{\mathfrak{L}}_+({z}_{\mathtt{p}}^{-1})(1-{z}_{\mathtt{p}}^2)})\big)\bigg).
\label{Cinfty}
\end{split}\end{equation}
Using \eqref{Cinfty} in \eqref{altu00}, after simplifying further,
\begin{equation}\begin{split}
{({\mathtt{u}}_{0, 0}+{\mathtt{u}}^{{\mathit A}_{\textrm{in}}}_{0, 0})}&={{\mathtt{u}}^{{\mathit A}_{\textrm{in}}}_{0,0}}\mathcal{U},\quad
\mathcal{U}:=-\frac{{{\mathfrak{L}}_+^{-1}({{z}}_{{\mathit A}}^{-1})}({z}_{\mathtt{p}}-{z}_{\mathtt{p}}^{-1})}{{\mathfrak{L}}_-({z}_{\mathtt{p}})(1+\mathtt{M}-{{z}_{\mathtt{p}}})-{\mathfrak{L}}_+^{-1}({z}_{\mathtt{p}}^{-1})(1+\mathtt{M}-{{z}_{\mathtt{p}}^{-1}})}.
\label{u00val}
\end{split}\end{equation}

\begin{remark}
In the context of Remark \ref{speccase2} (violation of \eqref{aAaBassumed}), assuming ${\alpha}_{{\mathit A}}={\alpha}_{{\mathit B}}$ but ${\mathit{m}}_{{\mathit A}}\ne{\mathit{m}}_{{\mathit B}}$, 
using the additive factorization \eqref{discCfacalleqeqaAB} of ${\mathfrak{C}}$
and \eqref{LpLmlim},
\begin{align}
\lim_{{z}\to\infty}{\mathfrak{C}}^+({{z}})
&={{{\mathtt{u}}^{{\mathit A}_{\textrm{in}}}_{0,0}}}(-\ell_\infty^{-1}+{\mathfrak{L}}_+^{-1}({{z}}_{{\mathit A}}^{-1}))+(\ell_{\infty}-\ell_\infty^{-1})\frac{{\mathit{m}}_{0}-{\mathit{m}}_{{\mathit A}}}{{\mathit{m}}_{{\mathit A}}-{\mathit{m}}_{{\mathit B}}}({\mathtt{u}}_{0, 0}+{\mathtt{u}}^{{\mathit A}_{\textrm{in}}}_{0, 0})
\end{align}
so that ${\mathtt{u}}_{0,0}={{{\mathtt{u}}^{{\mathit A}_{\textrm{in}}}_{0,0}}}(-1+\ell_\infty{\mathfrak{L}}_+^{-1}({{z}}_{{\mathit A}}^{-1}))+(\ell^2_\infty-1)\frac{{\mathit{m}}_{0}-{\mathit{m}}_{{\mathit A}}}{{\mathit{m}}_{{\mathit A}}-{\mathit{m}}_{{\mathit B}}}({\mathtt{u}}_{0, 0}+{\mathtt{u}}^{{\mathit A}_{\textrm{in}}}_{0, 0})$. 
Hence, $\mathcal{U}$ in \eqref{u00val} 
equals ${{\mathfrak{L}}_+^{-1}({{z}}_{{\mathit A}}^{-1})}\ell_\infty/(1+(1-\ell_\infty^2)({{\mathit{m}}_{0}-{\mathit{m}}_{{\mathit A}}})/({{\mathit{m}}_{{\mathit A}}-{\mathit{m}}_{{\mathit B}}})).$
\end{remark}

\begin{remark}
Last but not the least in this section, it is emphasized that, as ${\upomega}_2\to0+$, the case of real frequency ${\upomega}$ is recovered along with the scattering solution that satisfies the radiation conditions \cite{Shaban}.
\label{complexomg}
\end{remark}

\section{Incident energy flux leaked via bulk lattice waves}
\label{sec2leak}

At this point it is useful to recall the conditions described in the beginning of \S\ref{sec2dWH},
Remark \ref{complexomg}, and also Remark \ref{annulus} and definition \eqref{annulusA}.
Using the limiting absorption principle the scattering solutions of the physical problem can be obtained 
in the limit of vanishing damping, i.e. $\upomega_2\to0^+$.
In this limiting sense, 
with substitution of \eqref{discCfacalleq} in \eqref{discWHNsolhalf},
the excited (reflected and transmitted) surface wave amplitudes
defined in \eqref{reftranscoefftdef} are obtained as:
\begin{equation}\begin{split}
\mathcal{C}_{{\mathit A}_{\textrm{in}}}^{{\mathit A}}=\text{Res }\big|_{{{z}}={{z}}_{{\mathit A}}
}(+{{z}}^{-1}{\mathtt{u}}^{+}({{z}})),
\quad
\mathcal{C}_{{\mathit A}_{\textrm{in}}}^{{\mathit B}}=\text{Res }\big|_{{{z}}={{z}}_{{\mathit B}}^{-1}
}(-{{z}}^{-1}{\mathtt{u}}^{-}({{z}})).
\label{C0Cn1}
\end{split}\end{equation}
In view of Remarks \ref{annulus3}, \ref{annulus4}, \eqref{discCfacalleq}, and definition \eqref{defArat}, using \eqref{discWHLpLm}${}_1$, 
 \eqref{discWHNsolhalf}${}_1$, the reflection coefficient \eqref{C0Cn1}${}_1$ is found to be
\begin{align}
\mathcal{C}_{{\mathit A}_{\textrm{in}}}^{{\mathit A}}
&=-\frac{(-{{z}}_{{\mathit A}}^{-1}){{\mathfrak{K}}_{{{\mathit B}}{}+}({{z}}_{{\mathit A}})}}{{\mathfrak{K}}_{{{\mathit A}}{}+}'({{z}}_{{\mathit A}})}{\mathfrak{L}}_+^{-1}({{z}}_{{\mathit A}}^{-1})\bigg(-{\updelta}_{D}^-({{z}}_{{\mathit A}}^2)
+({\alpha}_{{\mathit A}}-{\alpha}_{{\mathit B}})\mathtt{C}_{00}\notag\\
&\bigg({\mathfrak{L}}_-({z}_{\mathtt{p}})\frac{(1+\mathtt{M}-{{z}_{\mathtt{p}}})}{{\mathtt{p}_+({z}_{{\mathit A}})}\mathtt{p}_-({z}_{\mathtt{p}})}
-{\mathfrak{L}}_+^{-1}({z}_{\mathtt{p}}^{-1})\frac{(1+\mathtt{M}-{{z}_{\mathtt{p}}^{-1}})}{\mathtt{p}_+({z}_{\mathtt{p}}^{-1})}(\frac{1}{\mathtt{p}_-(0)}-\frac{1}{\mathtt{p}_-({z}_{{\mathit A}})})\bigg)\bigg),
\label{CAhalfincA}
\end{align}
whereas,
using \eqref{discWHLpLm}${}_2$, \eqref{discWHNsolhalf}${}_2$,
the transmission coefficient \eqref{C0Cn1}${}_2$ is found to be
\begin{align}
\mathcal{C}_{{\mathit A}_{\textrm{in}}}^{{\mathit B}}
&=-\frac{(-{{z}}_{{\mathit B}}){{\mathfrak{K}}_{{{\mathit A}}{}-}({{z}}_{{\mathit B}}^{-1})}}{{\mathfrak{K}}_{{{\mathit B}}{}-}'({{z}}_{{\mathit B}}^{-1})}{\mathfrak{L}}_+^{-1}({{z}}_{{\mathit A}}^{-1})\bigg(-{\updelta}_{D}^-({{z}}_{{\mathit B}}^{-1}{{z}}_{{\mathit A}})
+({\alpha}_{{\mathit A}}-{\alpha}_{{\mathit B}})\mathtt{C}_{00}\notag\\
&\bigg({\mathfrak{L}}_-({z}_{\mathtt{p}})\frac{(1+\mathtt{M}-{{z}_{\mathtt{p}}})}{{\mathtt{p}_+({z}_{{\mathit B}}^{-1})}\mathtt{p}_-({z}_{\mathtt{p}})}
-{\mathfrak{L}}_+^{-1}({z}_{\mathtt{p}}^{-1})\frac{(1+\mathtt{M}-{{z}_{\mathtt{p}}^{-1}})}{\mathtt{p}_+({z}_{\mathtt{p}}^{-1})}(\frac{1}{\mathtt{p}_-(0)}-\frac{1}{\mathtt{p}_-({z}_{{\mathit B}}^{-1})})\bigg)\bigg).
\label{CBhalfincA}
\end{align}
Some details behind above expressions \eqref{CAhalfincA}, \eqref{CBhalfincA} are included in Appendix \ref{appB}.

In general, 
a certain amount of energy contained in the incident surface wave is converted into the bulk waves during process of transmission on the surface. The expressions \eqref{reftransance}, i.e. the surface wave reflectance and transmittance, already provide a formula for the reflected and transmitted energy flux in the surface wave, $\mathcal{E}_{{\mathit A}}, \mathcal{E}_{{\mathit B}}$, respectively, per unit energy flux of incident surface wave $\mathcal{E}_{{\mathit A}_{\textrm{in}}}$ \eqref{Efluxall}. The physically important entity ${\mathscr{D}}_{{surf}}:=\mathcal{E}_{\textrm{lk}}/\mathcal{E}_{{\mathit A}_{\textrm{in}}}$ as fraction of incident energy flux that is `leaked' at the interface in the form of bulk waves can be obtained too. 
Using a viewpoint of work-energy relation \cite{Brillouin}, an expression for ${\mathscr{D}}_{{surf}}$ can be obtained.
In fact, as $\mathtt{y}\to\infty$,
\begin{equation}\begin{split}
{\mathscr{D}}_{{surf}}&=\frac{1}{\mathcal{E}_{{\mathit A}_{\textrm{in}}}}
\sum\nolimits_{\mathtt{x}\in\mathbb{Z}}({\mathtt{u}}_{\mathtt{x},\mathtt{y}+1}-{\mathtt{u}}_{\mathtt{x},\mathtt{y}})\overline{(-i\upomega){\mathtt{u}}_{\mathtt{x},\mathtt{y}}}
=\frac{\upomega}{2\pi {\mathcal{E}_{{\mathit A}_{\textrm{in}}}}}\oint_{{\mathbb{T}}}({\lambda^{-1}({z})-1})|{\mathtt{u}}({z})|^2d{z}.
\label{dissancewaves}
\end{split}\end{equation}
Using \eqref{unpm} and the properties of Fourier transform \eqref{dissancewaves} can simplified further though the details are omitted as there is an alternative to the expression \eqref{dissancewaves}.
In the limit $\upomega_2\to0^+$, by the conservation of energy $\mathcal{E}_{{\mathit A}_{\textrm{in}}}=\mathcal{E}_{{\mathit A}_{\textrm{in}}}+\mathcal{E}_{{\mathit B}}+\mathcal{E}_{\textrm{lk}}$, so that the expression for such a fraction of energy influx that is `leaked' is 
simply
\begin{equation}\begin{split}
{\mathscr{D}}_{{surf}}=1-{\mathscr{R}}_{{surf}}-{\mathscr{T}}_{{surf}},
\label{leakE}
\end{split}\end{equation}
using \eqref{reftransance}.
It is a fundamental question how ${\mathscr{D}}_{{surf}}$ depends on the wavenumber of the surface wave as well as material parameters associated with the surface structure on both sides. Indeed, in \eqref{reftransance}, 
${\mathscr{R}}_{{surf}}\equiv {\mathscr{R}}_{{surf}}(\upomega; {\mathit{m}}_{{\mathit A}}, {\mathit{m}}_{{\mathit B}}, {\mathit{m}}_0, {\alpha}_{{\mathit A}}, {\alpha}_{{\mathit B}}, {\alpha}_{{}}),$
${\mathscr{T}}_{{surf}}\equiv {\mathscr{R}}_{{surf}}(\upomega; {\mathit{m}}_{{\mathit A}}, {\mathit{m}}_{{\mathit B}}, {\mathit{m}}_0, {\alpha}_{{\mathit A}}, {\alpha}_{{\mathit B}}, {\alpha}_{{}}),$
and, therefore, in \eqref{leakE},
\begin{equation}\begin{split}
{\mathscr{D}}_{{surf}}\equiv {\mathscr{R}}_{{surf}}(\upomega; {\mathit{m}}_{{\mathit A}}, {\mathit{m}}_{{\mathit B}}, {\mathit{m}}_0, {\alpha}_{{\mathit A}}, {\alpha}_{{\mathit B}}, {\alpha}_{{}}),
\label{leakEexp}
\end{split}\end{equation}
where the surface material parameters are additionally listed, after the semi-colon, for an emphasis.
With reference to \eqref{surfgpvel}, let
\begin{align}
{{\mathtt{v}}}(\exp(\pm i{{\upxi}}_{{\mathit A}})):=\pm{{\mathit v}}_{{\mathit A}}({{\upxi}}_{{\mathit A}}), \quad {\upxi}_{{\mathit A}}\in (0, \pi),\quad
{{\mathtt{v}}}(\exp(\pm i{{\upxi}}_{{\mathit B}})):=\pm{{\mathit v}}_{{\mathit B}}({{\upxi}}_{{\mathit B}}), \quad {\upxi}_{{\mathit B}}\in (0, \pi).
\label{surfgpvel2}
\end{align}
since ${{{\mathtt{v}}}}({{z}}_{{\mathit A}}^{-1})=-{{{\mathtt{v}}}}({{z}}_{{\mathit A}}).$
Using the expressions 
\eqref{reftransance}, \eqref{leakE}, 
thus,
\begin{equation}\begin{split}
{\mathscr{D}}_{{surf}}=1-|\mathcal{C}_{{\mathit A}_{\textrm{in}}}^{{\mathit A}}|^2-|{\mathcal{C}_{{\mathit A}_{\textrm{in}}}^{{\mathit B}}}|^2\frac{{{{\mathtt{v}}}}({{z}}_{{\mathit B}})}{{{{\mathtt{v}}}}({{z}}_{{\mathit A}})},
\label{dissance}
\end{split}\end{equation}
where \eqref{CAhalfincA} and \eqref{CBhalfincA} as well as \eqref{surfgpvel2}, \eqref{surfgpvel} need to be substituted to see a more explicit expression in line with \eqref{leakEexp}.

\begin{remark}
In view of Remark \ref{altcase2}, the reflection coefficient and transmission coefficients for the incident wave \eqref{uincbwave2d}, travelling from $\mathtt{x}=-\infty$ to $\mathtt{x}\to+\infty$, are given by
\begin{equation}\begin{split}
\mathcal{C}_{{\mathit B}_{\textrm{in}}}^{{\mathit B}}&=-\frac{(-{{z}}_{{\mathit B}}){{\mathfrak{K}}_{{{\mathit A}}{}{-}}({{z}}^{-1}_{{\mathit B}})}}{{\mathfrak{K}}_{{{\mathit B}}{}{-}}'({{z}}^{-1}_{{\mathit B}})}\bigg(-({\updelta}_{D}^{+}({{z}}^{-2}_{{\mathit B}})-1)
{\mathfrak{L}}_{-}({{z}}_{{\mathit B}})
+({\alpha}_{{\mathit B}}-{\alpha}_{{\mathit A}})\frac{\mathtt{C}_{00}}{{\mathfrak{L}}^{-1}_{-}({{z}}_{{\mathit B}})}\\
&\bigg({\mathfrak{L}}^{-1}_{+}({z}^{-1}_{\mathtt{p}})\frac{(1+\mathtt{M}-{{z}^{-1}_{\mathtt{p}}})}{{\mathtt{p}_{-}({z}^{-1}_{{\mathit B}})}\mathtt{p}_{+}({z}^{-1}_{\mathtt{p}})}
-{\mathfrak{L}}_{-}({z}_{\mathtt{p}})\frac{(1+\mathtt{M}-{{z}_{\mathtt{p}}})}{\mathtt{p}_{-}({z}_{\mathtt{p}})}(\frac{1}{\mathtt{p}_{+}(\infty)}-\frac{1}{\mathtt{p}_{+}({z}^{-1}_{{\mathit B}})})\bigg)\bigg),
\label{CBhalfincB}
\end{split}\end{equation}
\begin{equation}\begin{split}
\text{and }
\mathcal{C}_{{\mathit B}_{\textrm{in}}}^{{\mathit A}}&=-\frac{(-{{z}}_{{\mathit A}}^{-1}){{\mathfrak{K}}_{{{\mathit B}}{}{+}}({{z}}_{{\mathit A}})}}{{\mathfrak{K}}_{{{\mathit A}}{}{+}}'({{z}}_{{\mathit A}})}\bigg(-({\updelta}_{D}^{+}({{z}}_{{\mathit A}}{{z}}_{{\mathit B}}^{-1})-1)
{\mathfrak{L}}_{-}({{z}}_{{\mathit B}})
+({\alpha}_{{\mathit B}}-{\alpha}_{{\mathit A}})\frac{\mathtt{C}_{00}}{{\mathfrak{L}}^{-1}_{-}({{z}}_{{\mathit B}})}\\
&\bigg({\mathfrak{L}}^{-1}_{+}({z}^{-1}_{\mathtt{p}})\frac{(1+\mathtt{M}-{{z}^{-1}_{\mathtt{p}}})}{{\mathtt{p}_{-}({z}_{{\mathit A}})}\mathtt{p}_{+}({z}^{-1}_{\mathtt{p}})}
-{\mathfrak{L}}_{-}({z}_{\mathtt{p}})\frac{(1+\mathtt{M}-{{z}_{\mathtt{p}}})}{\mathtt{p}_{-}({z}_{\mathtt{p}})}(\frac{1}{\mathtt{p}_{+}(\infty)}-\frac{1}{\mathtt{p}_{+}({z}_{{\mathit A}})})\bigg)\bigg),
\label{CBhalfincB}
\end{split}\end{equation}
respectively.
An expression similar to \eqref{dissance} can be 
found for ${\mathscr{D}}_{{surf}}$ in this case of incident surface wave.
\end{remark}

\begin{remark}
In the simpler situation mentioned in Remark \ref{speccase2},
i.e. assuming ${\alpha}_{{\mathit A}}={\alpha}_{{\mathit B}}$ and ${\mathit{m}}_{0}={\mathit{m}}_{{\mathit A}}$ but ${\mathit{m}}_{{\mathit A}}\ne{\mathit{m}}_{{\mathit B}}$,
using \eqref{discWHkerfacs},
the expressions \eqref{CAhalfincA}, \eqref{CBhalfincA}, respectively, can be simplified to obtain
\begin{align}
\mathcal{C}_{{\mathit A}_{\textrm{in}}}^{{\mathit A}}
=-{{z}}_{{\mathit A}}^{-1}{\updelta}_{D}^-({{z}}_{{\mathit A}}^2)\frac{{\mathfrak{K}}_{{{\mathit B}}{}+}({{z}}_{{\mathit A}})}{{\mathfrak{K}}_{{{\mathit A}}{}+}'({{z}}_{{\mathit A}})}
\frac{{\mathfrak{K}}_{{{\mathit A}}{}+}({{z}}_{{\mathit A}}^{-1})}{{\mathfrak{K}}_{{{\mathit B}}{}+}({{z}}_{{\mathit A}}^{-1})},
\mathcal{C}_{{\mathit A}_{\textrm{in}}}^{{\mathit B}}
=-{{z}}_{{\mathit B}}{\updelta}_{D}^-({{z}}_{{\mathit B}}^{-1}{{z}}_{{\mathit A}})\frac{{\mathfrak{K}}_{{{\mathit A}}{}-}({{z}}_{{\mathit B}}^{-1})}{{\mathfrak{K}}_{{{\mathit B}}{}-}'({{z}}_{{\mathit B}}^{-1})}
\frac{{\mathfrak{K}}_{{{\mathit A}}{}+}({{z}}_{{\mathit A}}^{-1})}{{\mathfrak{K}}_{{{\mathit B}}{}+}({{z}}_{{\mathit A}}^{-1})}.
\label{CAhalfincABspl}
\end{align}
According to \eqref{reftransance}, using \eqref{CAhalfincABspl}, \eqref{surfgpvel}, therefore,
\begin{align}
{\mathscr{R}}_{{surf}}
&=(\frac{{{z}}_{{\mathit A}}^{-2}}{({{z}}^{-1}_{{\mathit A}}-{{z}}_{{\mathit A}})^2})C_{RT}\frac{\overline{{\mathfrak{K}}_{{{\mathit A}}{}-}({{z}}_{{\mathit A}}){\mathfrak{K}}_{{{\mathit B}}{}+}({{z}}_{{\mathit A}})}}{{\mathfrak{K}}_{{{\mathit A}}{}+}'({{z}}_{{\mathit A}}){\mathfrak{K}}_{{{\mathit B}}{}-}({{z}}_{{\mathit A}})},\\
{\mathscr{T}}_{{surf}}
&=C_T(\frac{{{z}}_{{\mathit B}}^2}{({{z}}_{{\mathit B}}-{{z}}_{{\mathit A}})^2})C_{RT}\frac{\overline{{\mathfrak{K}}_{{{\mathit A}}{}-}({{z}}_{{\mathit B}}^{-1}){\mathfrak{K}}_{{{\mathit B}}{}+}({{z}}_{{\mathit B}}^{-1})}}{{\mathfrak{K}}_{{{\mathit A}}{}+}({{z}}_{{\mathit B}}^{-1}){\mathfrak{K}}_{{{\mathit B}}{}-}'({{z}}_{{\mathit B}}^{-1})},
\label{reftransancespl}\\
\text{where }
C_{RT}&:=\frac{{\mathfrak{K}}_{{{\mathit A}}{}+}({{z}}_{{\mathit A}}^{-1}){\mathfrak{K}}_{{{\mathit B}}{}-}({{z}}_{{\mathit A}}^{-1})}{\overline{{\mathfrak{K}}_{{{\mathit A}}{}-}'({{z}}_{{\mathit A}}^{-1}){\mathfrak{K}}_{{{\mathit B}}{}+}({{z}}_{{\mathit A}}^{-1})}},\quad
C_T:=\frac{({\mathit{m}}_{{\mathit A}}-1)(1-\lambda_{{\mathit A}}^2)+1}{({\mathit{m}}_{{\mathit B}}-1)(1-\lambda_{{\mathit B}}^2)+1}\frac{1-\lambda_{{\mathit B}}^2}{1-\lambda_{{\mathit A}}^2},\\
&\lambda_{{\mathit A}}:=\lambda({z}_{{\mathit A}})\text{ and }\lambda_{{\mathit B}}:=\lambda({z}_{{\mathit B}}),\notag
\label{reftransancesplCRT}
\end{align}
as reminiscent of the expressions of the reflectance and transmittance in waveguides \cite{Blsbif,Blsstep} except for the presence of $C_T$ which is peculiar to the half-plane problem with surface structure on its boundary.
To obtain above expressions \eqref{reftransancespl} from \eqref{reftransance}, recall \eqref{surfgpvel}, \eqref{surfgpvel2}, and it is useful to note ${\mathfrak{K}}_{{{\mathit A}}}'({{z}}_{{\mathit A}})=(-{{z}}_{{\mathit A}}^{-2}){\mathfrak{K}}_{{{\mathit A}}}'({{z}}_{{\mathit A}}^{-1}),$ ${\mathfrak{K}}_{{{\mathit B}}}'({{z}}_{{\mathit B}})=(-{{z}}_{{\mathit B}}^{-2}){\mathfrak{K}}_{{{\mathit B}}}'({{z}}_{{\mathit B}}^{-1}),$
${\mathfrak{K}}_{{{\mathit B}}}({{z}}_{{\mathit A}})={\mathfrak{K}}_{{{\mathit B}}}({{z}}_{{\mathit A}}^{-1})=-{\mathfrak{K}}_{{{\mathit A}}}({{z}}_{{\mathit B}})=-{\mathfrak{K}}_{{{\mathit A}}}({{z}}_{{\mathit B}}^{-1})=
-({\mathit{m}}_{{\mathit A}}-{\mathit{m}}_{{\mathit B}}){\upomega}^2$,
as well as
${\mathfrak{K}}_{{{\mathit A}}}'({{z}}_{{\mathit A}})
=(\upomega^{-1}({z}_{{\mathit A}}-{{z}}_{{\mathit A}}^{-1})({{\alpha}_{{}}}+({\alpha}_{{\mathit A}}-{{\alpha}_{{}}})(1-\lambda_{{\mathit A}}^2))){{z}}_{{\mathit A}}^{-1}\upomega(1-\lambda_{{\mathit A}}^2)^{-1}$, 
and
${\mathfrak{K}}_{{{\mathit B}}}'({{z}}_{{\mathit B}})=(\upomega^{-1}({z}_{{\mathit B}}-{{z}}_{{\mathit B}}^{-1})({{\alpha}_{{}}}+({\alpha}_{{\mathit B}}-{{\alpha}_{{}}})(1-\lambda_{{\mathit B}}^2))){{z}}_{{\mathit B}}^{-1}\upomega(1-\lambda_{{\mathit B}}^2)^{-1}$.
\end{remark}

\begin{remark}
The reflectance and the transmittance of surface waves (for a wave incident from the side ahead of interface) are given by 
\eqref{reftransance}.
Within the surface wave propagation equation assumed for the lattice structure, accounting for the evanascent and propagating wave modes excited by the presence of an interface, the surface transmission conductance ${\mathscr{G}}$ can be obtained.
This is the sum of transmittance for each incident surface wave from one side to the other side at given ${\upomega}$. 
The maximum value, so called ballistic limit, is the minimum of $N_{{{\mathit A}}}$ and $N_{{{\mathit B}}}$, i.e. belongs to $\{0, 1\}$ since
the number of propagating surface wave modes is $N_{{{\mathit A}}}\in\{0, 1\}$ on the side behind and on the side ahead $N_{{{\mathit B}}}\in\{0,1\}$, at a given frequency ${\upomega}$. In particular, it is easy to see that
${\mathscr{G}}={\mathscr{G}}_{{{\mathit A}}\leftarrow{{\mathit B}}}
={\mathscr{G}}_{{{\mathit B}}\leftarrow{{\mathit A}}}={\mathscr{T}}_{{surf}}$ where ${\mathscr{T}}_{{surf}}$ equals $0$ whenever surface wave mode is absent on either side of the interface at given frequency ${\upomega}$.
\label{conductancesurf}
\end{remark}

\begin{figure}[!ht]
\center
\includegraphics[width=\textwidth]{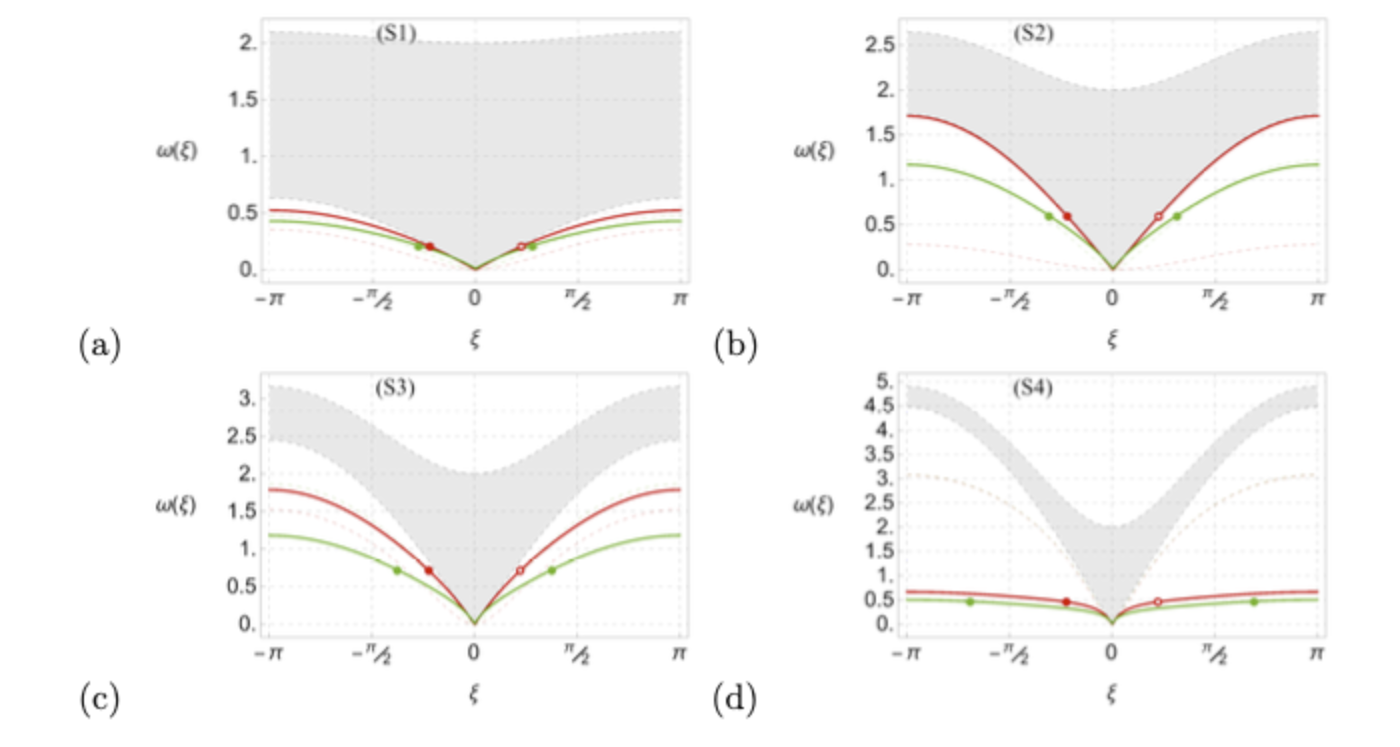}
\caption[]{Dispersion relation for $\upomega({\upxi})$ (solid curves) and attenuation coefficient $\eta({\upxi})$ (dashed curves) vs ${\upxi}$, \eqref{funcomega}, \eqref{funcomegaeq}, for the choice of parameters (S$1$)--(S$4$) stated in \eqref{S14choice}.
The solid gray shaded region indicates the pass band for the bulk lattice waves and surface wave dispersion is in red for surface structure ahead of the interface while surface wave dispersion is in green for 
structure behind the interface. 
Red and green dots demonstrate constant frequency surface wave modes of both signs of energy flux velocity.
}
\label{S14surfacestructure}
\end{figure}

\begin{figure}[!ht]
\center
\includegraphics[width=.75\textwidth]{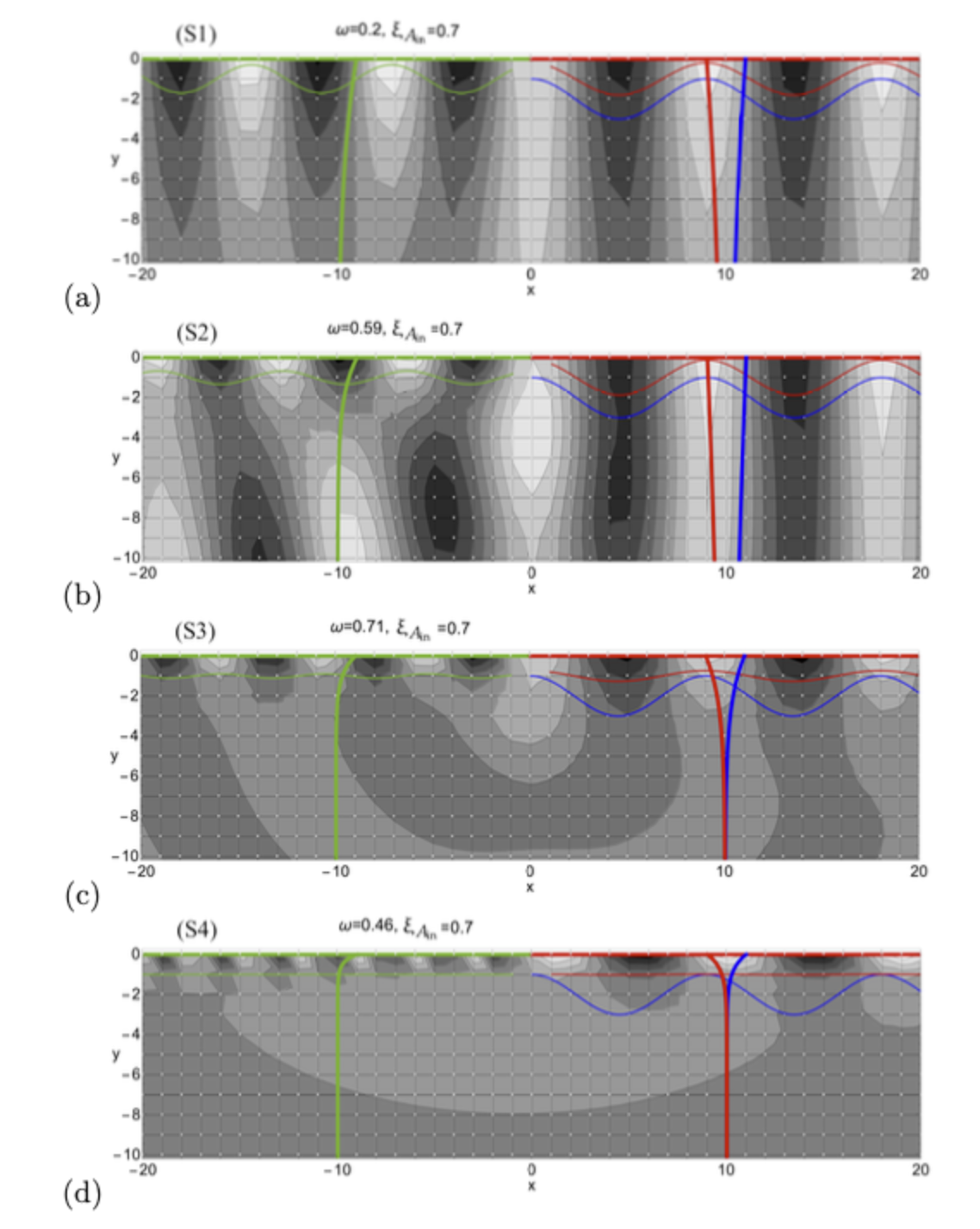}
\caption[]{Contour plot illustration of time snapshot of $\Re{\mathtt{u}}^{\textrm{to}}_{{\mathtt{x}}, {\mathtt{y}}}(t)$ (with lattice $\mathbb{H}$ overlaid on it schematically) with respective parameters same as in Fig. \ref{S14surfacestructure} and specific frequency as listed on top.
Vertical blue (resp. red, green) curve illustrates the decaying amplitude of the incident (resp. reflected, transmitted) surface wave shown as horizontal wavy curve on top of each plot.
Red (resp. green) dots show the lattice sites ahead (resp. behind) the interface.}
\label{ContourS14}
\end{figure}

\begin{figure}[!ht]
\center
\includegraphics[width=\textwidth]{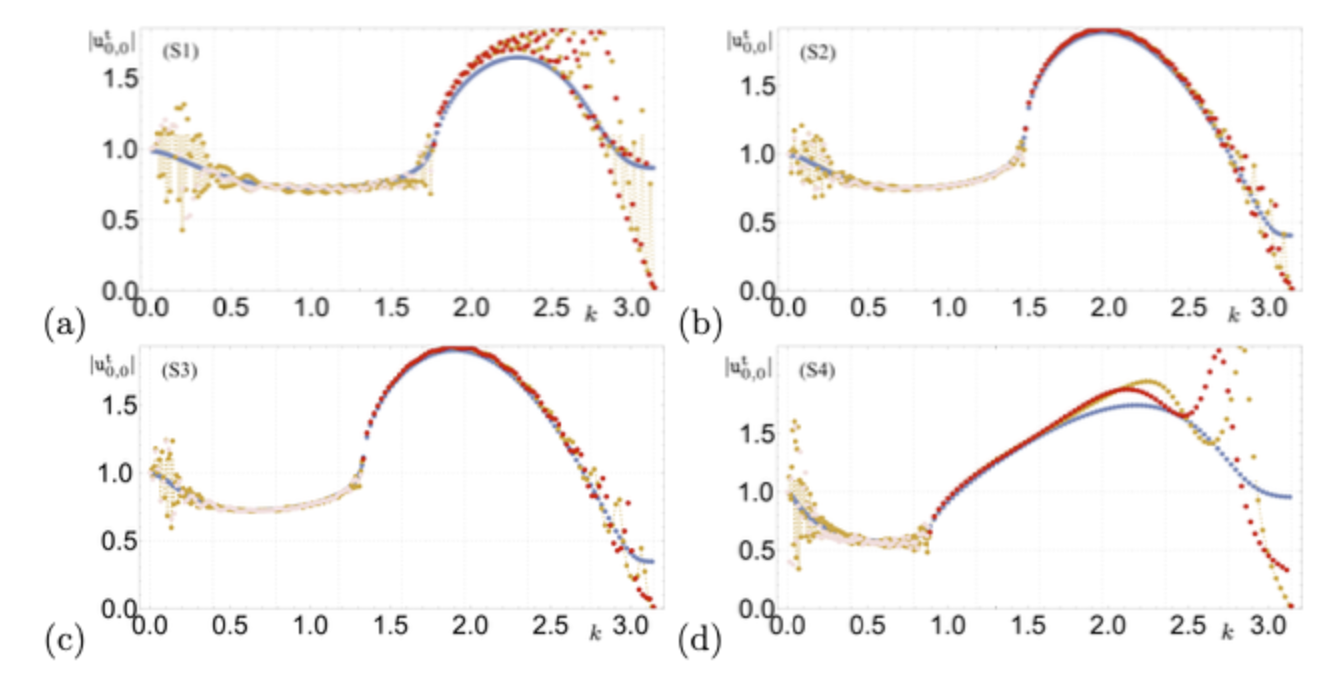}
\caption[]{$|{\mathtt{u}}^{\mathrm t}_{0, 0}|=|{({\mathtt{u}}_{0, 0}+{\mathtt{u}}^{{\mathit A}_{\textrm{in}}}_{0, 0})}
|$ 
vs incident surface wavenumber $k={{\upxi}}_{{\mathit A}}$ for ${\mathtt{u}}^{{\mathit A}_{\textrm{in}}}_{0, 0}=1$.
The dots in blue represent the analytical solution \eqref{u00val} while other colored dots represent the numerical solutions.
The parameters for surface structure correspond (S$1$)--(S$4$) stated in \eqref{S14choice} and shown in Fig. \ref{S14surfacestructure}(a)-(d), respectively.}
\label{ut00_surfacestructure}
\end{figure}

\begin{figure}[!ht]
\center
\includegraphics[width=.7\textwidth]{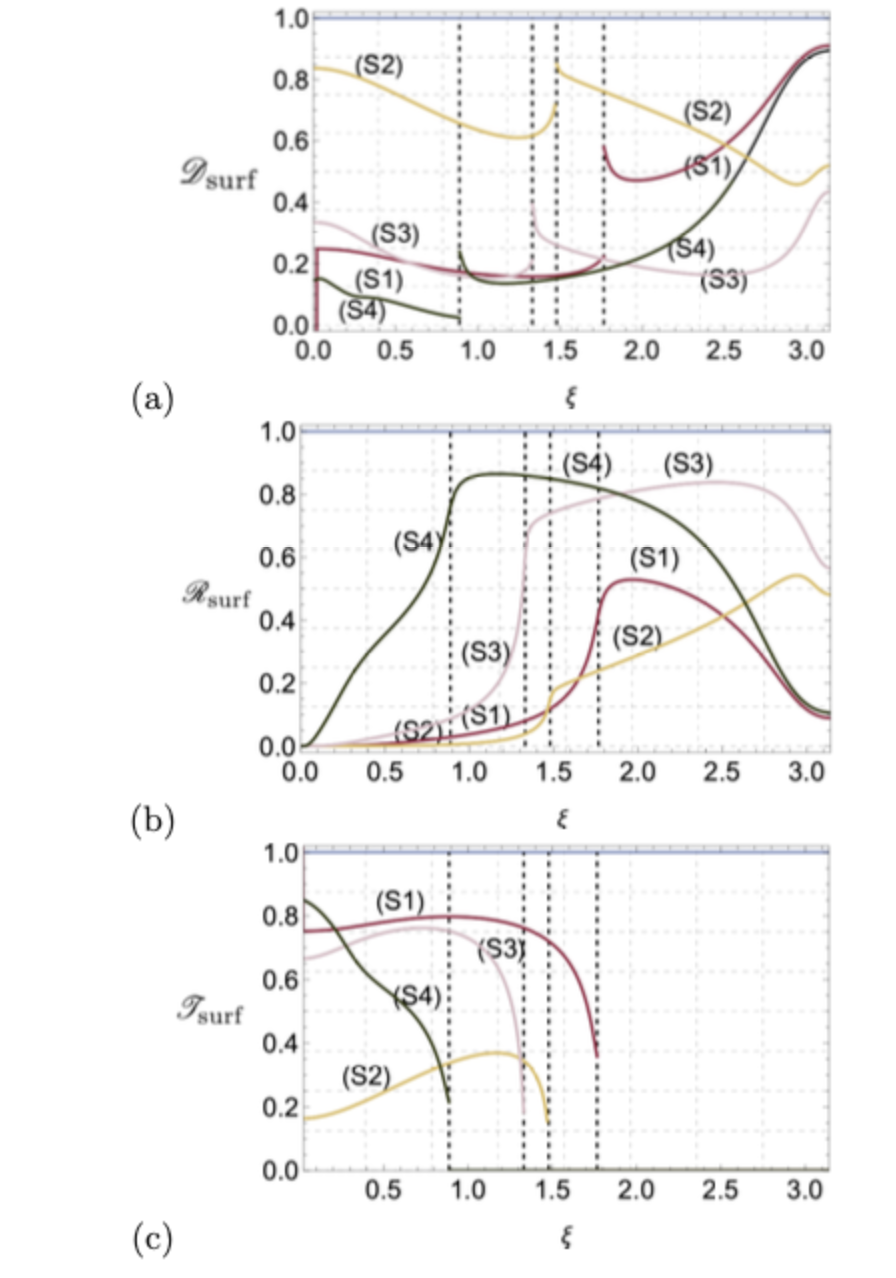}
\caption[]{Results using \eqref{CAhalfincA},
\eqref{CBhalfincA} for (a) energy leakage \eqref{leakE} into half-plane via bulk waves ${\mathscr{D}}_{{surf}}=\mathcal{E}_{\textrm{lk}}/\mathcal{E}_{{\mathit A}_{\textrm{in}}}$ and (b) reflectance 
\eqref{reftransance}${}_1$
${\mathscr{R}}_{{surf}}=\mathcal{E}_{\textrm{rf}}/\mathcal{E}_{{\mathit A}_{\textrm{in}}}$ and (c) transmittance 
\eqref{reftransance}${}_2$ ${\mathscr{T}}_{{surf}}=\mathcal{E}_{\textrm{tr}}/\mathcal{E}_{{\mathit A}_{\textrm{in}}}$ vs incident surface wavenumber $k={{\upxi}}_{{\mathit A}_{\textrm{in}}}$ for the choice of the surface structure parameters ${\mathit{m}}_{{\mathit A}}, {\mathit{m}}_{{\mathit B}}, {\alpha}_{{\mathit A}}, {\alpha}_{{\mathit B}}$ and ${\alpha}_{{}}$ corresponding to Fig. \ref{S14surfacestructure}(a)-(d), respectively.
The dashed black lines represent the limit of the surface wave band on the other side of the interface for each choice of parameters (S$1$)--(S$4$) \eqref{S14choice}.}
\label{ErefA_surfacestructure}
\end{figure}

\section{Numerical results}
\label{secNum}

In Fig. \ref{S14surfacestructure}, an illustration is provided for the 
dispersion relation \eqref{funcomega},\eqref{funcomegaeq} for $\upomega({\upxi})$ (solid curves) and attenuation coefficient $\eta({\upxi})$ (dashed curves) vs ${\upxi}$ using the choice of parameters:
\begin{equation}\begin{split}
({\mathrm S}1)&: {\mathit{m}}_{{\mathit A}}={\mathit{m}}_{0}=4, {\mathit{m}}_{{\mathit B}}=9, {\alpha}_{{\mathit A}}=0.2, {\alpha}_{{\mathit B}}=0.3, {\alpha}_{{}}=0.1,\\
({\mathrm S}2)&: {\mathit{m}}_{{\mathit A}}={\mathit{m}}_{0}=2, {\mathit{m}}_{{\mathit B}}=5, {\alpha}_{{\mathit A}}=1.4, {\alpha}_{{\mathit B}}=1.5, {\alpha}_{{}}=0.75,\\
({\mathrm S}3)&: {\mathit{m}}_{{\mathit A}}={\mathit{m}}_{0}=2, {\mathit{m}}_{{\mathit B}}=5, {\alpha}_{{\mathit A}}=1.4, {\alpha}_{{\mathit B}}=1.5, {\alpha}_{{}}=1.5,\\
({\mathrm S}4)&: {\mathit{m}}_{{\mathit A}}={\mathit{m}}_{0}=4, {\mathit{m}}_{{\mathit B}}=9, {\alpha}_{{\mathit A}}=0.2, {\alpha}_{{\mathit B}}=0.3, {\alpha}_{{}}=5.
\label{S14choice}
\end{split}\end{equation}
These four choices (S$1$)--(S$4$) are samples of surface structure, while the pairs $({\mathrm S}1)$, $({\mathrm S}4)$ and $({\mathrm S}2)$, $({\mathrm S}3)$ employ an identical set of surface structure parameters. 
Notice that ${\alpha}_{{}}$ varies across all four choices in \eqref{S14choice} and affects the pass band of the bulk lattice as indicated by region in gray shade in Fig. \ref{S14surfacestructure}. In particular, small values of ${\alpha}_{{}}$ (horizontal bonds in bulk) in comparison with $1$ (vertical bonds in bulk) lead to widening of the pass band while large values lead to its narrowing. This is expected as ${\alpha}_{{}}\gg1$ tends to approach effectively a one dimensional lattice model while $0<{\alpha}_{{}}\ll1$ softens the propagation parallel to surface and reduces surface wave band.

Using the four samples (S$1$)--(S$4$) of surface structure parameters in \eqref{S14choice}, Fig. \ref{ContourS14} presents a contour plot illustration of a time snapshot of $\Re{\mathtt{u}}^{\textrm{to}}_{{\mathtt{x}}, {\mathtt{y}}}(t)$, at some fixed $t$, say $t=0$, based on a scheme similar to that explained in Appendix D of \cite{sK} and Appendix E of \cite{Blsbif}.
The choice of incident surface wave
frequency and wavenumber is listed on top in these plots (this choice is the same as red dot present in $(0, \pi)$ as shown in corresponding parts of Fig. \ref{S14surfacestructure}). 
The decaying amplitudes of the incident (resp. reflected, transmitted) surface wave are also shown schematically in the same figure. Notice that large value of ${\alpha}_{{}}$ in the case of $({\mathrm S}4)$ is associated with highly localized, almost one dimensional wave motion in clear contrast to small value of ${\alpha}_{{}}$ in $({\mathrm S}1)$.

As a testimony to the exact solution presented in this article, particularly the closed form expression \eqref{u00val},
Fig. \ref{ut00_surfacestructure} presents a comparison of analytical solution with numerical solution for ${({\mathtt{u}}_{0, 0}+{\mathtt{u}}^{{\mathit A}_{\textrm{in}}}_{0, 0})}$ (using the same numerical scheme that led to the solution behind contourplots shown in Fig. \ref{ContourS14}).
The parameters for surface structure in Fig. \ref{ut00_surfacestructure}(a)--(d) correspond to Fig. \ref{S14surfacestructure}(a)--(d), respectively.
The oscillatory behavior of numerical solution for ${({\mathtt{u}}_{0, 0}+{\mathtt{u}}^{{\mathit A}_{\textrm{in}}}_{0, 0})}$ near ${\upxi}=0$ and ${\upxi}=\pi$ is due to finite size of grid and a rudimentary absorbing boundary condition at the grid boundary \cite{sK,Blsbif}. The waves reflected from the boundary enhance the numerical error in the simulation of time harmonic lattice dynamics.

\begin{figure}[!ht]
\center
{\includegraphics[width=.9\textwidth]{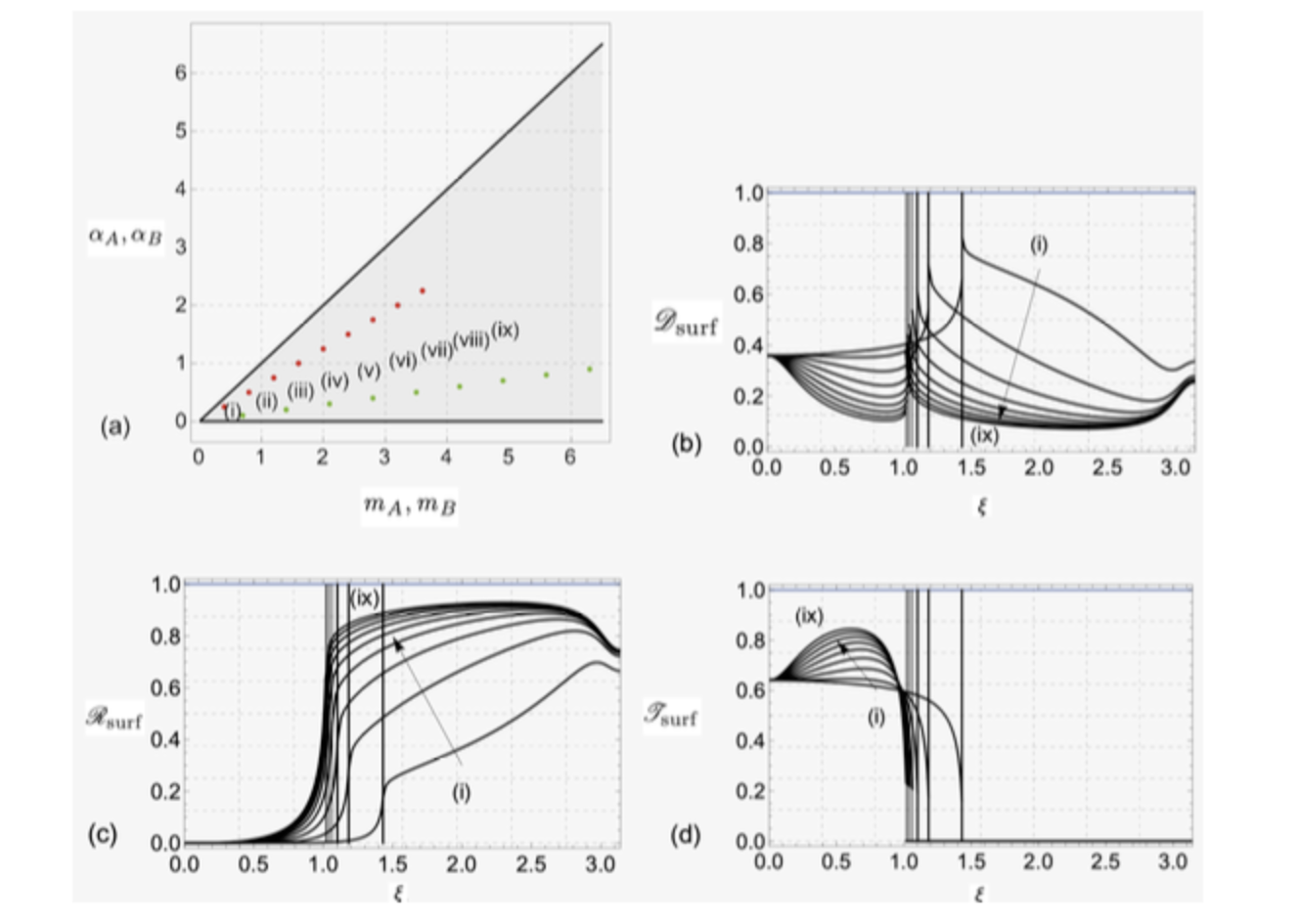}}
\caption[]{Corresponding to the $9$ choices (i)-(ix) of parameters in (a) ${\mathit{m}}_{{\mathit A}}, {\alpha}_{{\mathit A}}$ as red dots, ${\mathit{m}}_{{\mathit B}}, {\alpha}_{{\mathit B}}$ as green dots (with (i)-(ix) corresponding to $1$-$9$,
${\mathit{m}}_{{\mathit A}}={\mathit{m}}_{0}
=\{1, 2, \dotsc, 9\}0.4, {\mathit{m}}_{{\mathit B}}
=7{\mathit{m}}_{{\mathit A}}/4, {\alpha}_{{\mathit A}}
={\mathit{m}}_{{\mathit A}}5/8, {\alpha}_{{\mathit B}}
={\mathit{m}}_{{\mathit A}}/4$), with ${\alpha}_{{}}=1$, graphical results on (b) energy leakage into half-plane via bulk waves ${\mathscr{D}}_{S}=\mathcal{E}_{\textrm{lk}}/\mathcal{E}_{{\mathit A}_{\textrm{in}}}$ and (c) reflectance ${\mathscr{R}}_{S}=\mathcal{E}_{\textrm{rf}}/\mathcal{E}_{{\mathit A}_{\textrm{in}}}$ and (d) transmittance ${\mathscr{T}}_{S}=\mathcal{E}_{\textrm{tr}}/\mathcal{E}_{{\mathit A}_{\textrm{in}}}$ vs incident surface wavenumber $k={{\upxi}}_{{\mathit A}_{\textrm{in}}}$.
Compare the plot (b) with Fig. \ref{ErefA_surfacestructure}(a),
the plot (c) with Fig. \ref{ErefA_surfacestructure}(b),
and
the plot (d) with Fig. \ref{ErefA_surfacestructure}(c).
}
\label{ErefAEleakA}
\end{figure}

Using \eqref{CAhalfincA},
\eqref{CBhalfincA} in
\eqref{reftransance},
Fig. \ref{ErefA_surfacestructure}(b)--(c) presents the results for the surface reflectance and transmittance, ${\mathscr{R}}_{{surf}}=\mathcal{E}_{\textrm{rf}}/\mathcal{E}_{{\mathit A}_{\textrm{in}}}$ and ${\mathscr{T}}_{{surf}}=\mathcal{E}_{\textrm{tr}}/\mathcal{E}_{{\mathit A}_{\textrm{in}}}$ vs incident surface wavenumber ${{\upxi}}_{{\mathit A}_{\textrm{in}}}$ using the surface structure parameters $({\mathrm S}1)$-$({\mathrm S}4)$ in \eqref{S14choice}, i.e. ${\mathit{m}}_{{\mathit A}}, {\mathit{m}}_{{\mathit B}}, {\alpha}_{{\mathit A}}, {\alpha}_{{\mathit B}}$ and ${\alpha}_{{}}$, corresponding to Fig. \ref{S14surfacestructure}(a)-(d), respectively. The change in behavior around dashed black lines occurs due to the absence of surface waves in that regime of excited frequencies. For example, the surface transmittance shown in Fig. \ref{ErefA_surfacestructure}(c) is identically zero when the surface wave is not admissible on the other side of interface. It is to be noted that in the regime where the surface waves are admissible on both sides ${\mathscr{R}}_{{surf}}$ and ${\mathscr{T}}_{{surf}}$ depend on the incident wavenumber (equivalently $\upomega$) in a non-monotone manner, nature of which further depends on a choice of  specific surface structure parameters.
However, an illustration of some pattern in the dependence of these results on surface structure parameters is provided in Fig. \ref{ErefAEleakA} which is self-explanatory. In all the plots of Fig. \ref{ErefAEleakA}, ${\alpha}_{{}}=1$ while the surface structure parameters are scaled incrementally as labels increase from (i) to (ix). The large reduction in leaked energy flux corresponding to (ix) occurs as the lattice half-plane motion becomes increasingly dominated by the surface motion in accordance with large values of surface structure parameters in comparison with values in bulk.

\section{Discussion}
\label{secDisc}

\subsection{Relation with Green's function on $\mathbb{H}$}
\label{sec2dgreen}

Recall the assumption ${\alpha}_{{\mathit A}}\ne{\alpha}_{{\mathit B}}$ in \S\ref{sec2dWH2} and the cases in Remark 
\ref{speccase2}.
The remaining case concerns the consideration of the equations \eqref{eqmotionandBCs} 
when 
${\mathit{m}}_{{\mathit A}}={\mathit{m}}_{{\mathit B}}\ne {\mathit{m}}_{0}, {\alpha}_{{\mathit A}}={\alpha}_{{\mathit B}}, {\alpha}_{{}}>0$.
Then the equations \eqref{eqmotionsca},
\eqref{eqmotionBCFT} get considerably simplified and it is noted that the scattering of a surface wave by the interface is in fact replaced by that due to mass defect at the origin $(0, 0)$ located on
$\partial{\mathbb{H}}$ (recall definition \eqref{hlfplane}). A moment's reflection reveals that the solution can be found precisely in terms of certain Green's function, denoted by $\mathcal{G}_{\mathtt{x},\mathtt{y}}$ for $({\mathtt{x}}, {\mathtt{y}})\in\mathbb{H}$, for the assumed square lattice half-plane $\mathbb{H}$ with a point source (at the origin) on its boundary $\partial{\mathbb{H}}$.

\begin{remark}
The Green's function $\mathcal{G}$ satisfies the following equations:
\begin{subequations}
\begin{equation}\begin{split}\label{eqbulk}
-\upomega^2\mathcal{G}_{{\mathtt{x}},{\mathtt{y}}}={{\alpha}_{{}}}(\mathcal{G}_{{\mathtt{x}}+1,{\mathtt{y}}}+\mathcal{G}_{{\mathtt{x}}-1,{\mathtt{y}}})+\mathcal{G}_{{\mathtt{x}},{\mathtt{y}}+1}+\mathcal{G}_{{\mathtt{x}},{\mathtt{y}}-1}-2(1+{{\alpha}_{{}}})\mathcal{G}_{{\mathtt{x}},{\mathtt{y}}},\quad ({\mathtt{x}}, {\mathtt{y}})\in\mathring{\mathbb{H}},
\end{split}\end{equation}
\begin{align}
\label{eqsurf}
-\upomega^2{\mathit{m}}_{{{\mathit A}}}\mathcal{G}_{{\mathtt{x}},{0}}&=({\alpha}_{{\mathit A}} \left(\mathcal{G}_{{\mathtt{x}}+1,{0}}+\mathcal{G}_{{\mathtt{x}}-1,{0}}-2\mathcal{G}_{{\mathtt{x}},{0}}\right)+\mathcal{G}_{{\mathtt{x}},{}-1}-\mathcal{G}_{{\mathtt{x}},{0}})+\delta_{\mathtt{x}, 0}, \quad({\mathtt{x}}, {0})\in\partial{\mathbb{H}}.
\end{align}
\label{discGreenalleq}
\end{subequations}
Recall the conditions described in 
the beginning of \S\ref{sec2dWH}
and \eqref{annulusA},
Remark \ref{annulus}.
Using the Fourier transform \eqref{unpm} and its inverse,
$\mathcal{G}$ is obtained 
as
$\mathcal{G}_{\mathtt{x},\mathtt{y}}={1}/({2\pi i})\oint_{\mathbb{T}}{{\lambda}^{-\mathtt{y}}({{z}}){z}^{|\mathtt{x}|-1}}/{{{\mathfrak{K}}}_{{\mathit A}}({{z}})}~d{z}, ({\mathtt{x}}, {\mathtt{y}})\in\mathbb{H},$
and ${{\mathfrak{K}}}_{{\mathit A}}$ is defined by \eqref{discWHker}.
In particular, it useful to note that
\begin{equation}\begin{split}
\mathcal{G}_{|\mathtt{x}|,0}=\frac{1}{2\pi i}\oint_{\mathbb{T}}\frac{{z}^{|\mathtt{x}|-1}}{{{\mathfrak{K}}}_{{\mathit A}}({{z}})}d{z},
\label{greenfn2}
\end{split}\end{equation}
which is the evaluation of Green's function 
on $\partial{\mathbb{H}}$.
\label{remGreen}
\end{remark}

In the scattering problem associated with a point mass defect,
at $\mathtt{y}=0$, 
\begin{equation}\begin{split}
{\alpha}_{{{\mathit A}}}({\mathtt{u}}_{{\mathtt{x}}+1, {\mathtt{y}}}+{\mathtt{u}}_{{\mathtt{x}}-1, {\mathtt{y}}}-2{\mathtt{u}}_{{\mathtt{x}}, {\mathtt{y}}})+{\mathtt{u}}_{{\mathtt{x}}, {\mathtt{y}}-1}-{\mathtt{u}}_{{\mathtt{x}}, {\mathtt{y}}}+{\mathit{m}}_{{{\mathit A}}}\upomega^2{\mathtt{u}}_{{\mathtt{x}}, {\mathtt{y}}}=-\upomega^2({\mathit{m}}_{0}-{\mathit{m}}_{{{\mathit A}}}){({\mathtt{u}}_{0, 0}+{\mathtt{u}}^{{\mathit A}_{\textrm{in}}}_{0, 0})}\delta_{{\mathtt{x}}, {0}}.
\label{eqsurf1}
\end{split}\end{equation}
Using the Green's function 
(see Remark \ref{remGreen}) and splitting into incident and scattered, the scattered field is given by
${\mathtt{u}}_{{\mathtt{x}}, {\mathtt{y}}}=-\upomega^2({\mathit{m}}_{0}-{\mathit{m}}_{{{\mathit A}}}){({\mathtt{u}}_{0, 0}+{\mathtt{u}}^{{\mathit A}_{\textrm{in}}}_{0, 0})}{\mathcal{G}}_{|\mathtt{x}|,\mathtt{y}},$
in particular, by \eqref{greenfn2},
\begin{equation}\begin{split}
{\mathtt{u}}_{{\mathtt{x}}, {0}}=-\upomega^2({\mathit{m}}_{0}-{\mathit{m}}_{{{\mathit A}}}){({\mathtt{u}}_{0, 0}+{\mathtt{u}}^{{\mathit A}_{\textrm{in}}}_{0, 0})}{\mathcal{G}}_{|\mathtt{x}|,0}.
\label{eqsurf2}
\end{split}\end{equation}
Using 
${z}_{{\mathit A}}^{-1}=\exp({-i\xi_{{\mathit A}}})$, (as $\xi_{{\mathit A}}$ is incident surface wavenumber, ${\mathtt{u}}^{{\mathit A}_{\textrm{in}}}_{{\mathtt{x}}, {0}}={z}_{{\mathit A}}^{-\mathtt{x}}$),
the question of scattered field can therefore be answered by the expression
\begin{equation}\begin{split}
{\mathtt{u}}_{0,0}=(1-{\mathcal{G}}_{0,0}C_0)^{-1}{\mathcal{G}}_{0,0}C_0{\mathtt{u}}_{0,0}^{{\mathit A}_{\textrm{in}}}~\text{ with }~ C_0:=-\upomega^2({\mathit{m}}_{0}-{\mathit{m}}_{{{\mathit A}}}).
\end{split}\end{equation}
For $\mathtt{x}\in{\mathbb{Z}}^+\setminus\{0\},$ as $|\mathtt{x}|\to\infty$ and due to \eqref{eqsurf2},
${\mathtt{u}}_{\mathtt{x},0}=C_0{({\mathtt{u}}_{0, 0}+{\mathtt{u}}^{{\mathit A}_{\textrm{in}}}_{0, 0})}{\mathcal{G}}_{-\mathtt{x},0}=\mathcal{C}_{{\mathit A}_{\textrm{in}}}^{{\mathit A}}{z}_{{\mathit A}}^{\mathtt{x}}+$ a contribution of continuous spectrum in the contour integral \eqref{greenfn2},
and the surface wave reflection coefficient is
$\mathcal{C}_{{\mathit A}_{\textrm{in}}}^{{\mathit A}}=C_0{({\mathtt{u}}_{0, 0}+{\mathtt{u}}^{{\mathit A}_{\textrm{in}}}_{0, 0})}{1}/({{z}{\mathfrak{K}}'_{{{\mathit A}}}({{z}})})\big|_{{z}={z}_{{\mathit A}}}.$
For $\mathtt{x}\in{\mathbb{Z}}^-,$ as $|\mathtt{x}|\to\infty$ and due to \eqref{eqsurf2},
${z}_{{\mathit A}}^{-\mathtt{x}}+C_0{({\mathtt{u}}_{0, 0}+{\mathtt{u}}^{{\mathit A}_{\textrm{in}}}_{0, 0})}{1}/({{z}{\mathfrak{K}}'_{{{\mathit A}}}({{z}})})\big|_{{z}={z}_{{\mathit A}}}{z}_{{\mathit A}}^{-\mathtt{x}}+$ a contribution of continuous spectrum in the contour integral \eqref{greenfn2}, so that the transmitted surface wave has the form $\mathcal{C}_{{\mathit A}_{\textrm{in}}}^{{\mathit B}}{z}_{{\mathit A}}^{-\mathtt{x}}$ and the corresponding transmission coefficient is 
$\mathcal{C}_{{\mathit A}_{\textrm{in}}}^{{\mathit B}}=1+\mathcal{C}_{{\mathit A}_{\textrm{in}}}^{{\mathit A}}$. However $|\mathcal{C}_{{\mathit A}_{\textrm{in}}}^{{\mathit A}}|^2+|\mathcal{C}_{{\mathit A}_{\textrm{in}}}^{{\mathit B}}|^2|{{{{\mathit v}}_{{\mathit B}}}({{\upxi}}_{{\mathit B}})}/{{{{\mathit v}}_{{\mathit A}}}({{\upxi}}_{{\mathit A}})}|<1$, in general, due to the generation of bulk lattice waves (contribution of continuous spectrum) at the point source.

\begin{remark}
The discussion above brings out the interpretation of coefficients appearing in \eqref{discWH} touted as the {Wiener-Hopf} equation on ${\mathscr{A}}$.
On the same lines, the Toeplitz structure of the relevant scattering problem in case of a sandwiched region of different surface parameters, essentially the case of a pair of interfaces, can be formulated in a way similar to that for finite cracks and rigid constraint \cite{sFK,sFC}; the corresponding analysis is omitted in the present article.
\end{remark}

\subsection{Scattering in one dimensional lattice model with an interface}
\label{secone}

Consider the physical lattice constituted by particles present on the lattice half-plane discussed in this article except one alteration that the interaction with half-plane is replaced by that with a substrate.
Let the sites in such one-dimensional lattice be described by
$\{\mathbf{x}\in{\mathbb{R}}: \mathbf{x}=\mathtt{x}{a}\text{ for some } \mathtt{x}\in{\mathbb{Z}}\}$.
Analogous to \eqref{eqmotionBCa},
\eqref{eqmotionBCb},
\eqref{eqmotionBC0},
the equation of motion is
\begin{subequations}
\begin{align}
{\alpha}_{\mathfrak{s}}({\mathtt{u}}^{\textrm{to}}_{{\mathtt{x}}+1}+{\mathtt{u}}^{\textrm{to}}_{{\mathtt{x}}-1}-2{\mathtt{u}}^{\textrm{to}}_{\mathtt{x}})-{\upkappa_{s}}_{\mathfrak{s}}{\mathtt{u}}^{\textrm{to}}_{\mathtt{x}}
&={\mathit{m}}_{\mathfrak{s}}\ddot{\mathtt{u}}^{\textrm{to}}_{\mathtt{x}},
\notag\\
\text{where }
\text{for $\mathfrak{s}={{\mathit B}}$}, \mathtt{x}\in{\mathbb{Z}}^-,\quad
\text{for $\mathfrak{s}={{\mathit A}}$}, \mathtt{x}\in{\mathbb{Z}}^+\setminus\{0\}, \label{bc_sq_AB}\\
{\alpha}_{{\mathit B}}({\mathtt{u}}^{\textrm{to}}_{{\mathtt{x}}-1}-{\mathtt{u}}^{\textrm{to}}_{\mathtt{x}})+{\alpha}_{{\mathit A}}({\mathtt{u}}^{\textrm{to}}_{{\mathtt{x}}+1}-{\mathtt{u}}^{\textrm{to}}_{\mathtt{x}})-{\upkappa_{s}}_{0}{\mathtt{u}}^{\textrm{to}}_{\mathtt{x}}
&={\mathit{m}}_{0}\ddot{\mathtt{u}}^{\textrm{to}}_{\mathtt{x}},
\quad
{\mathtt{x}}=0\label{bc_sq_0}.
\end{align}
\end{subequations}
Notice that above lattice model possesses two more material parameters than the case of lattice half-plane, namely ${\upkappa_{s}}_{{\mathit A}}, {\upkappa_{s}}_{{\mathit B}}, {\upkappa_{s}}_{0}$ in place of ${\alpha}_{{}}$ in addition to the mutually common list of ${\mathit{m}}_{{\mathit A}}, {\mathit{m}}_{{\mathit B}}, {\mathit{m}}_{0}, {\alpha}_{{\mathit A}}, {\alpha}_{{\mathit B}}$.

Recall the incident wave \eqref{uincawave2d}
for half-plane and ensuing discussion of wave dispersion.
Consider the following incident wave in one dimensional model:
\begin{equation}\begin{split}
{\mathtt{u}}^{{\mathit A}_{\textrm{in}}}_{\mathtt{x}}={{{\mathtt{u}}^{{\mathit A}_{\textrm{in}}}_{0}}}\exp({-i{{\upxi}}_{{\mathit A}} {\mathtt{x}}-i\upomega t}),
\label{uincawave1}
\end{split}\end{equation}
where ${{\upxi}}_{{\mathit A}}$ is the discrete wavenumber from the side ahead such that ${{{\mathit v}}}({{\upxi}}_{{\mathit A}})>0$, ${{\upxi}}_{{\mathit A}}\in(0,\pi)$.
It is easily found that ${\upomega}={\upomega}_{{\mathit A}}({{\upxi}}_{{\mathit A}})$ satisfies
$-{\upomega}^2{\mathit{m}}_{{\mathit A}}
=\left(2{\alpha}_{{\mathit A}}\cos{{\upxi}}_{{\mathit A}}-(2{\alpha}_{{\mathit A}}+{\upkappa_{s}}_{{\mathit A}})\right).$
The group velocity is $\upomega^{-1}({{\alpha}_{{\mathit A}}}/{{\mathit{m}}_{{\mathit A}}})\sin{{\upxi}}_{{\mathit A}}$.
\begin{remark}
An incident wave traveling towards the interface from the portion behind is given by
${\mathtt{u}}^{{\mathit B}_{\textrm{in}}}_{\mathtt{x}}={{{\mathtt{u}}^{{\mathit B}_{\textrm{in}}}_{0}}}\exp({+i{{\upxi}}_{{\mathit B}} {\mathtt{x}}-i\upomega t})$,
where $-{\mathit{m}}_{{\mathit B}}{\upomega}^2
=\left(2{\alpha}_{{\mathit B}}\cos{{\upxi}}_{{\mathit B}}-(2{\alpha}_{{\mathit B}}+{\upkappa_{s}}_{{\mathit B}})\right).$
An arbitrary incident wave considered is a combination of both, i.e.
${\mathtt{u}}^{\textrm{in}}_{\mathtt{x}}={\mathtt{u}}^{{\mathit B}_{\textrm{in}}}_{\mathtt{x}}+{\mathtt{u}}^{{\mathit A}_{\textrm{in}}}_{\mathtt{x}}.$
\end{remark}
\begin{remark}
Following \S\ref{sec2dsca} on the 
problem formulation of surface wave scattering on $\mathbb{H}$,
the discrete Helmholtz equation on each semi-infinite piece of the lattice modulo ${\mathtt{x}}=0$ and at the site ${\mathtt{x}}=0$
is given by
\eqref{BCeqa},\eqref{deffinca}. The analysis of above {Wiener-Hopf} equation follows that for the lattice half-plane except a matter of convenience in this case that the calculation of multiplicative factors of {Wiener-Hopf} kernel is explicit.
See Appendix \ref{seconeappendix} for some details regarding the above equation.
\end{remark}

\begin{figure}[!ht]
\center
(a){\includegraphics[width=.7\textwidth]{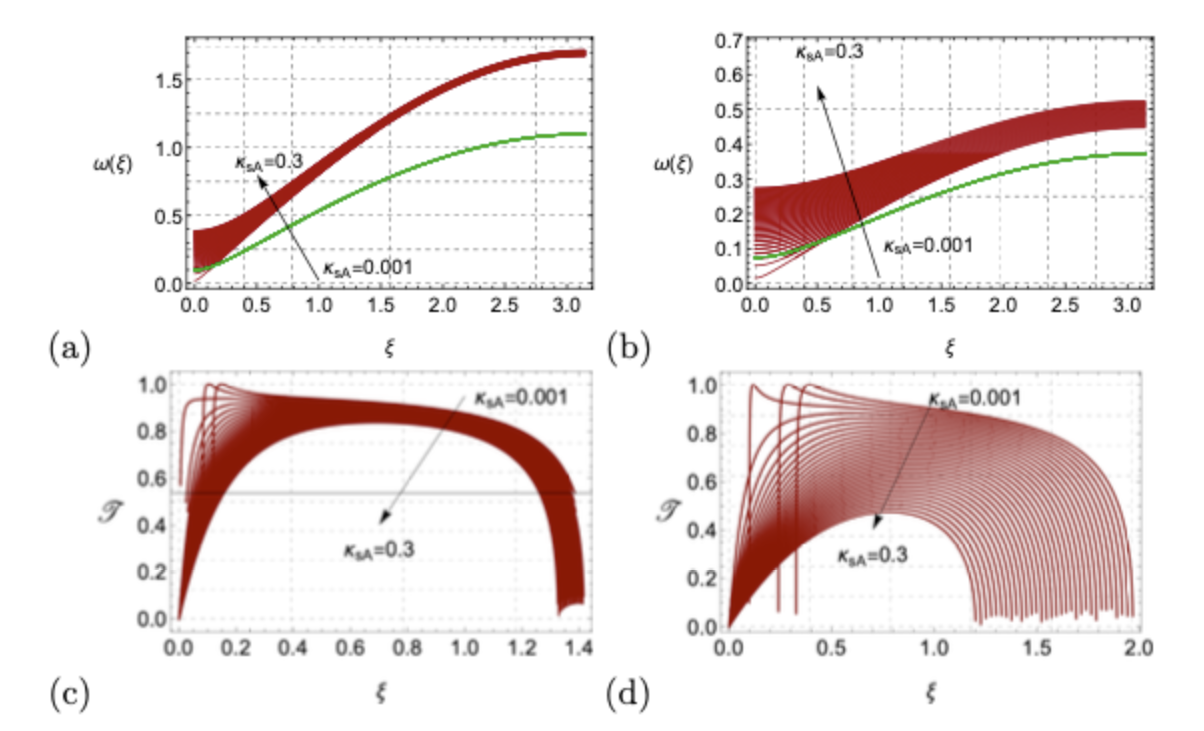}}
\caption[]{(Top) Dispersion relation for $\upomega({\upxi})$ vs ${\upxi}$ when ${\upxi}$ belongs to the common passing band sub-interval of $(0,\pi)$. Green curve corresponds to ${\alpha}_{{\mathit B}}=0.05$. 
(Bottom) Transmittance ${\mathscr{T}}$ for incidence from the side ahead. The choice of parameters in (a), (c) is (S$1$) and in (b), (d) is (S$2$) stated in \eqref{S14choice} except for ${\alpha}_{{}}$ which is not assigned in the one-dimensional case. Compare (c) and (d) with Fig. \ref{ErefA_surfacestructure}(c).}
\label{S14onedimsurfacestructure}
\end{figure}

In the simpler situation
assuming ${\alpha}_{{\mathit A}}={\alpha}_{{\mathit B}}$ and ${\mathit{m}}_{0}={\mathit{m}}_{{\mathit A}}$ but ${\mathit{m}}_{{\mathit A}}\ne{\mathit{m}}_{{\mathit B}}$,
(reall Remark \ref{speccase2}) and assuming ${\upkappa_{s}}_{0}={\upkappa_{s}}_{{\mathit A}}$,
following the reasoning based on \eqref{totincsplit}, \eqref{uscadef},
as an example of calculation on the back-of-the-envelope in this case,
let 
\begin{equation}\begin{split}
\mathtt{x}\in{\mathbb{Z}}^+: {\mathtt{u}}_{\mathtt{x}}&=\exp({-i {\upxi}_{{\mathit A}} \mathtt{x}})+\mathcal{C}_{{\mathit A}_{\textrm{in}}}^{{\mathit A}} \exp({i {\upxi}_{{\mathit A}} \mathtt{x}}),\\
\mathtt{x}\in{\mathbb{Z}}^-: {\mathtt{u}}_{\mathtt{x}}&=\mathcal{C}_{{\mathit A}_{\textrm{in}}}^{{\mathit B}} \exp({-i {\upxi}_{{\mathit B}} \mathtt{x}}).
\label{solspeccaseone}
\end{split}\end{equation}
Due to `continuity' at $\mathtt{x}=0,$ i.e. using the discrete Helmholtz equation with above ansatz from left and right of $\mathtt{x}=0$,
$1+\mathcal{C}_{{\mathit A}_{\textrm{in}}}^{{\mathit A}}=\mathcal{C}_{{\mathit A}_{\textrm{in}}}^{{\mathit B}}$
and 
$-{\alpha}_{{\mathit A}}(\exp({i {\upxi}_{{\mathit A}}})+\mathcal{C}_{{\mathit A}_{\textrm{in}}}^{{\mathit A}} \exp({-i {\upxi}_{{\mathit A}}}))+{\alpha}_{{\mathit A}}\mathcal{C}_{{\mathit A}_{\textrm{in}}}^{{\mathit B}} \exp({i {\upxi}_{{\mathit B}}})=0,$
whose solution gives $\mathcal{C}_{{\mathit A}_{\textrm{in}}}^{{\mathit A}}$ and $\mathcal{C}_{{\mathit A}_{\textrm{in}}}^{{\mathit B}}$ same as the expressions stated in \eqref{discWHNsolhalf2p},\eqref{discWHNsolhalf2n};
see Remark \ref{speccaseone} for some details.
In general, \eqref{solspeccaseone} continues to hold and the back-of-another-envelope can be used in the same way,
the expressions of the reflection and transmission coefficients can be found to be
\begin{align}
\mathcal{C}_{{\mathit A}_{\textrm{in}}}^{{\mathit A}}
=\mathcal{C}_{{\mathit A}_{\textrm{in}}}^{{\mathit B}}-1,\quad
\mathcal{C}_{{\mathit A}_{\textrm{in}}}^{{\mathit B}}=\frac{{\alpha}_{{\mathit A}}({{z}}_{{\mathit A}}-{{z}}_{{\mathit A}}^{-1})}{({\mathtt{M}_0}+1) ({\alpha}_{{\mathit A}}-{\alpha}_{{\mathit B}})-{\alpha}_{{\mathit A}}{{z}}_{{\mathit A}}^{-1}+{\alpha}_{{\mathit B}}{{z}}_{{\mathit B}}},
\label{CABhalfincAone}
\end{align}
where ${\mathtt{M}_0}$ is defined in \eqref{defAratone}.
It can be verified that
${\mathscr{R}}+{\mathscr{T}}=1$, where ${\mathscr{R}}:=|\mathcal{C}_{{\mathit A}_{\textrm{in}}}^{{\mathit A}}|^2, {\mathscr{T}}:=|\mathcal{C}_{{\mathit A}_{\textrm{in}}}^{{\mathit B}}|^2
({{\alpha}_{{\mathit B}}\sin{{\upxi}}_{{\mathit B}}})/({{\alpha}_{{\mathit A}}\sin{{\upxi}}_{{\mathit A}}})
$,
due to absence of any damping or energy leakage.

Fig. \ref{S14onedimsurfacestructure}(a),(b) illustrates the dispersion relation for $\upomega({\upxi})$ vs ${\upxi}$, mentioned after \eqref{uincawave1}, when ${\upxi}$ belongs to the common passing band sub-interval of $(0,\pi)$; Fig. \ref{S14onedimsurfacestructure}(c),(d) presents the corresponding transmittance. The green curve in Fig. \ref{S14onedimsurfacestructure}(a),(b) corresponds to
the portion behind the interface. The choice of parameters in (a), (b) is (S$1$) and in (c), (d) is (S$2$) stated in \eqref{S14choice} except for ${\alpha}_{{}}$ which is not relevant in the one-dimensional model as stated, however additional parameters ${\upkappa_{s}}_{{\mathit A}}, {\upkappa_{s}}_{{\mathit B}}$ appear. On comparison of Fig. \ref{S14onedimsurfacestructure}(c) and Fig. \ref{S14onedimsurfacestructure}(d) with Fig. \ref{ErefA_surfacestructure}(c), it seen that large values of surface structure parameters correspond to transmittance of similar nature as that with higher values of ${\alpha}_{{\mathit A}}$ in the choices used (${\alpha}_{{\mathit B}}$ is fixed in the plot). However, it is not possible to generalize this observation  to a great extent as they are inherently different models; for example, the one-dimensional model does not exhibit any energy leakage but that is an important feature of surface wave motion in the lattice half-plane problem with interface.

\subsection{Continuum limit}
\label{sec2dcont}

A continuum limit can be obtained following \cite{Victor_Bls_surf1} (when ${\alpha}_{{}}=1$, for simplicity, while the results for arbitrary ${\alpha}_{{}}>0$ can be obtained in a similar manner).
An exact solution for the continuous surface interface is also possible using Wiener-Hopf formulation;
this has been also found out independent of the analysis presented in the present article. The details of the continuous problem await a forthcoming manuscript.
It is clear that the lattice model serves as a regularization of the corresponding continuum model. 
In other words, the regime $0<\upomega\ll1$ can be approximated by the continuous problem and vice versa. Thus the problem solved in this article also helps in providing a very good estimate of the transmittance and reflectance in the continuous problem by restricting to lower values of frequency, $\upomega\in(0, 0.1 \upomega_{\max})$, for example ($\upomega_{\max}$ represents the maximum of surface wave band of either side of interface).

\subsection{Relation between two dimensional lattice half-plane with an interface vs lattice strip}
\label{sec1strip}

In this article, an expression of exact solution has been found for the lattice half-plane problem that was shelved in \cite{Victor_Bls_surf2}, although lattice strip problem was (deemed to be) presented as a very good approximation for the half-plane when the width of the strip is large enough.
In fact, the exact solution of the Wiener-Hopf equation for the scattering problem posed and solved in \cite{Victor_Bls_surf2} coincides with the solution presented in this article provided $\lambda$ in \eqref{discWHker} is replaced by $\Lambda$ defined in (32) of \cite{Victor_Bls_surf2}, ${z}_{{\mathit A}}^{-1}$ is replaced by ${{z}}_{\text{in}}$ from \cite{Victor_Bls_surf2}, and ${{\mathtt{u}}^{{\mathit A}_{\textrm{in}}}_{0,0}}$ is replaced by appropriate factor to account for the incident wave mode amplitude in the case of lattice strip. As the width of lattice strip increases the limiting behaviour coincides with that obtained in the present article.
In fact, the surface structure parameters in $({\mathrm S}2)$, $({\mathrm S}3)$ are also same as those in several graphical results of \cite{Victor_Bls_surf2}, for example, the choices accompanying Fig 4(b), Fig. 5, Fig. 6, Fig. 7, and Fig. 8. Thus, a visual comparison between graphical results of \cite{Victor_Bls_surf2} and some results presented in the previous section may prepare a case to convince the reader about the above statements concerning the relation between two dimensional lattice half-plane with an interface vs lattice strip.

\section*{Concluding remarks}

As structured surfaces and interfaces begin to play a significant role in technology, the problem of  optimization of  structure parameters paves way, to enhance the desired effects, for tuning related to particular applications. 
In the scattering problem analyzed in this article, from a viewpoint of classical mechanics,
the scalar field at a site
in the lattice half-plane with boundary corresponds to the dimensionless out-of-plane displacement 
of that particle at that site.
The specific model adopted 
incorporates an anisotropy due to unequal force-constants parallel to surface versus normal to surface.
The surface interface appears as a boundary between material particles of two types which both are different from the ones in the bulk. 
 In other words, the 
 half-space boundary consists of two parts with different surface wave properties.

The wave transmission and reflection along the interface is analyzed, as well as energy leakage ${\mathscr{D}}_{{surf}}$ into the bulk.
In the context of the present article, tunable energy `damping', or `activation' of surface waves, by an appropriate modification of surface material properties and structure of interfaces appears plausible in the entire surface wave band.
As a preliminary step towards this goal, a prototype problem is solved exactly in this article which allows closed form expression for transmittance in surface wave band. An exact expression for the fraction ${\mathscr{D}}_{{surf}}$ of surface energy influx that enters the bulk of the lattice half-plane, per unit energy flux of the incident surface wave, during the process of wave transmission on the surface is given.
The question about the dependence of ${\mathscr{D}}_{{surf}}$ on the surface material parameters is, thus, answered
after solving for the scattering solution using the Wiener-Hopf technique and the limiting absorption principle.

A quantitative comparison of the limiting behaviour of wide enough lattice strip 
with that of 
the present article as advertized in \cite{Victor_Bls_surf2} is not difficult but details have been omitted.
Besides the surface wave propagation across interfaces, the problem of excitation of surface waves due to incident bulk wave can be solved following \cite{sharma2017scattering}, however this has been omitted too in the present article.
The continuum limit has been treated briefly and the complete solution of the continuum problem has been relegated to another forum.
The related problems of multiple scattering due to a finite number of such interfaces is currently under investigaton.

\section*{Acknowledgments}
{Thanks to Isaac Newton Institute for Mathematical Sciences (INI), Cambridge, for support and hospitality during the programme `Mathematical theory and applications of multiple wave scattering' (MWS) where partial work on this article was undertaken and was supported by EPSRC grant no EP/R014604/1. Same visit to INI during January-June 2023 for participation in MWS was partially supported by a grant from the Simons Foundation.}

\appendix

\renewcommand{\theequation}{A.\arabic{equation}}
\section{Energy flux in the incident surface wave}
\label{appenfixsurfwaveflux}

Consider the incident wave
\eqref{uincawave2d} as an example.
Following Appendix B of \cite{Victor_Bls_surf2}, consider the energy flux across, say $\mathtt{x}=\mathtt{x}_0-1/2, \mathtt{x}_0=0$ from side behind ($\mathtt{x}-\mathtt{x}_0\in{\mathbb{Z}}^-$) to side ahead ($\mathtt{x}-\mathtt{x}_0{\mathbb{Z}}^+\setminus\{0\}$).
Then 
from the side $\mathtt{x}-\mathtt{x}_0\in{\mathbb{Z}}^-$ on particle at $\mathtt{x}_0$, the rate of work done (average over a large time period $T$) is,
using \eqref{uincawave2d}${}_2$, a sum of two terms, the first of which is
\begin{align}
{{\alpha}_{{}}}\frac{1}{2}\Re\sum\nolimits_{\mathtt{y}=0}^\infty({\mathtt{u}}^{{\mathit A}_{\textrm{in}}}_{\mathtt{x}_0-1,\mathtt{y}}-{\mathtt{u}}^{{\mathit A}_{\textrm{in}}}_{\mathtt{x}_0,\mathtt{y}})\overline{\dot{\mathtt{u}}^{{\mathit A}_{\textrm{in}}}_{\mathtt{x}_0,\mathtt{y}}}
&=
{{\alpha}_{{}}}\frac{1}{2}\frac{|{{\mathtt{u}}^{{\mathit A}_{\textrm{in}}}_{0}}|^2\upomega\sin{{\upxi}}_{{\mathit A}}}{1-\exp(-2{{\eta}({\upxi}_{{\mathit A}})})},
\end{align}
and the second is
\begin{align}
({\alpha}_{{\mathit A}}-{{\alpha}_{{}}})\frac{1}{2}\Re({\mathtt{u}}^{{\mathit A}_{\textrm{in}}}_{\mathtt{x}_0-1,0}-{\mathtt{u}}^{{\mathit A}_{\textrm{in}}}_{\mathtt{x}_0,0})\overline{\dot{\mathtt{u}}^{{\mathit A}_{\textrm{in}}}_{\mathtt{x}_0,0}}=\frac{1}{2}({\alpha}_{{\mathit A}}-{{\alpha}_{{}}})|{{\mathtt{u}}^{{\mathit A}_{\textrm{in}}}_{0}}|^2\upomega\sin{{\upxi}}_{{\mathit A}}\frac{1-\exp(-2{{\eta}({\upxi}_{{\mathit A}})})}{1-\exp(-2{{\eta}({\upxi}_{{\mathit A}})})}.
\end{align}
The kinetic energy (average over $T$) between $\mathtt{x}_0-1/2$ and $\mathtt{x}_0+1/2$ is
\begin{align}
(\frac{1}{2}({\mathit{m}}_{{\mathit A}}-1)|\dot{\mathtt{u}}^{{\mathit A}_{\textrm{in}}}_{\mathtt{x}_0,0}|^2+\frac{1}{2}\sum\nolimits_{\mathtt{y}=0}^\infty|\dot{\mathtt{u}}^{{\mathit A}_{\textrm{in}}}_{\mathtt{x}_0,\mathtt{y}}|^2)=\frac{1}{2}\upomega^2({\mathit{m}}_{{\mathit A}}-1)|{{\mathtt{u}}^{{\mathit A}_{\textrm{in}}}_{0}}|^2
+\frac{\frac{1}{2}\upomega^2|{{\mathtt{u}}^{{\mathit A}_{\textrm{in}}}_{0}}|^2}{1-\exp(-2{{\eta}({\upxi}_{{\mathit A}})})}.
\end{align}
Threfore, the energy velocity
is
\begin{align}
{{\mathit v}}({{\upxi}}_{{\mathit A}}):=
\frac{\sin{{\upxi}}_{{\mathit A}}}{\upomega}\frac{({\alpha}_{{\mathit A}}-{{\alpha}_{{}}})(1-\exp(-2{{\eta}({\upxi}_{{\mathit A}})}))+{{\alpha}_{{}}}}{({\mathit{m}}_{{\mathit A}}-1)(1-\exp(-2{{\eta}({\upxi}_{{\mathit A}})}))+1}.
\label{energyvel}\end{align}
Clearly, ${{\mathit v}}({{\upxi}}_{{\mathit A}})=-{{\mathtt{v}}}(-{{\upxi}}_{{\mathit A}})$.
It is not surprising \cite{Brillouin} that the energy velocity \eqref{energyvel} coincides with group velocity \eqref{surfgpvel}.

\renewcommand{\theequation}{B.\arabic{equation}}
\section{Application of Fourier transform}
\label{appA}

The Fourier transform of \eqref{BCeq} leads to
\begin{equation}\begin{split}\label{discWHkerpre}
{\alpha}_{{\mathit A}}({{z}}{\mathtt{u}}^+-{{z}}{\mathtt{u}}_{0, 0}-{\mathtt{u}}^+)+{\alpha}_{{\mathit A}}({{z}}^{-1}{\mathtt{u}}^++{\mathtt{u}}_{0, 0}-{\mathtt{u}}^+)\\
+{\alpha}_{{\mathit B}}({{z}}{\mathtt{u}}^-+{{z}}{\mathtt{u}}_{0, 0}-{\mathtt{u}}^-)+{\alpha}_{{\mathit B}}({{z}}^{-1}{\mathtt{u}}^--{\mathtt{u}}_{0, 0}-{\mathtt{u}}^-)\\
+({\mathtt{u}}^++{\mathtt{u}}^-)({\lambda}-1)+{\mathit{m}}_{{\mathit A}}{\upomega}^2{\mathtt{u}}^++{\mathit{m}}_{{\mathit B}}{\upomega}^2{\mathtt{u}}^-=-\sum\nolimits_{{\mathtt{x}}\in{\mathbb{Z}}}f^{{\mathit A}_{\textrm{in}}}_{\mathtt{x}}{{z}}^{-{\mathtt{x}}}-({\mathit{m}}_{0}-{\mathit{m}}_{{\mathit A}})\upomega^2{({\mathtt{u}}_{0, 0}+{\mathtt{u}}^{{\mathit A}_{\textrm{in}}}_{0, 0})},
\end{split}\end{equation}
where, using \eqref{deffinc} and \eqref{uincF}, the first term in right hand side has the expanded form
\begin{equation}\begin{split}
-({\alpha}_{{\mathit B}}-{\alpha}_{{\mathit A}})({{z}}{\mathtt{u}}^{{\mathit A}_{\textrm{in}}}{}^-+{{z}}{\mathtt{u}}^{{\mathit A}_{\textrm{in}}}_{0, 0}-{\mathtt{u}}^{{\mathit A}_{\textrm{in}}}{}^-+{{z}}^{-1}{\mathtt{u}}^{{\mathit A}_{\textrm{in}}}{}^--{\mathtt{u}}^{{\mathit A}_{\textrm{in}}}_{0, 0}-{\mathtt{u}}^{{\mathit A}_{\textrm{in}}}{}^-)-({\mathit{m}}_{{\mathit B}}-{\mathit{m}}_{{\mathit A}}){\upomega}^2{\mathtt{u}}^{{\mathit A}_{\textrm{in}}}{}^-\\=
-({\alpha}_{{\mathit B}}-{\alpha}_{{\mathit A}})(({{z}}+{{z}}^{-1}-2){\mathtt{u}}^{{\mathit A}_{\textrm{in}}}{}^-+({{z}}-1){\mathtt{u}}^{{\mathit A}_{\textrm{in}}}_{0, 0})-({\mathit{m}}_{{\mathit B}}-{\mathit{m}}_{{\mathit A}}){\upomega}^2{\mathtt{u}}^{{\mathit A}_{\textrm{in}}}{}^-,
\label{BCeqF}
\end{split}\end{equation}
where ${\mathtt{u}}^{{\mathit A}_{\textrm{in}}}{}^-:={{{\mathtt{u}}^{{\mathit A}_{\textrm{in}}}_{0,0}}}{\updelta}_{D}^-({{z}}{{z}}_{{\mathit A}})$.
Due to \eqref{uincawave2d}${}_2$, the expression \eqref{uincF} appears as a result of
${\mathtt{u}}^{{\mathit A}_{\textrm{in}}}{}^-=\sum\nolimits_{{\mathtt{x}}\in{\mathbb{Z}}^-}{\mathtt{u}}^{{\mathit A}_{\textrm{in}}}_{{\mathtt{x}},{\mathtt{y}}}{{z}}^{-{\mathtt{x}}}={{{\mathtt{u}}^{{\mathit A}_{\textrm{in}}}_{0,0}}}{\updelta}_{D}^-({{z}}{{z}}_{{\mathit A}})=-{{{\mathtt{u}}^{{\mathit A}_{\textrm{in}}}_{0,0}}}\frac{{{z}}}{{{z}}-{{z}}_{{\mathit A}}^{-1}}, \quad |{{z}}{{z}}_{{\mathit A}}|<1.$

\renewcommand{\theequation}{C.\arabic{equation}}
\section{Details for transmission and reflection coefficients}
\label{appB}

Using \eqref{discWHNsolhalf}${}_2$, \eqref{discCfacalleq},
\begin{align}
{\mathtt{u}}^{-}({{z}})
&={{{\mathtt{u}}^{{\mathit A}_{\textrm{in}}}_{0,0}}}(1-{\mathfrak{L}}_-^{-1}({{z}}){\mathfrak{L}}_+^{-1}({{z}}_{{\mathit A}}^{-1}))\frac{{{z}}}{{{z}}-{{z}}_{{\mathit A}}^{-1}}+({\alpha}_{{\mathit A}}-{\alpha}_{{\mathit B}}){({\mathtt{u}}_{0, 0}+{\mathtt{u}}^{{\mathit A}_{\textrm{in}}}_{0, 0})}{\mathfrak{L}}_-^{-1}({{z}})\notag\\
&\bigg(({\mathfrak{L}}_-({z})\frac{(1+\mathtt{M}-{{z}})}{{\mathtt{p}_+({z})}\mathtt{p}_-({z})}
-{\mathfrak{L}}_-({z}_{\mathtt{p}})\frac{(1+\mathtt{M}-{{z}_{\mathtt{p}}})}{{\mathtt{p}_+({z})}\mathtt{p}_-({z}_{\mathtt{p}})})
+{\mathfrak{L}}_+^{-1}({z}_{\mathtt{p}}^{-1})\frac{(1+\mathtt{M}-{{z}_{\mathtt{p}}^{-1}})}{\mathtt{p}_+({z}_{\mathtt{p}}^{-1})}(\frac{1}{\mathtt{p}_-(0)}-\frac{1}{\mathtt{p}_-({z})})\bigg),
\end{align}
so that
as ${{z}}\to{{z}}_{{\mathit B}}^{-1}$, ${\mathfrak{K}}_{{{\mathit B}}{}-}({{z}})= {\mathfrak{K}}_{{{\mathit B}}{}-}'({{z}}_{{\mathit B}}^{-1})({{z}}-{{z}}_{{\mathit B}}^{-1})+\mathpzc{o}({{z}}-{{z}}_{{\mathit B}}^{-1}),$
\begin{align}
{\mathtt{u}}^{-}({{z}})
&=({{z}}-{{z}}_{{\mathit B}}^{-1})^{-1}\bigg(\frac{({{z}}_{{\mathit B}}^{-1}{{z}}_{{\mathit A}})}{1-({{z}}_{{\mathit B}}^{-1}{{z}}_{{\mathit A}})}
\frac{{\mathfrak{K}}_{{{\mathit A}}{}-}({{z}}_{{\mathit B}}^{-1})}{{\mathfrak{K}}_{{{\mathit B}}{}-}'({{z}}_{{\mathit B}}^{-1})}\frac{{\mathfrak{K}}_{{{\mathit A}}{}+}({{z}}_{{\mathit A}}^{-1})}{{\mathfrak{K}}_{{{\mathit B}}{}+}({{z}}_{{\mathit A}}^{-1})}{\mathtt{u}}^{{\mathit A}_{\textrm{in}}}_{{0,0}}\notag\\
&-\frac{{\mathfrak{K}}_{{{\mathit A}}{}-}({{z}}_{{\mathit B}}^{-1})}{{\mathfrak{K}}_{{{\mathit B}}{}-}'({{z}}_{{\mathit B}}^{-1})}({\alpha}_{{\mathit A}}-{\alpha}_{{\mathit B}}){({\mathtt{u}}_{0, 0}+{\mathtt{u}}^{{\mathit A}_{\textrm{in}}}_{0, 0})}\notag\\
&\bigg({\mathfrak{L}}_-({z}_{\mathtt{p}})\frac{(1+\mathtt{M}-{{z}_{\mathtt{p}}})}{{\mathtt{p}_+({z}_{{\mathit B}}^{-1})}\mathtt{p}_-({z}_{\mathtt{p}})}-{\mathfrak{L}}_+^{-1}({z}_{\mathtt{p}}^{-1})\frac{(1+\mathtt{M}-{{z}_{\mathtt{p}}^{-1}})}{\mathtt{p}_+({z}_{\mathtt{p}}^{-1})}(\frac{1}{\mathtt{p}_-(0)}-\frac{1}{\mathtt{p}_-({z}_{{\mathit B}}^{-1})})\bigg)\bigg)+\mathpzc{O}(1),
\end{align}
which yields \eqref{CBhalfincA}.
Similarly, using \eqref{discWHNsolhalf}${}_1$, \eqref{discCfacalleq},
\begin{align}
{\mathtt{u}}^{+}({{z}})
&={{{\mathtt{u}}^{{\mathit A}_{\textrm{in}}}_{0,0}}}(-1+{\mathfrak{L}}_+({{z}}){\mathfrak{L}}_+^{-1}({{z}}_{{\mathit A}}^{-1}))\frac{1}{1-{{z}}_{{\mathit A}}^{-1}{{z}}^{-1}}+({\alpha}_{{\mathit A}}-{\alpha}_{{\mathit B}}){({\mathtt{u}}_{0, 0}+{\mathtt{u}}^{{\mathit A}_{\textrm{in}}}_{0, 0})}{\mathfrak{L}}_+({{z}})\notag\\
&\bigg({\mathfrak{L}}_-({z}_{\mathtt{p}})\frac{(1+\mathtt{M}-{{z}_{\mathtt{p}}})}{{\mathtt{p}_+({z})}\mathtt{p}_-({z}_{\mathtt{p}})}
-({\mathfrak{L}}_+^{-1}({z})\frac{(1+\mathtt{M}-{z})}{\mathtt{p}_+({z})\mathtt{p}_-({z})}
+{\mathfrak{L}}_+^{-1}({z}_{\mathtt{p}}^{-1})\frac{(1+\mathtt{M}-{{z}_{\mathtt{p}}^{-1}})}{\mathtt{p}_+({z}_{\mathtt{p}}^{-1})}(\frac{1}{\mathtt{p}_-(0)}-\frac{1}{\mathtt{p}_-({z})}))\bigg),
\end{align}
so that
as ${{z}}\to{{z}}_{{\mathit A}}$, ${\mathfrak{K}}_{{{\mathit B}}{}+}({{z}})= {\mathfrak{K}}_{{{\mathit B}}{}+}'({{z}}_{{\mathit A}})({{z}}-{{z}}_{{\mathit A}})+\mathpzc{o}({{z}}-{{z}}_{{\mathit A}}),$
\begin{align}
{\mathtt{u}}^{+}({{z}})
&=({{z}}-{{z}}_{{\mathit A}})^{-1}\bigg(-{\updelta}_{D}^-({{z}}_{{\mathit A}}{{z}}_{{\mathit A}})
\frac{{\mathfrak{K}}_{{{\mathit B}}{}+}({{z}}_{{\mathit A}})}{{\mathfrak{K}}_{{{\mathit A}}{}+}'({{z}}_{{\mathit A}})}\frac{{\mathfrak{K}}_{{{\mathit A}}{}+}({{z}}_{{\mathit A}}^{-1})}{{\mathfrak{K}}_{{{\mathit B}}{}+}({{z}}_{{\mathit A}}^{-1})}{\mathtt{u}}^{{\mathit A}_{\textrm{in}}}_{{0,0}}+\frac{{\mathfrak{K}}_{{{\mathit B}}{}+}({{z}}_{{\mathit A}})}{{\mathfrak{K}}_{{{\mathit A}}{}+}'({{z}}_{{\mathit A}})}({\alpha}_{{\mathit A}}-{\alpha}_{{\mathit B}}){({\mathtt{u}}_{0, 0}+{\mathtt{u}}^{{\mathit A}_{\textrm{in}}}_{0, 0})}\notag\\
&\bigg({\mathfrak{L}}_-({z}_{\mathtt{p}})\frac{(1+\mathtt{M}-{{z}_{\mathtt{p}}})}{{\mathtt{p}_+({z})}\mathtt{p}_-({z}_{\mathtt{p}})}
-{\mathfrak{L}}_+^{-1}({z}_{\mathtt{p}}^{-1})\frac{(1+\mathtt{M}-{{z}_{\mathtt{p}}^{-1}})}{\mathtt{p}_+({z}_{\mathtt{p}}^{-1})}(\frac{1}{\mathtt{p}_-(0)}-\frac{1}{\mathtt{p}_-({z})})\bigg)\bigg)+\mathpzc{O}(1),
\end{align}
which yields \eqref{CAhalfincA}.

\renewcommand{\theequation}{D.\arabic{equation}}
\section{Auxiliary details for one dimensional case}
\label{seconeappendix}

Since ${\mathtt{u}}^{{\mathit A}_{\textrm{in}}}_{\mathtt{x}}$ \eqref{uincawave1} satisfies the equation of motion in the portion with boundary with index ${\mathfrak{s}}={{\mathit A}}$,
\begin{align}
&({\mathtt{u}}_{{\mathtt{x}}+1}-{\mathtt{u}}_{\mathtt{x}})({\alpha}_{{\mathit A}}{{\mathscr{H}}}({\mathtt{x}})+{\alpha}_{{\mathit B}}{{\mathscr{H}}}(-{\mathtt{x}}-1))+({\mathtt{u}}_{{\mathtt{x}}-1}-{\mathtt{u}}_{\mathtt{x}})({\alpha}_{{\mathit A}}{{\mathscr{H}}}({\mathtt{x}}-1)+{\alpha}_{{\mathit B}}{{\mathscr{H}}}(-{\mathtt{x}}))\notag\\&
-({\upkappa_{s}}_{{\mathit A}}{{\mathscr{H}}}({\mathtt{x}})+{\upkappa_{s}}_{{\mathit B}}{{\mathscr{H}}}(-{\mathtt{x}}-1)){\mathtt{u}}_{\mathtt{x}}+{\upomega}^2{\mathtt{u}}_{\mathtt{x}}({\mathit{m}}_{{\mathit A}}{{\mathscr{H}}}({\mathtt{x}})+{\mathit{m}}_{{\mathit B}}{{\mathscr{H}}}(-{\mathtt{x}}-1))\notag\\
&=-f^{{\mathit A}_{\textrm{in}}}_{\mathtt{x}}-({\mathit{m}}_{0}-{\mathit{m}}_{{\mathit A}}){\upomega}^2{\mathtt{u}}^{\textrm{to}}_{\mathtt{x}}\delta_{\mathtt{x},0}+({\upkappa_{s}}_0-{\upkappa_{s}}_{{\mathit A}}){\mathtt{u}}^{\textrm{to}}_{\mathtt{x}}\delta_{\mathtt{x},0},\label{BCeqa}\\
\text{where }f^{{\mathit A}_{\textrm{in}}}_{\mathtt{x}}&=({\alpha}_{{\mathit B}}-{\alpha}_{{\mathit A}})({\mathtt{u}}^{{\mathit A}_{\textrm{in}}}_{{\mathtt{x}}+1}-{\mathtt{u}}^{{\mathit A}_{\textrm{in}}}_{\mathtt{x}}){{\mathscr{H}}}(-{\mathtt{x}}-1)+({\alpha}_{{\mathit B}}-{\alpha}_{{\mathit A}})({\mathtt{u}}^{{\mathit A}_{\textrm{in}}}_{{\mathtt{x}}-1}-{\mathtt{u}}^{{\mathit A}_{\textrm{in}}}_{\mathtt{x}}){{\mathscr{H}}}(-{\mathtt{x}})\notag\\
&+({\mathit{m}}_{{\mathit B}}-{\mathit{m}}_{{\mathit A}}){\upomega}^2{\mathtt{u}}^{{\mathit A}_{\textrm{in}}}_{\mathtt{x}}{{\mathscr{H}}}(-{\mathtt{x}}-1)-({\upkappa_{s}}_{{\mathit B}}-{\upkappa_{s}}_{{\mathit A}}){\mathtt{u}}^{{\mathit A}_{\textrm{in}}}_{\mathtt{x}}{{\mathscr{H}}}(-{\mathtt{x}}-1).\label{deffinca}
\end{align}

With
${\mathfrak{s}}={{\mathit A}}, {{\mathit B}}$,
and as an analog of \eqref{discWHker},
\begin{equation}\begin{split}
{\mathfrak{K}}_{\mathfrak{s}}:={\mathit{m}}_{\mathfrak{s}}{\upomega}^2-{\upkappa_{s}}_{\mathfrak{s}}+{\alpha}_{\mathfrak{s}}({{z}}+{{z}}^{-1}-2),
\label{discWHkera}\end{split}\end{equation}
collecting the terms accompanying ${\mathtt{u}}_{\pm}$ after taking the Fourier transform (see \eqref{BCeqa}) leads to 
\begin{equation}\begin{split}
{\mathfrak{K}}_{{\mathit A}}{\mathtt{u}}^++{\mathfrak{K}}_{{\mathit B}}{\mathtt{u}}{}^-
&=
\mathtt{p}({z}){\mathtt{u}}^{{\mathit A}_{\textrm{in}}}{}^- -({\alpha}_{{\mathit A}}-{\alpha}_{{\mathit B}})(1-{{z}}){({\mathtt{u}}_{0}+{\mathtt{u}}^{{\mathit A}_{\textrm{in}}}_{0})}\\
&+(({\upkappa_{s}}_0-{\upkappa_{s}}_{{\mathit A}})-({\mathit{m}}_{0}-{\mathit{m}}_{{\mathit A}}){\upomega}^2){({\mathtt{u}}_{0}+{\mathtt{u}}^{{\mathit A}_{\textrm{in}}}_{0})},
\label{discWHa}
\end{split}\end{equation}
where (using \eqref{uincawave1})
\begin{align}
{\mathtt{u}}^{{\mathit A}_{\textrm{in}}}{}^-&={{{\mathtt{u}}^{{\mathit A}_{\textrm{in}}}_{0}}}{\updelta}_{D}^-({{z}}{{z}}_{{\mathit A}}),\quad {{z}}_{{\mathit A}}=\exp({i{{\upxi}}_{{\mathit A}}}),
\label{uincFa}\\
\mathtt{p}({z})&:=
({\alpha}_{{\mathit A}}-{\alpha}_{{\mathit B}})({{z}}+{{z}}^{-1}-2)+({\mathit{m}}_{{\mathit A}}-{\mathit{m}}_{{\mathit B}}){\upomega}^2+({\upkappa_{s}}_{{\mathit B}}-{\upkappa_{s}}_{{\mathit A}}).
\label{kAkBidentity1}
\end{align}

Note that $\mathtt{p}({z})={\mathfrak{K}}_{{\mathit A}}-{\mathfrak{K}}_{{\mathit B}}$ and ${\mathfrak{K}}_{{\mathit A}}({{z}}_{{\mathit A}}^{-1})=0$.
The factorization of $\mathtt{p}$ follows \eqref{defpppm}.
Recall \eqref{discWHkera}. 
With ${\alpha}_{\mathfrak{s}}>0,$ (assuming $|{z}_{\mathfrak{s}}|<1$)
$({\alpha}_{\mathfrak{s}})^{-1}{\mathfrak{K}}_{\mathfrak{s}}({z})=-{z}_{\mathfrak{s}}({z}_{\mathfrak{s}}^{-1}-{z})({z}_{\mathfrak{s}}^{-1}-{z}^{-1}).$
Thus,
\begin{equation}\begin{split}
{\mathfrak{K}}_{\mathfrak{s}{}+}({z})
={\mathtt{C}}_{k\mathfrak{s}}(1-{z}^{-1}{z}_{\mathfrak{s}}),
{\mathfrak{K}}_{\mathfrak{s}{}-}({z})
={\mathtt{C}}_{k\mathfrak{s}}(1-{z}{z}_{\mathfrak{s}}),\quad {\mathtt{C}}_{k\mathfrak{s}}:=\sqrt{-{z}_{\mathfrak{s}}^{-1}{\alpha}_{\mathfrak{s}}}.
\label{discWHkerfacs}
\end{split}\end{equation}
Then upon division of both sides in \eqref{discWHa} by ${\mathfrak{K}}_{{{\mathit B}}{}+}{\mathfrak{K}}_{{{\mathit A}}{}-}$, it becomes,
\begin{subequations}
\begin{align}
&{\mathfrak{L}}_+^{-1}({z}){\mathtt{u}}^+({z})+{\mathfrak{L}}_-({z}){\mathtt{u}}^-({z})={\mathfrak{C}}({z}),\label{discWHN2a}\\
\text{where }
{\mathfrak{L}}_+({z})&:=\frac{{\mathfrak{K}}_{{{\mathit B}}{}+}({z})}{{\mathfrak{K}}_{{{\mathit A}}{}+}({z})}=\frac{{\mathtt{C}}_{k{\mathit B}}(1-{z}^{-1}{z}_{{\mathit B}})}{{\mathtt{C}}_{k{\mathit A}}(1-{z}^{-1}{z}_{{\mathit A}})}, \quad
{\mathfrak{L}}_-({z}):=\frac{{\mathfrak{K}}_{{{\mathit B}}{}-}({z})}{{\mathfrak{K}}_{{{\mathit A}}{}-}({z})}
=\frac{{\mathtt{C}}_{k{\mathit B}}(1-{z}{z}_{{\mathit B}})}{{\mathtt{C}}_{k{\mathit A}}(1-{z}{z}_{{\mathit A}})},\label{discWHLfacs}\\
{\mathfrak{C}}({z})&:=({\mathfrak{L}}_--{\mathfrak{L}}_+^{-1})(-{{{\mathtt{u}}^{{\mathit A}_{\textrm{in}}}_{0}}}{\updelta}_{D}^-({{z}}{{z}}_{{\mathit A}})
+\frac{({\alpha}_{{\mathit A}}-{\alpha}_{{\mathit B}})(1+\mathtt{M}_0-{{z}})}{\mathtt{p}_+({z})\mathtt{p}_-({z})}{({\mathtt{u}}_{0}+{\mathtt{u}}^{{\mathit A}_{\textrm{in}}}_{0})}),\label{discWHdefCone}\\
\mathtt{M}_0&:=\frac{-(({\upkappa_{s}}_0-{\upkappa_{s}}_{{\mathit A}})-({\mathit{m}}_{0}-{\mathit{m}}_{{\mathit A}}){\upomega}^2)}{({\alpha}_{{\mathit A}}-{\alpha}_{{\mathit B}})}.\label{defAratone}
\end{align}
\end{subequations}

\begin{remark}

In the simpler situation
assuming ${\alpha}_{{\mathit A}}={\alpha}_{{\mathit B}}$ and ${\mathit{m}}_{0}={\mathit{m}}_{{\mathit A}}$ but ${\mathit{m}}_{{\mathit A}}\ne{\mathit{m}}_{{\mathit B}}$,
(reall Remark \ref{speccase2}) and assuming ${\upkappa_{s}}_{0}={\upkappa_{s}}_{{\mathit A}}$,
using \eqref{discWHNsolhalf}, \eqref{discCpmhalfcase2}, \eqref{discWHkerfacs}, and \eqref{discWHLfacs},
\begin{align}
{\mathtt{u}}^+
&={{{\mathtt{u}}^{{\mathit A}_{\textrm{in}}}_{0}}}\mathcal{C}_{{\mathit A}_{\textrm{in}}}^{{\mathit A}}{\updelta}_{D}^+({{z}}{{z}}_{{\mathit A}}^{-1}), 
\quad
\mathcal{C}_{{\mathit A}_{\textrm{in}}}^{{\mathit A}}
=\frac{-({z}_{{\mathit A}}-{z}_{{\mathit B}})}{({z}_{{\mathit A}}^{-1}-{z}_{{\mathit B}})},\label{discWHNsolhalf2p}\\
{\mathtt{u}}^-
&=-{\mathtt{u}}^{{\mathit A}_{\textrm{in}}}{}^-({{z}})+{{{\mathtt{u}}^{{\mathit A}_{\textrm{in}}}_{0}}}\mathcal{C}_{{\mathit A}_{\textrm{in}}}^{{\mathit B}}{\updelta}_{D}^-({{z}}{{z}}_{{\mathit B}}), 
\quad
\mathcal{C}_{{\mathit A}_{\textrm{in}}}^{{\mathit B}}
={{z}}_{{\mathit A}}{z}_{{\mathit B}}^{-1}\frac{{\mathtt{C}}^2_{k{\mathit A}}}{{\mathtt{C}}^2_{k{\mathit B}}}\frac{({z}_{{\mathit A}}^{-1}-{z}_{{\mathit A}})}{({z}_{{\mathit A}}^{-1}-{z}_{{\mathit B}})},
\label{discWHNsolhalf2n}
\end{align}
employing the definitions analogous to \eqref{defzazb}.
Further, since ${\mathtt{C}}^2_{k{\mathit A}}=-{z}_{{\mathit A}}^{-1}{\alpha}_{{\mathit A}}$ (recall ${\alpha}_{{\mathit B}}={\alpha}_{{\mathit A}}$, \eqref{discWHkerfacs}${}_3$), it is found that \eqref{discWHNsolhalf2n}${}_2$ simplifies to
$\mathcal{C}_{{\mathit A}_{\textrm{in}}}^{{\mathit B}}
=\frac{({z}_{{\mathit A}}^{-1}-{z}_{{\mathit A}})}{({z}_{{\mathit A}}^{-1}-{z}_{{\mathit B}})}.$\\
The exact solution is
\begin{equation}\begin{split}
{\mathtt{u}}_{\mathtt{x}}=(\exp({-i {\upxi}_{{\mathit A}} \mathtt{x}})+\frac{{z}_{{\mathit B}}-{z}_{{\mathit A}}}{{z}_{{\mathit A}}^{-1}-{z}_{{\mathit B}}} \exp({i {\upxi}_{{\mathit A}} \mathtt{x}})){{\mathscr{H}}}(\mathtt{x})+\frac{{z}_{{\mathit A}}^{-1}-{z}_{{\mathit A}}}{{z}_{{\mathit A}}^{-1}-{z}_{{\mathit B}}} \exp({-i {\upxi}_{{\mathit B}} \mathtt{x}}){{\mathscr{H}}}(-\mathtt{x}-1).
\end{split}\end{equation}
\label{speccaseone}
\end{remark}
It is useful to compare the above expression \eqref{discWHdefCone} of the right hand side in {Wiener-Hopf} equation, i.e. ${\mathfrak{C}}$, with that in half space \eqref{discWHdefC}; in other words the factorization \eqref{discCfacalleq} 
continues to hold in this form.
Also compare the above definition \eqref{defAratone} of $\mathtt{M}_0$ with that in half space \eqref{defArat}.
The expression \eqref{u00val} of ${\mathtt{u}}_{0,0}$ is found by substitution of \eqref{discWHLfacs} as
\begin{equation}\begin{split}
{({\mathtt{u}}_{0, 0}+{\mathtt{u}}^{{\mathit A}_{\textrm{in}}}_{0, 0})}
&=\frac{{\alpha}_{{\mathit A}} {z}_{{\mathit B}} ({z}_{\mathtt{p}}^2-1) \sqrt{-\frac{{\alpha}_{{\mathit B}}}{{z}_{{\mathit B}}}} ({z}_{{\mathit A}} {z}_{\mathtt{p}}-1) ({z}_{{\mathit B}} {z}_{\mathtt{p}}-1)}{\sqrt{-\frac{{\alpha}_{{\mathit A}}}{{z}_{{\mathit A}}}} ({\alpha}_{{\mathit A}} {z}_{{\mathit B}} ((\mathtt{M}_0+1){z}_{\mathtt{p}}-1) ({z}_{{\mathit A}} {z}_{\mathtt{p}}-1)^2-{\alpha}_{{\mathit B}}{z}_{{\mathit A}} {z}_{\mathtt{p}} (\mathtt{M}_0-{z}_{\mathtt{p}}+1) ({z}_{{\mathit B}} {z}_{\mathtt{p}}-1)^2)}.
\label{u0valone}
\end{split}\end{equation}
The expression \eqref{CBhalfincA} (recall \eqref{defpppm})
can be simplified to obtain \eqref{CABhalfincAone}${}_1$.
and \eqref{u0valone} simplifies to ${({\mathtt{u}}_{0, 0}+{\mathtt{u}}^{{\mathit A}_{\textrm{in}}}_{0, 0})}=\mathcal{C}_{{\mathit A}_{\textrm{in}}}^{{\mathit B}}$.
The expression \eqref{CAhalfincA} 
can be also simplified to obtain \eqref{CABhalfincAone}${}_2$.

\end{document}